\documentclass[aps,prb,reprint,onecolumn,11pt,groupedaddress]{revtex4-2}
\usepackage[colorlinks=true,linkcolor=blue,citecolor=blue,urlcolor=blue]{hyperref}
\usepackage{amsmath}
\usepackage{amssymb} 
\usepackage[pdftex]{graphicx}
\usepackage{braket} 

\newcounter{firstbib}

\begin{document}


\title{Approaching a fully-polarized state of nuclear spins in a semiconductor quantum dot}



\author{Peter Millington-Hotze}
\affiliation{Department of Physics and Astronomy, University of
Sheffield, Sheffield S3 7RH, United Kingdom}
\author{Harry E. Dyte}
\affiliation{Department of Physics and Astronomy, University of
Sheffield, Sheffield S3 7RH, United Kingdom}
\author{Santanu Manna}
\affiliation{Institute of Semiconductor and Solid State Physics,
Johannes Kepler University Linz, Altenberger Str. 69, 4040 Linz,
Austria}
\author{Saimon F. Covre da Silva}
\affiliation{Institute of Semiconductor and Solid State Physics,
Johannes Kepler University Linz, Altenberger Str. 69, 4040 Linz,
Austria}
\author{Armando Rastelli}
\affiliation{Institute of Semiconductor and Solid State Physics,
Johannes Kepler University Linz, Altenberger Str. 69, 4040 Linz,
Austria}
\author{Evgeny A. Chekhovich}
\email[]{e.chekhovich@sheffield.ac.uk} \affiliation{Department of
Physics and Astronomy, University of Sheffield, Sheffield S3 7RH,
United Kingdom}

\date{\today}

\begin{abstract}
Magnetic noise of atomic nuclear spins is a major problem for solid state spin qubits. Highly-polarized nuclei would not only overcome this obstacle, but also make nuclear spins a useful quantum information resource. However, achieving sufficiently high nuclear polarizations has remained an evasive goal. Here we implement a nuclear spin polarization protocol which combines strong optical pumping and fast electron tunneling. Polarizations well above 95\% are generated in GaAs semiconductor quantum dots on a timescale of 1~minute. The technique is compatible with standard quantum dot device designs, where highly-polarized nuclear spins can simplify implementations of quantum bits and memories, as well as offer a testbed for studies of many-body quantum dynamics and magnetism.
\end{abstract}


\maketitle

\begin{figure*}
\includegraphics[width=0.7\linewidth]{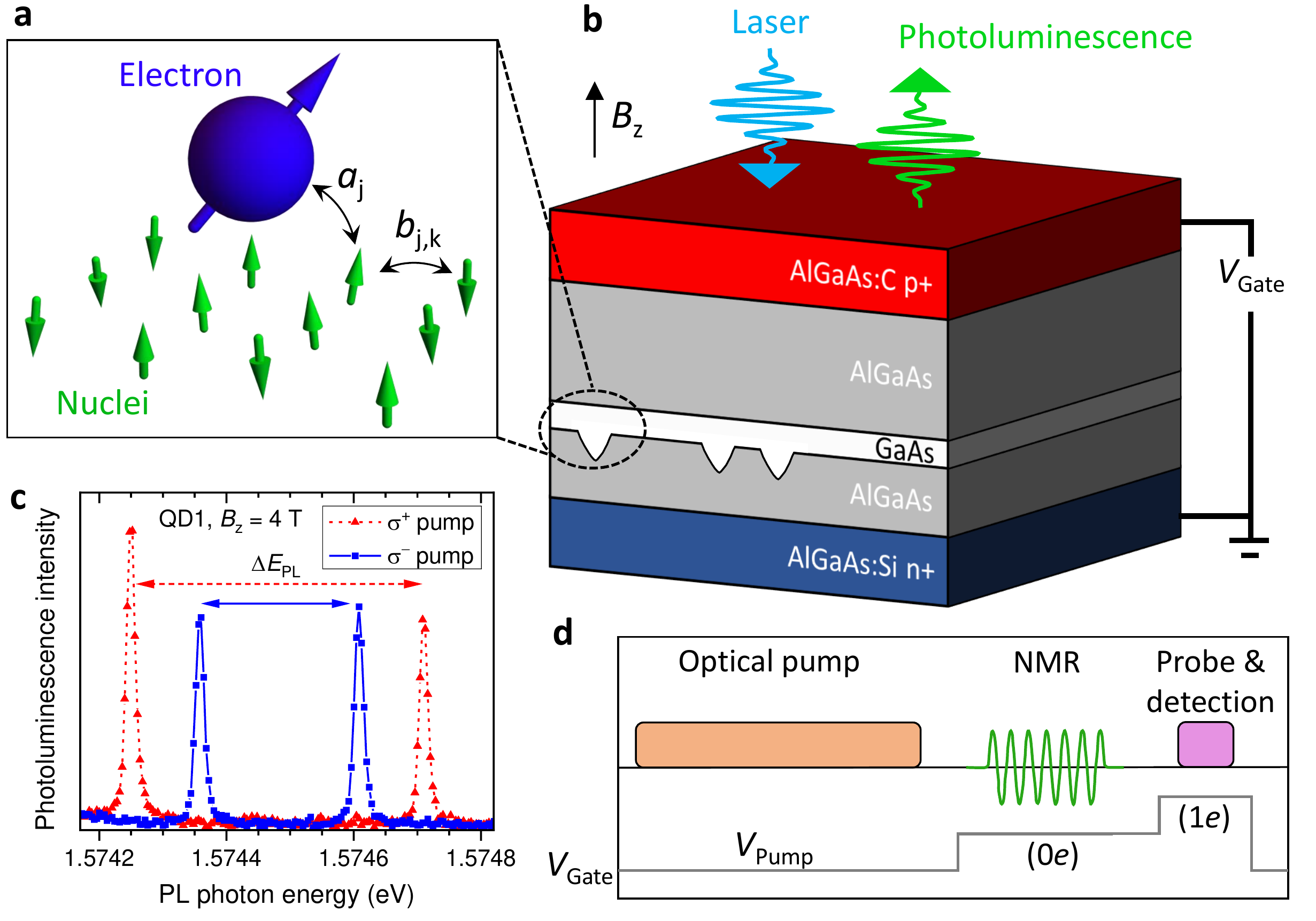}
\caption{\textbf{Optical control of quantum dot nuclear spins.} \textbf{a} Schematic of a central electron spin and an ensemble of nuclear spins coupled through hyperfine interaction with constants $a_j$. The nuclei are coupled through dipolar interaction with pairwise constants $b_{{j,k}}$ (see Supplementary Note 2).  \textbf{b} Schematic cross-section of a $p-i-n$ diode with embedded epitaxial GaAs quantum dots. Laser excitation, photoluminescence collection and external magnetic field are directed along the sample growth axis $z$. Doped semiconductor layers are used to apply the gate bias $V_{\rm{Gate}}$, resulting in a tunable electric field along $z$. \textbf{c} Typical photoluminescence spectra of a negatively charged trion $X^-$ in an individual QD. The spectral splitting $\Delta E_{\rm{PL}}$ depends both on $B_{\rm{z}}$ and the helicity of the optical pumping ($\sigma^\pm$) due to the buildup of the nuclear spin polarization. \textbf{d} Experimental cycle consisting of nuclear spin optical pumping, nuclear magnetic resonance (NMR) excitation, and optical probing of the photoluminescence spectrum. $V_{\rm{Gate}}$ is varied to switch the QD state between electron-charged (1$e$) and neutral (0$e$) or is set to an arbitrary value $V_{\rm{Pump}}$ during pumping.} \label{fig:Intro}
\end{figure*}

\newcommand{\FigSpins}{Fig.~\ref{fig:Intro}a}
\newcommand{\FigQD}{Fig.~\ref{fig:Intro}b}
\newcommand{\FigPLOHS}{Fig.~\ref{fig:Intro}c}
\newcommand{\FigTDiagr}{Fig.~\ref{fig:Intro}d}

\begin{figure*}
\includegraphics[width=0.7\linewidth]{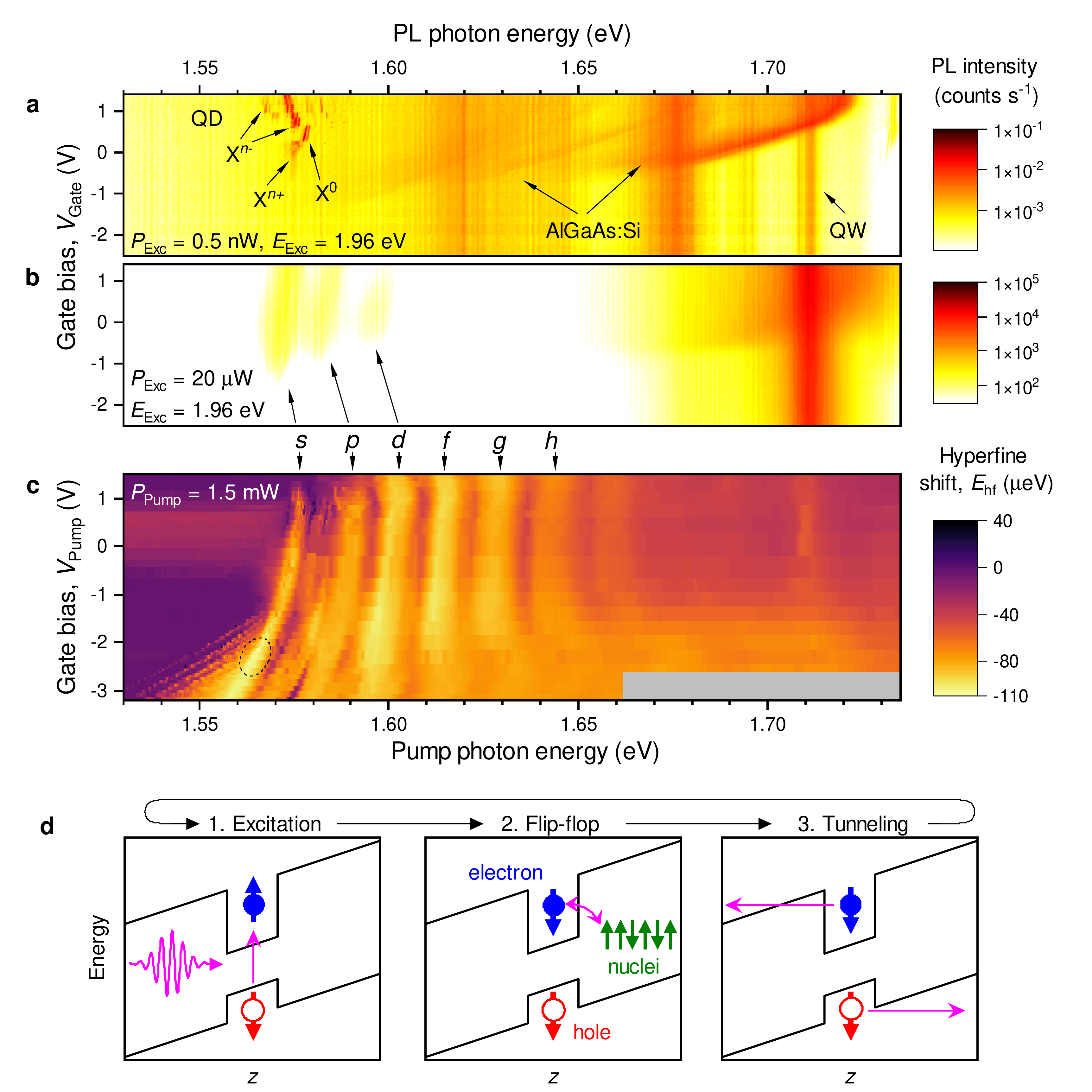}
\caption{\textbf{Tunneling-assisted optical nuclear spin pumping.} \textbf{a} Bias-dependent photoluminescence spectra of an individual dot QD1 measured at $B_{\rm{z}}=10$~T at low excitation power $P_{\rm{Exc}}=0.5$~nW and excitation photon energy $E_{\rm{Exc}}=1.96$~eV. Spectral features are attributed to: neutral (X$^{0}$), positively (X$^{n+}$) and negatively (X$^{n-}$) multi-charged QD excitons, Si doped AlGaAs, and the quantum well (QW). \textbf{b} Photoluminescence spectra at an increased power $P_{\rm{Exc}}=20~\mu$W reveal saturated QD exciton shell emission, labeled $s$, $p$, $d$. \textbf{c} Hyperfine shift measured in QD1 as a function of sample bias $V_{\rm{Pump}}$ and the photon energy $E_{\rm{Pump}}$ of the optical pump with power $P_{\rm{Pump}}=1.5$~mW. Grey shows parameter regions where no data has been measured. Excitonic spectral features are labeled up to the $h$ shell. The dashed ellipse highlights the parameters that result in the most efficient nuclear spin polarization. \textbf{d} Schematic of the conduction band electron (full circles) and valence band hole (open circles) energies along the $z$ direction. The three stages of the cyclic nuclear spin pumping process are shown schematically.} \label{fig:WlDep}
\end{figure*}

\newcommand{\FigPLLowP}{Fig.~\ref{fig:WlDep}a}
\newcommand{\FigPLHighP}{Fig.~\ref{fig:WlDep}b}
\newcommand{\FigOHSEV}{Fig.~\ref{fig:WlDep}c}
\newcommand{\FigDNPCycle}{Fig.~\ref{fig:WlDep}d}

\begin{figure*}
\includegraphics[width=0.95\linewidth]{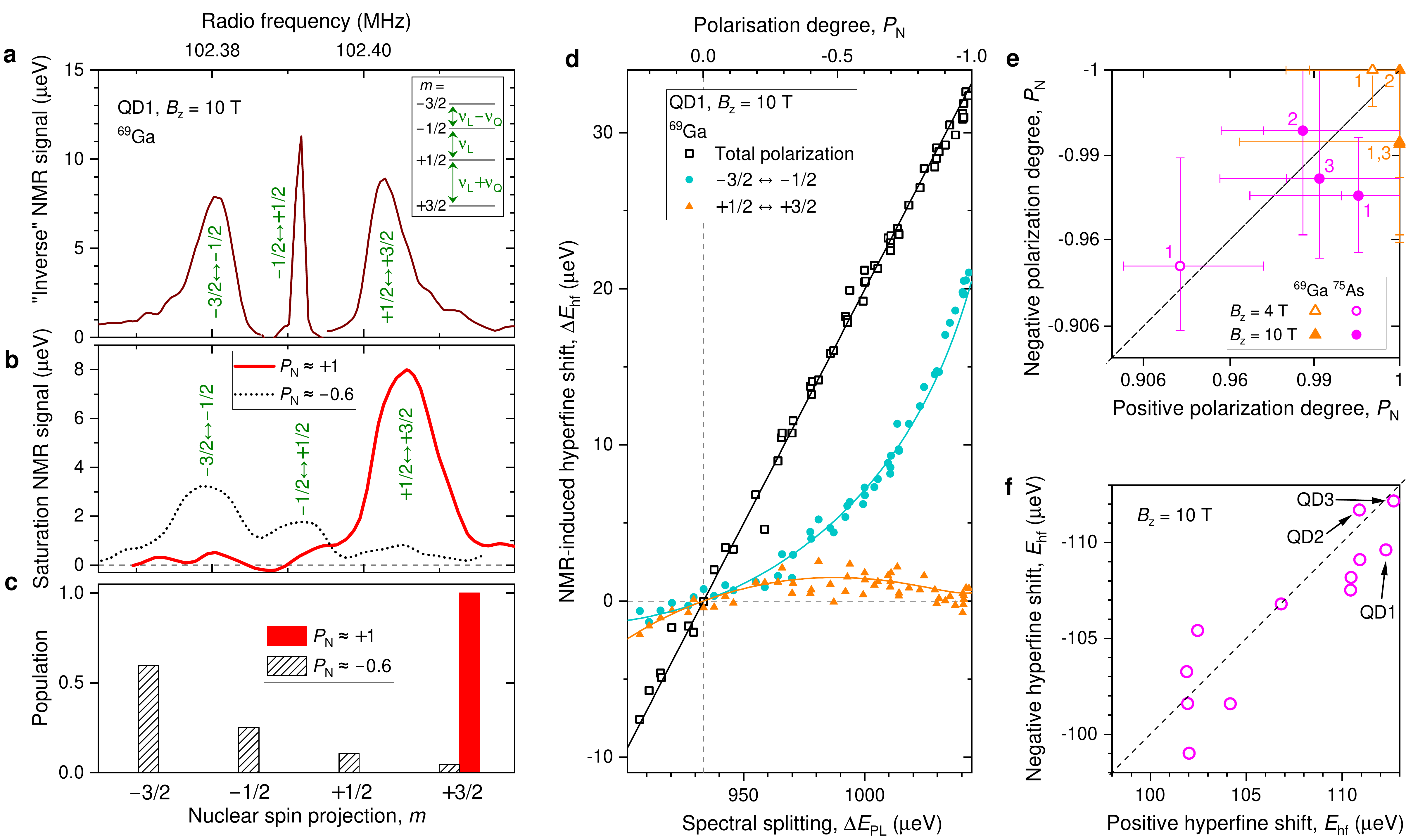}
\caption{\textbf{Nuclear magnetic resonance spin thermometry.} \textbf{a} High-resolution spectrum of $^{69}$Ga measured in QD1 at $B_{\rm{z}}=10$~T using ``inverse NMR'' signal enhancement technique \citep{Chekhovich2012}. Inset shows energy levels of a spin-3/2 nucleus. The resonant frequency of the central transition between $m=\pm1/2$ is $\nu_{\rm{L}}$, whereas the satellite transitions involving $m=\pm3/2$ are split off by the quadrupolar shifts $\pm\nu_{\rm{Q}}$. \textbf{b} Low-resolution spectrum of the same QD1, but measured using the saturation technique in order to reveal the population probabilities of the nuclear spin levels. \textbf{c} Populations of spin levels with different projection $m=\pm1/2,\pm3/2$, sketched for the two different levels of optical nuclear spin polarization corresponding to the data in (b). \textbf{d} Hyperfine shift variation arising from selective manipulation of the $^{69}$Ga nuclear spin plotted against the photoluminescence spectral splitting $\Delta E_{\rm{PL}}$ in measurements where the degree of optical nuclear spin pumping is varied. Squares show the total $^{69}$Ga hyperfine shift measured by broadband saturation of the entire NMR triplet, which equalizes populations $p_m$ for all $m$. Circles and triangles show the selective signals of the $\pm 1/2\leftrightarrow\pm 3/2$ resonances measured via frequency-swept adiabatic inversion. Lines show fitting, from which nuclear spin polarization degree is derived and plotted in the top horizontal scale (see Supplementary Note 5). \textbf{e} Maximum positive and minimum negative nuclear spin polarization degrees $P_{\rm{N}}$ derived for $^{69}$Ga (triangles) and $^{75}$As (circles) in individual dots QD1 - QD3 at $B_{\rm{z}}=10$~T (solid symbols and QD numbers) and $B_{\rm{z}}=4$~T (open symbols). The nonlinear scale $\propto\sqrt{1-P_{\rm{N}}^2}$ is used to highlight the areas around $\vert P_{\rm{N}}\vert\approx 1$. Error bars are 95\% confidence intervals. \textbf{f} Maximum positive and minimum negative hyperfine shifts measured on individual dots QD1 - QD12 at $B_{\rm{z}}=10$~T. } \label{fig:Temp}
\end{figure*}

\newcommand{\FigNMRInv}{Fig.~\ref{fig:Temp}a}
\newcommand{\FigNMRSat}{Fig.~\ref{fig:Temp}b}
\newcommand{\FigNMRpm}{Fig.~\ref{fig:Temp}c}
\newcommand{\FigNMRTemp}{Fig.~\ref{fig:Temp}d}
\newcommand{\FigPN}{Fig.~\ref{fig:Temp}e}
\newcommand{\FigEhf}{Fig.~\ref{fig:Temp}f}

\begin{figure*}
\includegraphics[width=0.99\linewidth]{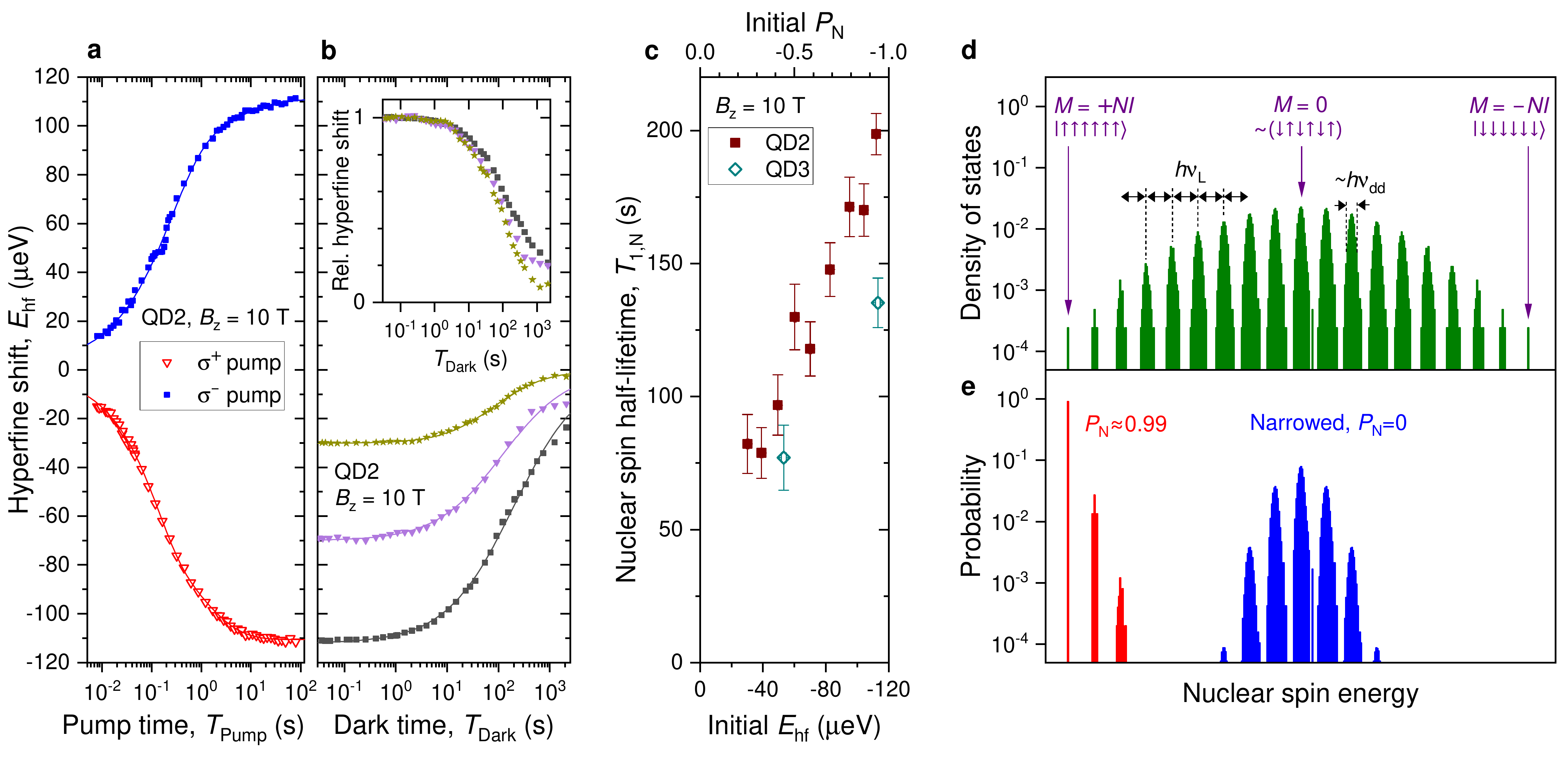}
\caption{\textbf{Nuclear spin dynamics.} \textbf{a} Nuclear spin buildup dynamics measured (symbols) in an individual QD2 at $B_{\rm{z}}=10$~T under $\sigma^+$ (triangles) and $\sigma^-$ (squares) optical pumping. Lines show biexponential fitting. \textbf{b} Nuclear spin relaxation dynamics in the dark measured in a neutral state following $\sigma^+$ optical pumping (squares). The same relaxation dynamics are also measured with partial NMR saturation after the optical pump, which reduces the initial $P_{\rm{N}}$ (triangles and stars). Lines show fitting used to derive the nuclear spin half-lifetimes $T_{\rm{1,N}}$. Inset shows the same data, but normalized by the hyperfine shift at short dark times. \textbf{c} Nuclear spin relaxation times $T_{\rm{1,N}}$ as a function of the initial hyperfine shift $E_{\rm{hf}}$. The corresponding approximate initial $P_{\rm{N}}$ is shown on the top axis. Error bars are 95\% confidence intervals. \textbf{d} Density of states calculated for $N=6$ dipolar-coupled $I=3/2$ nuclei (without the electron). Each band, broadened by dipolar couplings $h\nu_{\rm{dd}}\propto \max{\vert b_{{j,k}}\vert}$, corresponds to a well-defined total spin projection $M$. The adjacent bands are split by the Zeeman energy $h\nu_{\rm{L}}$. \textbf{e} Population probability of the eigenstates, calculated for the spectrum in (d) and for two types of mixed states: Boltzmann distribution of Zeeman energies with high polarization $P_{\rm{N}}\approx0.99$ (red) and a narrowed Gaussian distribution with $P_{\rm{N}}=0$ (blue). } \label{fig:Dyn}
\end{figure*}

\newcommand{\FigBuildup}{Fig.~\ref{fig:Dyn}a}
\newcommand{\FigDec}{Fig.~\ref{fig:Dyn}b}
\newcommand{\FigDecDep}{Fig.~\ref{fig:Dyn}c}
\newcommand{\FigDOS}{Fig.~\ref{fig:Dyn}d}
\newcommand{\FigNarrowed}{Fig.~\ref{fig:Dyn}e}

The capability of initializing a quantum system into a well-defined eigenstate is one of the fundamental requirements in quantum science and technology. This has been demonstrated for individual and dilute nuclear spins in the solid state \citep{Jacques2009,Falk2015},  but remains a long-standing challenge for dense three-dimensional lattices of nuclear spins. The spin-ensemble ground state is characterized by a polarization degree $P_{\rm{N}}=\pm 100\%$, which is equivalent to absolute zero spin temperature. Very high polarizations, $P_{\rm{N}}\approx 95-99\%$, have been demonstrated in bulk materials through brute-force cooling of the lattice, but the cooling cycle may take hours and even days \citep{Reichertz1994,Knuuttila2001}. More scalable approaches seek to use individual or dilute electron spins to polarize the dense nuclear ensembles. Microwave pumping of paramagnetic impurities in bulk solids \citep{Jacquinot1974,Goldman1976} provides polarizations up to $P_{\rm{N}}\approx 80-90\%$. In semiconductor nanostructures, $P_{\rm{N}}\approx 50-80\%$ is achieved either through electronic transport \citep{Petersen2013} or optical excitation \citep{Chekhovich2017}. At such polarization degrees, nuclear spin fluctuations are still similar to their thermal-equilibrium maximum, with $\sqrt{1-P_{\rm{N}}^2}$ being the figure of merit in reducing the nuclear spin noise \citep{Kloeffel2013}. Therefore, different techniques are needed to approach a fully-polarized nuclear state.

Extensive theoretical studies have been conducted to understand what limits nuclear spin pumping in a central-spin scenario, where the electron can be polarized on demand, while the ensemble of $N$ nuclei can only be accessed through hyperfine (magnetic) coupling with that central electron (\FigSpins). The formation of coherent ``dark'' states \citep{ImamogluPRL2003} has been shown to suppress the transfer of spin from the electron to nuclei \citep{Christ2007}. Thus an open question remains -- is it possible, even in principle, to reach a fully-polarized nuclear state in a real central-spin system?

Here we work with epitaxial GaAs/AlGaAs quantum dots (QDs) and use optical techniques to polarize nuclear spins. While the optical method is well known, its bottleneck is the slow, nanosecond-scale, recombination of the photo-generated electrons. We resolve this issue by introducing charge transport -- optical recombination is replaced with fast sub-picosecond electron tunneling. Moreover, no ``dark''-state limitation occurs, which we also attribute to the extremely short lifetime of the electron spin. As a result, we achieve nuclear polarization degrees well above $P_{\rm{N}}>95\%$. The maximum polarizations vary between individual QDs, which we ascribe to random QD anisotropies. For the best dots we derive $P_{\rm{N}}\gtrsim99\%$, limited only by the accuracy of the existing measurement techniques. These high polarizations surpass the predicted $P_{\rm{N}}\gtrsim90\%$ threshold for achieving non-trivial regimes, manifested in extended electron spin qubit coherence \citep{Khaetskii2002,Deng2008}, quantum memory operation \citep{Giedke2006,Deng2008}, superradiant electron-nuclear spin dynamics \citep{Kessler2010,Schuetz2012}, as well as magnetic-ordering phase transition \citep{Oja1997,Kotur2021}.

The semiconductor device, sketched in \FigQD, is a $p-i-n$ diode with epitaxial GaAs QDs embedded into the AlGaAs barrier layers (see Supplementary Note 1). By changing the gate bias $V_{\rm{Gate}}$ it is possible to charge the QD with individual resident electrons \citep{Miller1997,Warburton2000} and apply electric field. Each individual QD contains $N\approx10^5$ nuclei, with the three abundant isotopes $^{75}$As, $^{69}$Ga and $^{71}$Ga, all possessing spin momentum $I=3/2$. The sample is cooled to $\approx4.25$~K and placed in a magnetic field $B_{\rm{z}}$ parallel to electric field and sample growth direction (see Supplementary Note 3). Thanks to the selection rules \citep{Urbaszek2013}, optical laser excitation can create spin-polarized electron-hole pairs: photons with $+1$ ($-1$) angular momentum (in units of $\hbar$), corresponding to a $\sigma^+$ ($\sigma^-$) polarized beam, generate electrons with spin projection $s_{\rm{z}}=-1/2$ ($s_{\rm{z}}=+1/2$). Owing to the electron-nuclear hyperfine interaction (\FigSpins), a polarized electron can transfer its spin to one of the nuclei and, through repeated optical pumping, induce a substantial polarization $\vert P_{\rm{N}} \vert$. Conversely, the energy of the photon emitted from electron-hole recombination depends on the mutual alignment of $s_{\rm{z}}$ and the total magnetic field, which is a sum of $B_{\rm{z}}$ and the effective field of the polarized nuclear spins. The resulting optical spectrum is a doublet (\FigPLOHS), whose splitting $\Delta E_{\rm{PL}}$ is used as a sensitive probe of the nuclear spin polarization state. We define the exciton hyperfine shift $E_{\rm{hf}}=-(\Delta E_{\rm{PL}}-\Delta E_{\rm{PL,0}})$, where $\Delta E_{\rm{PL,0}}$ is the splitting measured for depolarized nuclei ($P_{\rm{N}}\approx0$).

The high resolution optical spectra (\FigPLOHS), required for accurate measurement of $E_{\rm{hf}}$, can only be observed for a narrow range of sample biases and optical excitation powers. In order to cover a wide range of pumping parameters we use a pump-probe technique (\FigTDiagr). We maximize the hyperfine shift $\vert E_{\rm{hf}} \vert$ by optimizing the following four parameters: the elliptical polarization of the optical pump, its power $P_{\rm{Pump}}$, photon energy $E_{\rm{Pump}}$ and the sample bias $V_{\rm{Pump}}$ during pumping. The corresponding results are interpreted with reference to the broad range photoluminescence spectra shown in Figs.~\ref{fig:WlDep}a, b. \FigPLLowP\: shows low power spectra, which reveal a well-known bias-controlled charging of the ground state ($s$-shell) exciton \citep{Warburton2000}. High optical power (\FigPLHighP) broadens the spectra, also populating the higher shells $p$ and $d$ \citep{Raymond2004,Babinski2006}. (See additional data in Supplementary Note 4.)

The dependence of $E_{\rm{hf}}$ on $E_{\rm{Pump}}$ and $V_{\rm{Pump}}$, shown in \FigOHSEV, reveals spectral bands that match the excitonic shells in \FigPLHighP, demonstrating that nuclear spin pumping proceeds through resonant optical driving of the QD exciton transitions. The largest $\vert E_{\rm{hf}} \vert$ is observed when resonant with the $s$ shell ($E_{\rm{Pump}}\approx1.565$~eV), and at a large reverse bias $V_{\rm{Pump}}=-2.3$~V, where photoluminescence is completely quenched. Moreover, the optimal pump laser power $P_{\rm{Pump}}=1.5$~mW is five orders of magnitude higher than the $s$-shell saturation power. Based on these observations, the nuclear spin pumping effect can be understood as a cyclic process sketched in \FigDNPCycle. First, circularly-polarized resonant optical excitation creates a spin-polarized electron-hole pair in the quantum dot. Then, the electron has a small but finite probability to undergo a flip-flop with one of the nuclei, increasing the ensemble polarization $\vert P_{\rm{N}} \vert$. Finally, in order to proceed to the next cycle, the electron is removed through tunneling. The tunneling time, estimated from bias-dependent photoluminescence in Supplementary Note 4, is $\lesssim 0.1$~ps, much shorter than the $\approx300$~ps radiative recombination time \citep{Schimpf2019}. The combination of high-power optical pumping and fast tunnel escape results in rapid cycling. This in turn leads to a high rate of nuclear spin pumping, which helps to outpace the inevitable nuclear spin relaxation. The cycling time is also much shorter than the period of coherent electron precession  $\gtrsim20$~ps, ensuring the spin-flipped electrons are removed before they can undergo a reverse flip-flop \citep{Taylor2003}. The ultimate result is a large steady-state $\vert E_{\rm{hf}} \vert$.  

Although the hyperfine shift $E_{\rm{hf}}$ scales linearly with polarization degree $P_{\rm{N}}$, its absolute value depends on the QD structure. The electron wavefunction leaks into the barriers where the fraction of Ga atoms replaced with Al atoms is not known precisely. A more reliable measurement of the $P_{\rm{N}}$ is achieved through nuclear magnetic resonance (NMR) spin thermometry (see Supplementary Note 5 for details). The method rests on the assumption that the probability $p_{m}$ for each nucleus to occupy a state with spin projections $m$ follows the Boltzmann distribution $p_m \propto e^{m \beta}$, where $\beta=h\nu_{\rm{L}}/k_\textrm{b} T_\textrm{N}$ is the dimensionless inverse spin temperature, expressed in terms of nuclear spin Larmor frequency $\nu_{\rm{L}}$ and spin temperature $T_\textrm{N}$ ($h$ is the Planck's constant and $k_\textrm{b}$
is the Boltzmann constant). For spin $I$=1/2 where $m=\pm1/2$, any statistical distribution has the Boltzmann form. By contrast, for $I>$1/2 Boltzmann distribution expresses the non-trivial nuclear spin temperature hypothesis \cite{GoldmanBook}, which was verified for epitaxial GaAs quantum dots previously \citep{Chekhovich2017}.

In order to perform spin thermometry, we first measure the single-QD NMR spectra \citep{Chekhovich2012}, as exemplified in \FigNMRInv\: for $^{69}$Ga spins. The three magnetic-dipole transitions of the 3/2 spins are well resolved thanks to the  quadrupolar shifts $\nu_{\rm{Q}}$, which originate from the lattice mismatch of GaAs and AlGaAs. On the other hand, these quadrupolar effects are too small to impede nuclear spin cooling -- a significant advantage over the highly-strained Stranski-Krastanov QDs, where quadrupolar shifts are large and disordered \citep{Chekhovich2012}. The resolved NMR triplet is essential, as it allows $\beta$ to be derived from the Boltzmann exponent, which then relates to $P_{\rm{N}}$ through the standard Brillouin function. Qualitatively this is demonstrated in \FigNMRSat\: with simple saturation NMR spectroscopy \citep{Bloch1946}. At moderate polarization $P_{\rm{N}}\approx-0.6$ (dashed lines) all three magnetic-dipole transitions $m \leftrightarrow m+1$ are observed, and their amplitudes are proportional to the differences $\vert p_{m+1}-p_m \vert$ (\FigNMRpm). At the maximum positive polarization (solid line) a single NMR peak $+1/2\leftrightarrow +3/2$ is observed, indicating that nearly all spins have been cooled to the $m=+3/2$ state. By changing the helicity of the optical pump it is possible to cool the nuclei towards the $m=-3/2$ state.

For quantitative spin thermometry we measure the peak areas of the $-3/2\leftrightarrow -1/2$ and $+1/2\leftrightarrow +3/2$ NMR transitions at different initial polarizations of the nuclei, quantified by $\Delta E_{\rm{PL}}$. The results are shown in \FigNMRTemp\: (circles and triangles), together with the total signal obtained by saturating all three NMR transitions (squares). Fitting with Boltzmann model is shown by the lines, together with the derived polarization degree $P_{\rm{N}}$ in the top axis. The model reproduces well both the linear dependence of the total NMR signal and the non-linear dependencies of the selective $\pm1/2\leftrightarrow \pm3/2$ signals, revealing a close approach to $P_{\rm{N}}\approx-1$. Qualitatively, at $P_{\rm{N}}=-1$ the $m=+1/2,+3/2$ states must be depopulated, resulting in a vanishing $+1/2\leftrightarrow +3/2$ signal, as indeed observed experimentally. Moreover, at $P_{\rm{N}}=-1$ the $-3/2\leftrightarrow -1/2$ signal must be $2/3$ of the total NMR signal, also in good agreement with experiment.

The largest positive and negative $P_{\rm{N}}$ derived from spin thermometry on individual dots QD1 - QD3, chosen for their highest $\vert E_{\rm{hf}} \vert$, are shown in \FigPN. At the highest static field $B_{\rm{z}}=10$~T the best fit estimates for $^{69}$Ga are around $\vert P_{\rm{N}} \vert \approx0.99$, with somewhat lower $\vert P_{\rm{N}} \vert \approx0.98$ for $^{75}$As. Spin thermometry conducted at $B_{\rm{z}}=4$~T for one of the QDs also yields high polarizations, although the measurement accuracy is reduced due to the less efficient optical probing.

A simpler measurement of the largest positive and negative $E_{\rm{hf}}$ is shown in \FigEhf\: for 12 randomly chosen dots. For some QDs, nuclear polarization is reduced to $P_{\rm{N}}\approx0.9$. We also observe for all studied QDs that the optimal optical polarization of the pump is not circular, having a randomly-oriented linearly-polarized contribution ranging between 0 and 0.4. This points to in-plane anisotropy of QDs. From a control measurement, with magnetic field tilted by $\approx12^{\circ}$ away from the growth axis, we find a reduction in maximum $\vert E_{\rm{hf}} \vert$. A reduction is also found in a piece of the same QD structure subject to a uniaxial stress in the sample plane. Therefore, the reduced $P_{\rm{N}}$ in some QDs is attribute to the random anisotropy of the confining potential or strain. Low-symmetry confinement is known to result in heavy-light hole mixing \citep{Huo2014,Csontosova2020}, which may explain why optimal electron and nuclear spin pumping requires elliptically-polarized light.

The buildup dynamics, measured under optimal nuclear spin pumping, are shown in \FigBuildup. The approach to the steady state is non-exponential since the nuclei that are further away from the center of the QD are less coupled to the electron and take longer to polarize. It takes on the order of $\approx60$~s to reach the steady-state $P_{\rm{N}}$ within the measurement accuracy. Once optical pumping is switched off, nuclear spins depolarize in the dark (squares in \FigDec) on a timescale of minutes, mainly through spin diffusion \citep{MillingtonHotze2022}. Such long lifetimes mean that a highly-polarized nuclear spin state can be prepared and used for the subsequent fast (nanosecond) control of the electron spin qubit. We further examine the effect of the initial $P_{\rm{N}}$ on the relaxation dynamics by augmenting the optically-pumped nuclear state with a short partially-depolarizing NMR pulse (triangles and stars  in \FigDec). When normalized by the initial polarization, the plot reveals accelerated nuclear spin relaxation under reduced initial polarization (inset in \FigDec). This is quantified in \FigDecDep, where at high polarization the nuclear spin relaxation half-lifetime $T_{\rm{1,N}}$ is seen to be a factor of $\approx 2-3$ larger than in case of low initial polarization (lowest initial polarization is limited by the accuracy of the $T_{\rm{1,N}}$ measurement). This is a non-trivial result: the spin diffusion model, as well as non-diffusion relaxation mechanisms are linear, so that scaling of initial $P_{\rm{N}}$ should not change $T_{\rm{1,N}}$.

In order to explain the non-linear relaxation, we consider the eigenspectrum of a nuclear spin ensemble, with an example shown in \FigDOS\: for $N=6$ spins $I=3/2$. The adjacent bands, separated by the Zeeman energy, typically $h\nu_{\rm{L}}\approx1 - 100$~MHz, correspond to a flip of a single nucleus, which changes the total ensemble spin projection $M$ by $\pm1$. Each band consists of all possible superpositions with a given $M$, with degeneracy lifted by the nuclear-nuclear dipolar magnetic interaction. For  $M\approx0$ (i.e. $P_{\rm{N}}\approx0$) the broadening of each band is maximal, characterized by the dipole-dipole energy $h\nu_{\rm{dd}}\approx1$~kHz. In the opposite limit, there are only two non-degenerate fully-polarized states with $M=\pm NI$ (i.e. $P_{\rm{N}}=\pm1$). Thus at $\vert P_{\rm{N}}\vert\rightarrow 1$, the distribution of the available dipolar energies is narrower than at $P_{\rm{N}}\approx0$. The dipolar reservoir can act as a source or sink of energy for a flip-flop spin exchange between two nuclei whose energy gaps are slightly different (for example due to the inhomogeneity of the quadrupolar shifts $\nu_{\rm{Q}}$). Therefore, the slow-down of nuclear spin diffusion, which proceeds through pairwise nuclear flip-flops, is interpreted as a witness of dipolar reservoir narrowing at high $\vert P_{\rm{N}}\vert$.

The aforementioned narrowing of the nuclear dipolar reservoir is conceptually similar to the state-narrowing technique, which aims to reduce the statistical dispersion of the nuclear Zeeman energies $\propto M$ in order to enhance the coherence of the electron spin qubit. An example of a narrowed mixed state is sketched in \FigNarrowed\: for $P_{\rm{N}}\approx0$, but with uncertainty in $M$ reduced down to a few units, as demonstrated experimentally previously \citep{Xu2009,Jackson2022}. The fundamental advantage of a polarized state (also sketched in \FigNarrowed), is that it both narrows the uncertainty in $M$ by a factor $\propto\sqrt{1-P_{\rm{N}}^2}$ and reduces the dipolar broadening. The ultimate limit of $P_{\rm{N}}=\pm1$ is the only case where electron spin qubit coherence is predicted to be essentially non-decaying \citep{Khaetskii2002,Deng2008}. By contrast, even if the dispersion of $M$ is reduced to zero, the dipolar energy uncertainty of a depolarized ensemble may still cause dynamics on the timescales of $1/\nu_{\rm{dd}}\approx1$~ms, leading in turn to electron spin qubit decoherence. Evaluation of electron spin coherence in a highly-polarized nuclear spin environment is an interesting subject for future work and may also provide a more sensitive tool for nuclear spin thermometry near $\vert P_{\rm{N}} \vert\approx1$. Alternatively, more accurate measurement of $P_{\rm{N}}$ can be sought through nuclear-nuclear interactions and the ``trigger'' detection method designed for dilute spins \citep{GoldmanBook} but applied to the few abundant nuclei occupying the thermally excited spin states. 

The nuclear spin cooling method reported here is applicable to a standard $p-i-n$ diode structure, fully compatible with high-quality electron spin qubit operation, as demonstrated recently in the same semiconductor structure \citep{Zaporski2022}. The technique is simple to implement and robust -- once optical pumping parameters are optimized for a certain QD, they do not require any correction over months of experiments. Even larger nuclear polarizations can be sought by combining QDs of high in-plane symmetry with biaxial strain in order to reduce the heavy-light hole mixing. Our nuclear spin cooling method uses the purity of the optical pump polarization as the final heat sink, ultimately limiting the achievable $P_{\rm{N}}$ even for a perfectly symmetric QD. This is different from the resonant ``dragging'' schemes \citep{Latta2009,Hoegele2012,Gangloff2021} where the ultimate heat sink is the photon number in the optical mode, offering in principle a much closer approach to $\vert P_{\rm{N}} \vert\approx1$, provided the dark-state bottleneck could be avoided. Combining the advantages of the two approaches in a two-stage cooling cycle can be a route towards the ultimate goal of initializing a nuclear spin ensemble into its fully-polarized quantum ground state. This would be a prerequisite for turning the enormously large Hilbert space of the $N\approx10^5$ QD nuclei into a high-capacity quantum information resource.

Acknowledgements: P.M-H. and H.D. were supported by EPSRC doctoral training grants. E.A.C. was supported by a Royal Society University Research Fellowship and EPSRC award EP/V048333/1. A.R. acknowledges support of the Austrian Science Fund (FWF) via the Research Group FG5, I 4320, I 4380, I 3762, the Linz Institute of Technology (LIT), and the LIT Secure and Correct Systems Lab, supported by the State of Upper Austria, the European Union's Horizon 2020 research and innovation program under Grant Agreements No. 899814 (Qurope), No. 871130 (Ascent+), the QuantERA II project QD-E-QKD and the FFG (grant No. 891366).

Authors' contributions: S.M., S.F.C.S and A.R. developed, grew, and processed the quantum dot samples. P.M-H. and E.A.C. conducted nuclear spin pumping experiments. H.D. and E.A.C. conducted supporting experiments on a stressed semiconductor sample. E.A.C. and P.M-H analyzed the data. E.A.C. drafted the manuscript with input from all authors. E.A.C. coordinated the project.


\newcommand{\RedText}[1]{{#1}}
\renewcommand{\thesection}{Supplementary Section \arabic{section}}
\setcounter{section}{0}
\renewcommand{\thefigure}{\arabic{figure}}
\renewcommand{\figurename}{Supplementary Figure}
\renewcommand{\theequation}{S\arabic{equation}}
\renewcommand{\thetable}{\arabic{table}}
\renewcommand{\tablename}{Supplementary Table}

\makeatletter
\def\l@subsection#1#2{}
\def\l@subsubsection#1#2{}
\makeatother

\pagebreak \pagenumbering{arabic}
\newpage


\section*{Supplementary Material}

\section{Sample structure}
\label{sec:Sample}

The sample is grown using molecular beam epitaxy (MBE) on a
semi-insulating GaAs (001) substrate. The layer sequence of the semiconductor structure is shown in Supplementary Fig.~\ref{Fig:SSample}. The growth starts with a
layer of Al$_{0.95}$Ga$_{0.05}$As followed by a single pair of
Al$_{0.2}$Ga$_{0.8}$As and Al$_{0.95}$Ga$_{0.05}$As layers acting
as a Bragg reflector in optical experiments. Then, a 95~nm thick
layer of Al$_{0.15}$Ga$_{0.85}$As is grown, followed by a 95~nm thick layer of Al$_{0.15}$Ga$_{0.85}$As
doped with Si at a volume concentration of $1.0\times10^{18}$~cm$^{-3}$. The low Al concentration of $0.15$
in the Si doped layer mitigates the issues caused by the deep DX
centers \citep{Oshiyama1986,Mooney1990,Zhai2020}. The $n$-type doped layer is followed by the
electron tunnel barrier layers: first a 5~nm thick
Al$_{0.15}$Ga$_{0.85}$As layer is grown at a reduced temperature of $560$~$^{\circ}$C to suppress Si segregation, followed by a 10~nm thick
Al$_{0.15}$Ga$_{0.85}$As and then a 15~nm thick
Al$_{0.33}$Ga$_{0.67}$As layer grown at $600$~$^{\circ}$C. Aluminium droplets are grown on the surface of the Al$_{0.33}$Ga$_{0.67}$As layer and are used to etch the nanoholes \citep{Heyn2009,Atkinson2012}. Atomic force
microscopy shows that typical nanoholes have a depth of $\approx6.5$~nm and are
$\approx70$~nm in diameter. Next, a 2.1~nm thick layer of GaAs is
grown to form QDs by infilling the nanoholes as well as to form
the quantum well (QW) layer. Thus, the maximum height of the QDs
in the growth $z$ direction is $\approx9$~nm. The GaAs layer is followed by a
268~nm thick Al$_{0.33}$Ga$_{0.67}$As barrier layer. Finally, the
$p$-type contact layers doped with C are grown: a 65~nm thick
layer of Al$_{0.15}$Ga$_{0.85}$As with a
$5\times10^{18}$~cm$^{-3}$ doping concentration, followed by a
5~nm thick layer of Al$_{0.15}$Ga$_{0.85}$As with a
$9\times10^{18}$~cm$^{-3}$ concentration, and a 10~nm thick layer
of GaAs with a $9\times10^{18}$~cm$^{-3}$ concentration.

\begin{figure}
\includegraphics[width=0.63\linewidth]{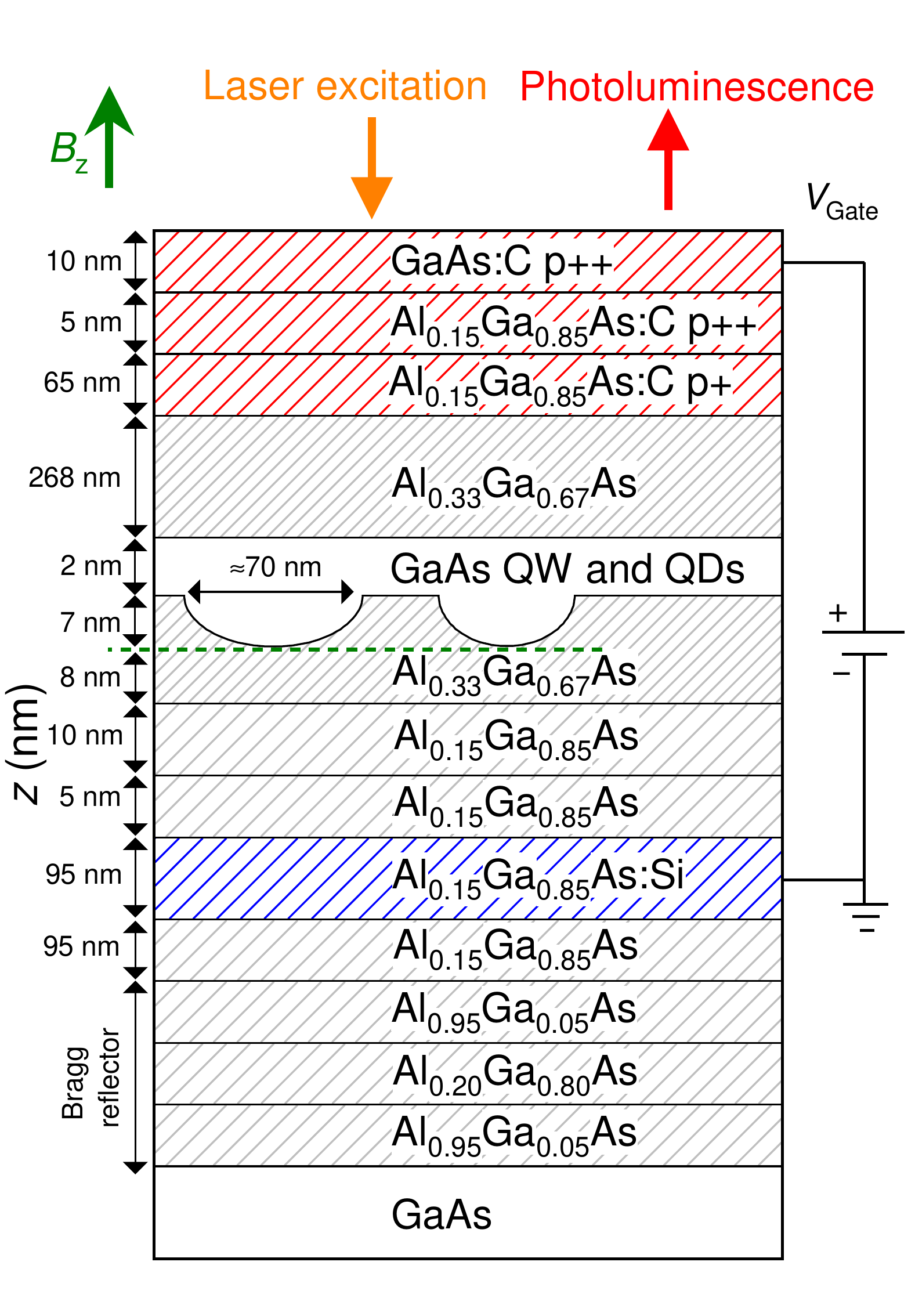}
\caption{\label{Fig:SSample} {\bf{Quantum dot sample structure.}} }
\end{figure}

The sample is processed into a $p-i-n$ diode structure. Mesa
structures with a height of 250~nm are formed by etching away the
$p$-doped layers and depositing Ni(10~nm)/AuGe(150~nm)/Ni(40~nm)/ Au(100~nm) on the etched areas. The sample is then annealed to
enable diffusion down to the $n$-doped layer to form the ohmic
back contact. The top gate contact is formed by depositing Ti(15~nm)/Au(100~nm) on to the $p$-type surface of the mesa areas. Quantum dot photoluminescence (PL) is excited and collected through the top of the sample. The
sample gate bias $V_{\rm{Gate}}$ is the bias of the $p$-type top
contact with respect to the grounded $n$-type back contact. Due to the
large thickness of the top Al$_{0.33}$Ga$_{0.67}$As layer, the
tunneling of the holes is suppressed, whereas tunnel
coupling to the $n$-type layer enables deterministic charging of
the quantum dots with electrons by
changing $V_{\rm{Gate}}$. 

\section{Electron-nuclear spin system of a quantum dot\label{sec:eNSystem}}

The band structure of the electrons and holes in a GaAs QD is sketched in Supplementary Fig.~\ref{Fig:SEnDiag} (see for example Ref.~\citep{Urbaszek2013} for a review).
The electron conduction band in GaAs has spin $s=1/2$, with two possible spin projections $s_{\rm{z}}=\pm1/2$ along the quantizing magnetic field. The valence band is four-fold degenerate at the center of the Brillouin zone in bulk GaAs. The confinement along the $z$ growth axis is sufficient to split the valence band into the heavy hole and light hole subbands with total momentum projections $j_{\rm{z}}=\pm 3/2$ and $j_{\rm{z}}=\pm 1/2$, respectively. The typical heavy-light hole splitting is $\Delta E_{\rm{hh-lh}}\approx10 - 15$~meV in GaAs/AlGaAs quantum wells \citep{ElKhalifi1989,Timofeev1996}. The selection rules for the ground state heavy-hole excitons are such that  $\sigma^+$ ($\sigma^-$)  circularly polarized light couples only to the $s_{\rm{z}}=-1/2$ ($s_{\rm{z}}=+1/2$) electron state in the conduction band. For the light hole excitons the selection rules are inverted. This means that high-fidelity initialization of the electron spin via circularly-polarized optical pumping is only possible for a sufficiently large spectral separation $\Delta E_{\rm{hh-lh}}$ of the heavy and light hole exciton transitions. 

\begin{figure}
\includegraphics[width=0.99\linewidth]{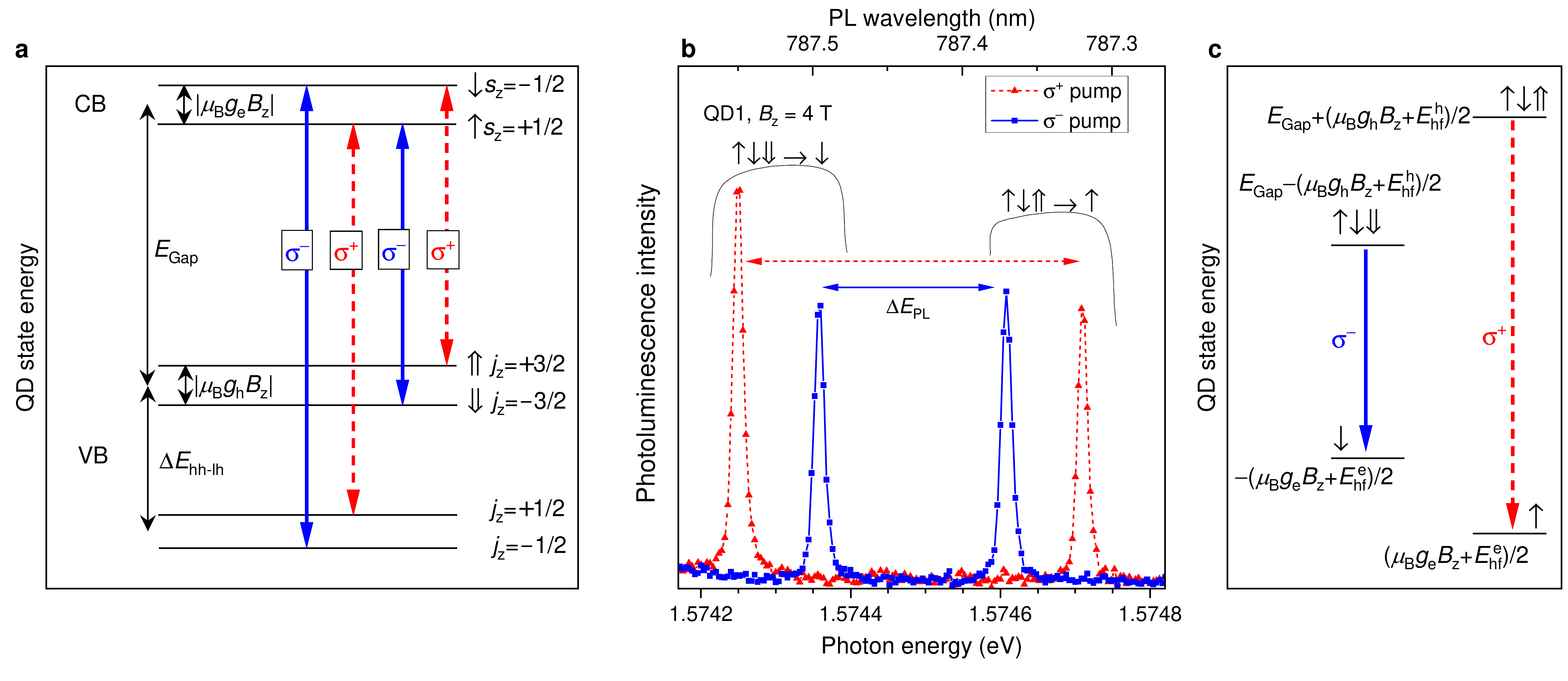}
\caption{\label{Fig:SEnDiag}{\bf{Optical transitions in a GaAs quantum dot.}} {\bf{a}} Electron energy levels in the conduction band (CB) and hole energy levels in the valence band (VB). Electron states with spin up ($\uparrow$) and down ($\downarrow$) have spin projections $s_{\rm{z}}=+1/2$ and $s_{\rm{z}}=-1/2$, respectively. The heavy hole states with pseudospin up ($\Uparrow$) and down ($\Downarrow$) have momentum projections $j_{\rm{z}}=+3/2$ and $j_{\rm{z}}=-3/2$, respectively. The light hole states have momentum projections $j_{\rm{z}}=\pm1/2$. Dashed (solid) arrows show $\sigma^+$ ($\sigma^-$) polarized optical transitions.  {\bf{b}} Typical photoluminescence spectra of an individual QD. The spectral splitting $\Delta E_{\rm{PL}}$ depends on the helicity of the optical pumping ($\sigma^\pm$) due to the buildup of the nuclear spin polarization. {\bf{c}} Energy level diagram of a negatively charged trion in a GaAs QD. The electron ground state is split by the Zeeman energy $\mu_{\rm{B}}g_{\rm{e}}B_{\rm{z}}$ and the hyperfine shift $E_{\rm{hf}}^{\rm{e}}$. The $X^-$ trion energy includes the QD bandgap energy $E_{\rm{Gap}}$, the Zeeman splitting and a small \citep{Chekhovich2013NPhys} hyperfine shift $E_{\rm{hf}}^{\rm{h}}$. The electron and hole $g$-factors are $g_{\rm{e}}$ and $g_{\rm{h}}$, respectively, with
$|g_{\rm{h}}|\gg|g_{\rm{e}}|$ in the studied QDs. Arrows depict the two optically allowed transitions responsible for the spectral doublet in (b). }
\end{figure}

Apart from the quantum-well type of confinement along the $z$ axis, carriers in a QD are also confined in the orthogonal $xy$ plane. In a real semiconductor structure there is always some breaking of the symmetry in the $xy$ plane. Such in-plane anisotropy can mix the heavy and light holes, so that the eigenstates are no longer described by pure $j_{\rm{z}}=\pm 3/2$ or $j_{\rm{z}}=\pm 1/2$ projections. As a result of such mixing the selection rules change, and the optical transitions in general become elliptically polarized.

In the pump-probe experiments we use photoluminescence of a negatively charged trion $X^{-}$, where the electron-hole recombination occurs in presence of another resident electron. The $X^{-}$ spectra, such as shown in Supplementary Fig.~\ref{Fig:SEnDiag}b tend to have the narrowest linewidths and their Zeeman splittings are free from the non-linearity which is presented for neutral excitons $X^{0}$ due to the fine structure splitting. The energies of the states involved in $X^{-}$ photoluminescence are shown in Supplementary Fig.~\ref{Fig:SEnDiag}c. The energy of the ground state resident electron is $(\mu_{\rm{B}}g_{\rm{e}}B_{\rm{z}}+E_{\rm{hf}}^{\rm{e}})s_{\rm{z}}$, whereas the energy of the optically excited trion is $E_{\rm{Gap}}+(\mu_{\rm{B}}g_{\rm{h}}B_{\rm{z}}+E_{\rm{hf}}^{\rm{h}})j_{\rm{z}}/3$. Taking the differences and substituting the momentum projections allowed by the selection rules $s_{\rm{z}}+j_{\rm{z}}=\pm1$, we find the photon energies of the two optically-allowed transitions $E_{\rm{Gap}}\pm\left(\mu_{\rm{B}}(g_{\rm{h}}-g_{\rm{e}})B_{\rm{z}}+(E_{\rm{hf}}^{\rm{h}}-E_{\rm{hf}}^{\rm{e}})\right)/2$, where $g_{\rm{e}}$ ($g_{\rm{h}}$) is the electron (heavy hole) $g$-factor, $E_{\rm{hf}}^{\rm{e}}$ ($E_{\rm{hf}}^{\rm{h}}$) is the electron (heavy hole) hyperfine shift and $E_{\rm{Gap}}$ is the difference between the conduction band and valence band ground states in a QD. The splitting of the spectral doublet is then
\begin{eqnarray}
\begin{aligned}
& \Delta E_{\rm{PL}}=\mu_{\rm{B}}(g_{\rm{h}}-g_{\rm{e}})B_{\rm{z}}+(E_{\rm{hf}}^{\rm{h}}-E_{\rm{hf}}^{\rm{e}}).
\label{eq:DEPL}
\end{aligned}
\end{eqnarray}
Next we eliminate the Zeeman contribution and define the excitonic hyperfine shift:
\begin{eqnarray}
\begin{aligned}
& E_{\rm{hf}}=-(\Delta E_{\rm{PL}}-\Delta E_{\rm{PL,0}})=E_{\rm{hf}}^{\rm{e}}-E_{\rm{hf}}^{\rm{h}},
\label{eq:DEPLhf}
\end{aligned}
\end{eqnarray}
where $\Delta E_{\rm{PL,0}}$ is the photoluminescence doublet splitting at zero nuclear spin polarization. Valence band hole hyperfine interaction is of the order of $10\%$ of the electron hyperfine interaction \citep{Chekhovich2013NPhys}. Consequently, the excitonic hyperfine shift $E_{\rm{hf}}$ is dominated by the electronic contribution $E_{\rm{hf}}^{\rm{e}}$.

The Hamiltonian describing the nuclear spin system alone includes the Zeeman, the quadrupolar and the dipole-dipole terms.
The Zeeman term accounts for the coupling of the QD nuclear
spins ${\bf{I}}_j$ to the static magnetic field $B_{\rm{z}}$ directed along the $z$ axis:
\begin{align}
    \mathcal{H}_{\rm{Z,N}} = -\sum_{j=1}^{N} \hbar\gamma_j B_{\rm{z}}\hat{I}_{{\rm{z}},j},\label{Eq:HZN}
\end{align}
where the summation goes over all individual nuclei $1\leq j \leq N$, $\hbar=h/(2\pi)$ is the reduced Planck's constant,
$\gamma_{j}$ is the gyromagnetic ratio of the $j$-th nuclear spin
and $\hat{\bf{I}}_j$ is a vector of spin operators with Cartesian
components $(\hat{I}_{{\rm{x}},j},\hat{I}_{{\rm{y}},j},\hat{I}_{{\rm{z}},j})$.
The result of the Zeeman term alone is a spectrum of equidistant
single-spin eigenenergies $m \hbar \gamma_{j}B_{\rm{z}}$,
corresponding to $2I+1$ eigenstates with $\hat{I}_{{\rm{z}}}$ projections $m$ satisfying $-I\leq m\leq +I$. 

The interaction of the nuclear electric quadrupolar moment
with the electric field gradients is described by the term (Ch.~10
in Ref.~\citep{SlichterBook}):
\begin{align}
\mathcal{H}_{\rm{Q,N}} = \sum_{j=1}^N
\frac{q_{j}}{6}[3\hat{I}_{{\rm{z'}},j}^2-I_{j}^2+\eta_{j}(\hat{I}_{{\rm{x'}},j}^2-\hat{I}_{{\rm{y'}},j}^2)],\label{Eq:HQN}
\end{align}
where $q_{j}$ and $\eta_{j}$ describe the magnitude and asymmetry
of the electric field gradient tensor, whose principal axes are
$x'y'z'$. The strain is inhomogeneous
within the QD volume, so that $q_{j}$ and $\eta_{j}$ vary
between the individual nuclei. The axes $x'y'z'$ are different for
each nucleus and generally do not coincide with crystallographic
axes or magnetic field direction. In lattice-matched GaAs/AlGaAs QDs the electric field gradients at
the nuclear sites do not exceed $q_{j}/h\approx100$~kHz, as witnessed via NMR spectroscopy. At sufficiently strong magnetic
fields $|\hbar\gamma_jB_{\rm{z}}|\gg |q_{j}|$, quadrupolar effects
can be treated perturbatively -- the main effect is the
anharmonicity of the nuclear spin energies and the resulting
quadrupolar NMR multiplet of $2I$
magnetic-dipole transitions, split by $\nu_{\rm{Q}}\approx q_{j}/h$. The $m=\pm1/2$ states of a
half-integer nuclear spin are influenced by quadrupolar effects
only in the second order, resulting in a smaller inhomogeneous
broadening, which scales as $\propto\nu_{\rm{Q}}^2/\nu_{\rm{L}}$ with nuclear spin Larmor frequency $\nu_{\rm{L}}=\gamma B_{\rm{z}}/(2\pi)$.

Direct interaction between the nuclei is described by the
dipole-dipole Hamiltonian:
\begin{align}
\mathcal{H}_{\rm{DD}}=\sum_{1\leq j<k\leq N}b_{j,k}\left(3\hat{I}_{{\rm{z}},j}\hat{I}_{{\rm{z}},k}-\hat{\bf{I}}_j{\bf{\cdot}}\hat{\bf{I}}_k\right){\rm{,}}\nonumber\\
b_{j,k}=\frac{\mu_0 \hbar^2}{4\pi}\frac{\gamma_{j}\gamma_{k}}{2}\frac{1-3\cos^2{\theta_{j,k}}}{r_{j,k}^3}\label{Eq:HDD}
\end{align}
Here, $\mu_0=4\pi\times 10^{-7}\;{\rm{N A}}^{-2}$ is the magnetic
constant and $r_{j,k}$ denotes the length of the vector, which
forms an angle $\theta$ with the $z$ axis and connects the two
spins $j$ and $k$. The typical magnitude of the interaction
constants for the nearby nuclei in GaAs is
$\max{(|b_{j,k}|)}/h\approx100$~Hz. The Hamiltonian of
Supplementary Eq.~(\ref{Eq:HDD}) has been truncated to eliminate
all spin non-conserving terms -- this is justified for static
magnetic field exceeding $\gtrsim1$~mT. While the eigenstates of an isolated nucleus have well-defined spin projections $m$, the presence of the dipole-dipole interactions means that the true eigenstates of the nuclear spin ensemble in general cannot be written as product states of the single-nucleus states. The only two states where the nuclei are not entangled are the fully-polarized states, where all individual spins occupy the states with $m=-I$ or  $m=+I$. On the other hand, at high magnetic field the total $z$-projection operator $\sum_{j} \hat{I}_{{\rm{z}},j}$ approximately commutes with the nuclear spin Hamiltonian. Therefore, the nuclear ensemble eigenstates can be described by the well defined total spin projections $M$. An example of an eigenenergy spectrum, calculated for $N=6$  nuclei of $^{75}$As, is shown in Fig.~4d of the main text. In this calculation we use $\nu_{\rm{L}}=1.8$~kHz, $\nu_{\rm{Q}}=0$ and the nuclei are taken from a single cubic cell of the group-III face-centered-cubic sublattice of GaAs. The bands observed in the spectrum correspond to the different values of $M$, ranging between $-NI$ and $+NI$. The broadening of each band is due to the dipole-dipole interactions, which lifts the degeneracy of the of the different states with the same $M$.

The interaction of the conduction band electron spin $\bf{s}$ with
the ensemble of the QD nuclear spins is dominated by the contact
(Fermi) hyperfine interaction, with the following Hamiltonian:
\begin{align}
\mathcal{H}_{\rm{hf}}^{\rm{e}}=\sum_{j=1}^N{a_j(\hat{s}_{\rm{x}}\hat{I}_{{\rm{x}},j}+\hat{s}_{\rm{y}}\hat{I}_{{\rm{y}},j}+\hat{s}_{\rm{z}}\hat{I}_{{\rm{z}},j})},\label{Eq:Hhfe}
\end{align}
where the hyperfine constant of an individual nucleus $j$ is
$a_j=A^{(j)}|\psi({\bf{r}}_j)|^2{\it{v}}$. Unlike $a_j$, the
$A^{(j)}$ hyperfine constant is a parameter describing only the
material and the isotope type to which nucleus $j$ belongs,
$|\psi({\bf{r}}_j)|^2$ is the density of the electron envelope
wavefunction at the nuclear site ${\bf{r}}_j$ of the
crystal lattice, and ${\it{v}}$ is the crystal volume per one
cation or one anion. The definitions of the hyperfine constants
differ between different sources. With the definition adopted
here, a fully polarized isotope with spin $I$, hyperfine constant
$A$ and a 100\% abundance (e.g. $^{75}$As), would shift the
energies of the electron spin states $s_{\rm{z}}=\pm1/2$ by $\pm
AI/2$, irrespective of the shape of $|\psi({\bf{r}}_j)|^2$. With such definition, the typical values in GaAs are $A\approx50$~$\mu$eV
(Ref.~\citep{Chekhovich2017}).

For valence band holes the contact (Fermi) contribution vanishes, leaving the weaker dipole-dipole terms to dominate the hyperfine interaction. Compared to the valence band electrons, the coupling has a more complicated non-Ising form \citep{Chekhovich2013NPhys}. The effect of the net nuclear polarization on the heavy-hole spin splitting can be captured by a simplified expression:
\begin{align}
\mathcal{H}_{\rm{hf}}^{\rm{h}}\approx\sum_{j=1}^N\frac{1}{3}{C^{(j)}|\psi({\bf{r}}_j)|^2{\it{v}} \hat{j}_{\rm{z}}\hat{I}_{{\rm{z}},j}},\label{Eq:Hhfh}
\end{align}
where $\hat{j}_{\rm{z}}$ is the $z$ component of the hole spin momentum operator. The valence band hyperfine material constants $C^{(j)}$ are sensitive to heavy-light hole mixing and both their signs and magnitudes depend on the material \citep{Chekhovich2013NPhys}. 

Owing to the flip-flop term $\propto(\hat{s}_{\rm{x}}\hat{I}_{{\rm{x}},j}+\hat{s}_{\rm{y}}\hat{I}_{{\rm{y}},j})$ of the hyperfine Hamiltonian (Supplementary Eq.~\ref{Eq:Hhfe}) the eigenstates of the electron-nuclear central spin system are in general entangled, i.e. they cannot be written as a direct product of the electron spin single-particle state and the nuclear spin ensemble state. Consequently, when such product state is generated through optical injection of a spin-polarized electron into the quantum dot, the wavefunction of the central spin system starts evolving. We estimate the rate of coherent evolution using the Rabi frequency $\propto\sqrt{\sum_{j}{a_j^2}}/h\approx A/(h\sqrt{N})$ derived previously in Ref.~\citep{Taylor2003} for the limit of vanishing electron spin splitting. For a fully polarized nuclear spin ensemble coupled to an electron spin polarized in the opposite direction, this Rabi frequency describes the exact solution of periodic spin exchange between the electron and the collective nuclear spin state. Therefore, in order for dynamic nuclear spin polarization to be efficient, the polarized electron spins need to be removed and injected much faster than the hyperfine-induced Rabi rotations (otherwise the electron spin will periodically polarize and depolarize the nuclei, without any net spin transfer). For a typical GaAs QD with $N\approx10^5$ nuclear spins we have $h\sqrt{N}/A\approx25$~ns. When electron spin splitting is not zero, there is an increase in the frequency of coherent oscillations that follow initialization into a product electron-nuclear state. In the limit of large electron spin splitting this is approximately the electron spin resonance frequency. For experimental conditions used in our work the maximum sum of the net hyperfine shift and the electron Zeeman splitting at $B_{\rm{z}}=10$~T  is within $\lesssim180~\mu$eV, which corresponds to electron Larmor period of $\gtrsim 20$~ps. From these basic derivations, we arrive to a rough estimate that electron spin recycling must occur on a sub-picosecond timescale in order to achieve near-unity nuclear spin polarization.

\section{Experimental methods and techniques\label{sec:Exper}}

All measurements are performed in a liquid helium bath cryostat. The sample is placed in an insert tube filled with a low-pressure heat-exchange helium gas. The base temperature is $\approx4.25$~K. We use confocal microscopy configuration where QD
photoluminescence (PL) is excited by a laser beam focused by a cryo-compatible apochromatic objective with a focal length of 2.89~mm and a numerical aperture of 0.81. The excitation spot diameter is $\approx1~\mu$m. Both the optical excitation and a static magnetic field $B_{\rm{z}}$ up to 10~T are applied along the sample growth axis $z$ (Faraday geometry). Quantum dot photoluminescence is collected and collimated by the same cryo-compatible objective. The PL signal is dispersed in a two-stage grating spectrometer, followed by a pair of achromatic doublets, which transfers the spectral image onto a charge-coupled device (CCD) detector with a magnification of 3.75. The orientation of the semiconductor sample is verified by reflecting a collimated laser off the sample surface -- the small unintentional tilt of the sample is found to be $\approx0.7^{\circ}$. The laser used for optical pumping of the nuclear spins is a ring-cavity tunable titanium sapphire (Ti:Sa) laser, operating in a single-mode continuous-wave regime. This laser is coupled with a wavelength meter (30~MHz accuracy) for precise tuning and stabilization of the optical pumping wavelength. The sample gate bias is connected by a combination of a twisted pair (inside the cryostat) and a 50~$\Omega$ coaxial cable (outside the cryostat) to an arbitrary function generator through a low-pass LC filter with a 1.9~MHz cut-off frequency. Selective manipulation of the nuclear spins is achieved with a resonant radiofrequency oscillating magnetic field, generated by a small copper wire coil. This coil is placed to have its axis within the top surface of the semiconductor sample and perpendicular to the static magnetic field. A 50~$\Omega$ cryogenic coaxial cable is used to connect the coil to a radiofrequency amplifier with a maximum rated power of 100~W.

\subsection{Pump probe experiment timing}

\begin{figure}
\includegraphics[width=0.7\linewidth]{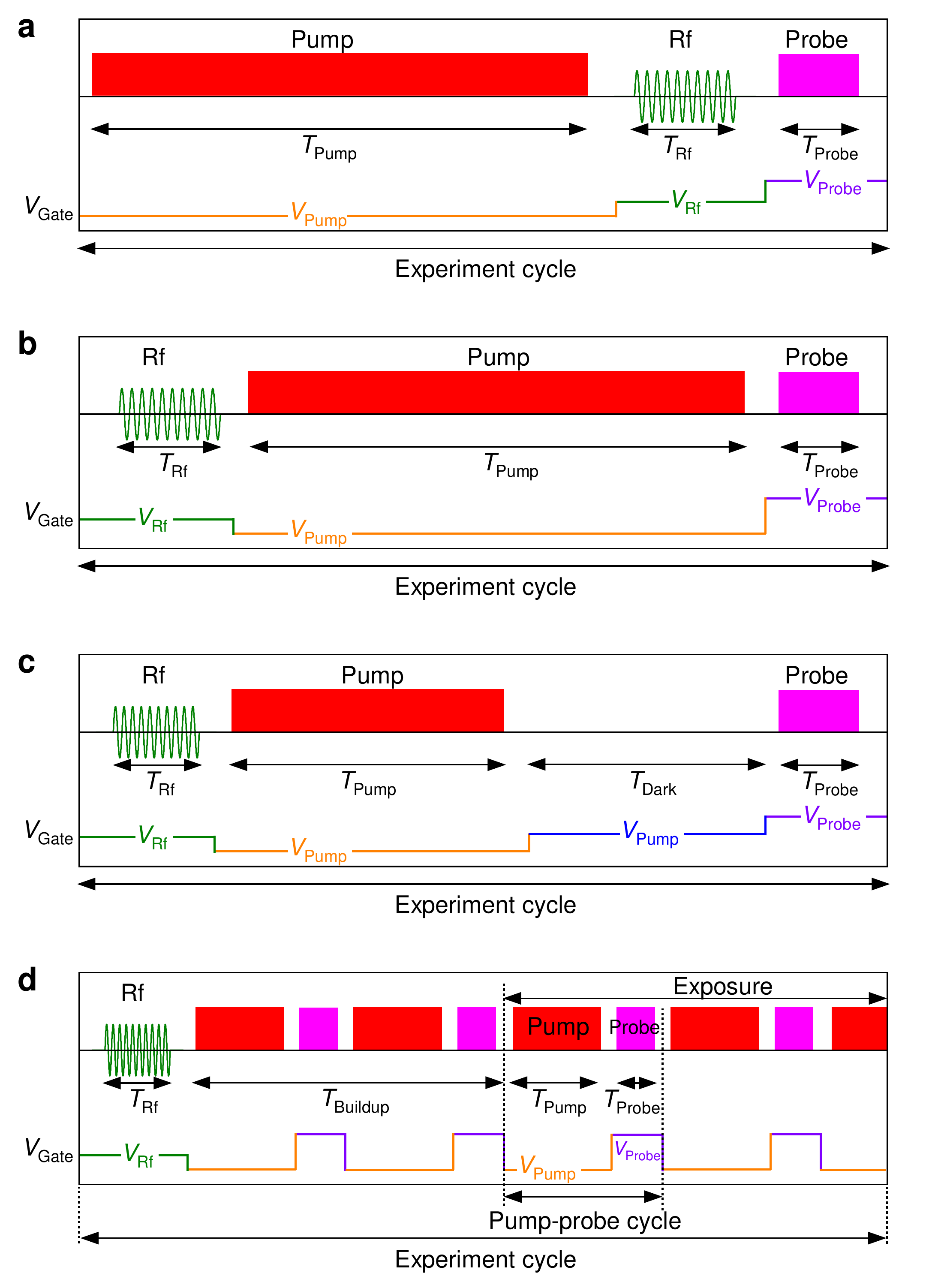}
\caption{\label{Fig:STDiagr}{\bf{Timing diagrams of different experiments.}} {\bf{a}} NMR spectroscopy and measurement of nuclear spin polarization.  {\bf{b}} Nuclear spin buildup dynamics. {\bf{c}} Nuclear spin relaxation in the dark. {\bf{d}} Optical pump power and wavelength dependence of the steady state nuclear spin polarization.}
\end{figure}

Supplementary Fig.~\ref{Fig:STDiagr} shows the timing diagrams for the different types of experiments. Supplementary Fig.~\ref{Fig:STDiagr}a shows the experimental cycle used in NMR spectroscopy, adiabatic sweep calibration and in nuclear spin polarization measurement (nuclear spin thermometry). The cycle consists of a radiofrequency burst between the pump and the probe optical pulses. During the cycle, the sample gate bias $V_{\rm{Gate}}$ is switched between the required levels by an arbitrary function generator. Both the pump and the probe optical pulses are implemented with mechanical shutters. A mechanical shutter on the spectrometer is synchronized with the probe laser shutter to prevent the pump laser reaching the detector. Multiple pump-probe cycles, typically between 5 and 15, are accumulated by the CCD detector in order to improve the signal to noise ratio. For inverse NMR spectroscopy and adiabatic sweep calibration the pump duration is reduced to $T_{\rm{Pump}}=5$~s to speed up the measurements. For saturation NMR spectroscopy and spin thermometry we use $T_{\rm{Pump}}$ between 25~s and 30~s in order to approach the steady-state of the nuclear spin polarization. The maximum $T_{\rm{Pump}}$ is limited by the need to collect photoluminescence from a sufficient number of probe pulses and the thermal noise of the CCD detector, which affects long exposures. While radiofrequency pulses can be applied at any bias, in this work we use $V_{\rm{Pump}}=-1.3$~V in order to keep the quantum dot free of charges during the radiofrequency manipulation of the nuclei.

Supplementary Fig.~\ref{Fig:STDiagr}b shows a cycle used in the measurements of the nuclear spin buildup dynamics. Each cycle starts from a radiofrequency pulse that saturates the resonances of $^{75}$As, $^{69}$Ga, $^{69}$Ga and $^{27}$Al in order to depolarize these nuclei in the entire sample. Next, the pump pulse of a variable duration $T_{\rm{Pump}}$ is applied, and the resulting hyperfine shift is measured with a probe pulse.

Supplementary Fig.~\ref{Fig:STDiagr}c shows a cycle used in the nuclear spin relaxation measurements. The cycles start with radiofrequency depolarization that is sufficiently long to eliminate any effect of the nuclear polarization left over from the previous measurement cycle. This is followed by optical pumping for $T_{\rm{Pump}}=60$~s. After the pump, the sample is kept in the dark for a time $T_{\rm{Dark}}$ under gate bias $V_{\rm{Dark}}$. After that, a probe pulse is applied to measure the fraction of the nuclear spin polarization that decayed during $T_{\rm{Dark}}$. In this type of experiments the number of the pump-probe cycles used to collect the probe photoluminescence signal varies between 1 and 10 -- single-shot probing is required when $T_{\rm{Dark}}$ exceeds a few hundred seconds. Furthermore, we perform measurements using the cycle of Supplementary Fig.~\ref{Fig:STDiagr}c, but with a second radiofrequency pulse added after the pump and before the dark interval. The duration of this second Rf pulse  $T_{\rm{Rf,2}}$ is varied between 0 and 5~s to control the degree of the initial nuclear spin polarization. In principle, there are multiple ways to control the degree of the initial nuclear spin polarization in the quantum dot, such as the duration $T_{\rm{Pump}}$ of the pump or its power and wavelength. However, any such changes in the optical pumping also affect the rate of nuclear spin diffusion into the barriers around the quantum dot \citep{MillingtonHotze2022}. The degree of nuclear polarization in the barriers then affects the rate of nuclear spin relaxation in the subsequent dark interval. The advantage of the second radiofrequency pulse is that it depolarizes the nuclei at the same rate in the entire sample. Therefore, the spatial profile of the nuclear spin polarization $P_{\rm{N}}$ after the second pulse is simply a scaled profile of the $P_{\rm{N}}$ profile produced by the optical pulse. Spin diffusion is described by a linear differential equation, so that proportional reduction of $P_{\rm{N}}$ in the entire sample should not affect the timescales of the subsequent nuclear spin diffusion and relaxation in the dark. Consequently, any dependence of the relaxation time on the degree of the initial nuclear spin polarization (left after the second radiofrequency pulse) is ascribed purely to the reduction (narrowing) of the energy that the dipole-dipole reservoir can supply or absorb during the nuclear flip-flop events of the spin diffusion process. We note that the second radiofrequency pulse has minimal effect on the measurement of the subsequent relaxation dynamics, since its duration is no more than a factor of 0.1 of the shortest measured nuclear spin relaxation time $T_{\rm{1,N}}$.

Supplementary Fig.~\ref{Fig:STDiagr}d shows the timing of the experiment used to study the dependence of the steady-state nuclear spin polarization on the optical pumping parameters such as sample bias $V_{\rm{Pump}}$, pump power and wavelength. The experiment cycle starts with a radiofrequency erase that eliminates any leftover nuclear polarization. Then the pump and probe pulses start, but the acquisition (CCD detector exposure) of the probe photoluminescence begins only after a delay $T_{\rm{Buildup}}=50$~s. This delay allows nuclear spin polarization to build up closer towards its steady state so that relatively short pump pulses $T_{\rm{Pump}}=5$~s can be used for faster acquisition of the photoluminescence signal. The probe pulses are kept short, in order to produce minimal nuclear spin depolarization during each pump-probe cycle (see details in~\ref{Subsec:Probe}). The probe pulses are also much shorter than the pump $T_{\rm{Probe}}/T_{\rm{Pump}}<0.003$, to ensure minimal effect on the steady-state nuclear spin polarization.

\subsection{Optical pumping of quantum dot nuclear spins}

The steady-state nuclear spin polarization depends on the wavelength of the pump laser. Maximum hyperfine shifts $\vert E_{\rm{hf}}\vert$ are found to occur when the laser is resonant with a certain optical transition of the quantum dot. Calibration of the optimal pumping parameters starts with a measurement of a broad-range wavelength dependence -- an example is shown in Supplementary Fig.~\ref{Fig:SDNPWl}. Once the individual spectral features, such as $s-$, $p-$, $d-$ and $f-$shell peaks, are identified, a more detailed optimization is performed. We focus on the $s-$shell pumping peak and measure more detailed dependencies on the wavelength (or equivalently the pump photon energy) at different values of pump power $P_{\rm{Pump}}$ and sample gate bias $V_{\rm{Pump}}$. The insert in Supplementary Fig.~\ref{Fig:SDNPWl} shows an example of such a detailed dependence at the optimum $P_{\rm{Pump}}=2.7$~mW, $V_{\rm{Pump}}=-2.7$~V. It can be seen that the pump laser needs to be tuned to within a narrow margin of $\approx0.2$~meV in order to achieve the highest possible $\vert E_{\rm{hf}} \vert$.

\begin{figure}
\includegraphics[width=0.6\linewidth]{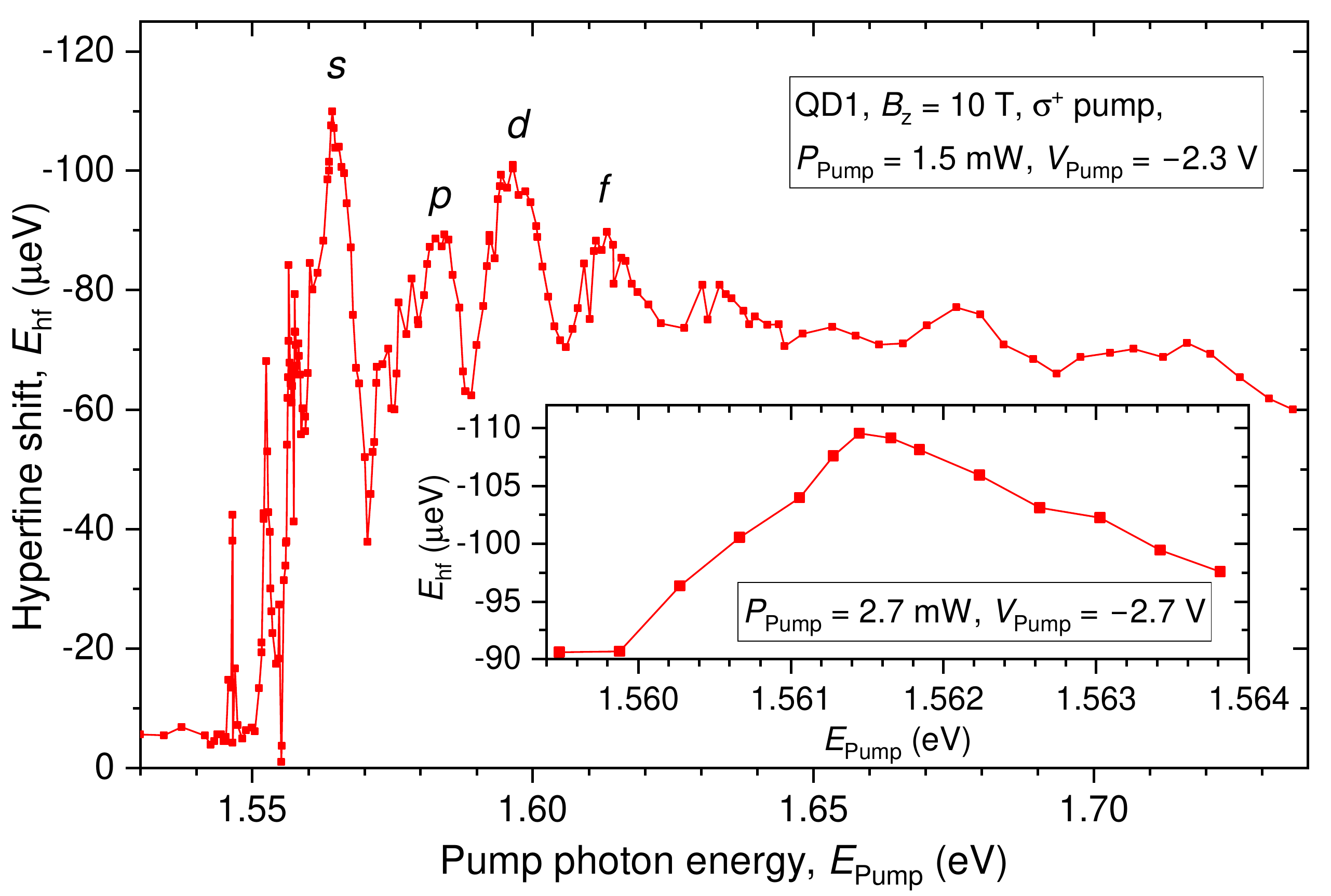}
\caption{\label{Fig:SDNPWl} {\bf{Calibration of the pump laser wavelength.}} Hyperfine shift measured as a function of the pump laser wavelength expressed in terms of the photon energies $E_{\rm{Pump}}$. Results are shown for QD1 at $B_{\rm{z}}=10$~T under $\sigma^+$ pumping. The pump power is $P_{\rm{Pump}}=1.5$~mW and the sample bias is $V_{\rm{Pump}}=-2.3$~V. Inset shows a zoomed in dependence for the $s$-shell peak measured at slightly different optimized parameters $P_{\rm{Pump}}=2.7$~mW, $V_{\rm{Pump}}=-2.7$~V.}
\end{figure}

In the experimental setup the collimated pump laser beam first passes through a linear polarizer and then a $\lambda/2$ waveplate installed in a motorized rotation mount. This way it is possible to create arbitrary orientation of the linearly polarized beam, which is then directed to a cube beamsplitter, followed by a $\lambda/4$ waveplate installed in another motorized rotation mount. By placing the $\lambda/4$ waveplate last, it is possible to compensate for any polarization imperfections of the nominally non-polarizing beamsplitter and obtain a beam with high degree of circular polarization. This beam is then directed through a quartz window of the cryostat insert and the cryogenic objective, which focuses it on the surface of the QD semiconductor sample. In order to account for any polarization imperfections in the optical path we perform calibration measurements where both the $\lambda/2$ and $\lambda/4$ waveplate orientations are scanned and the resulting hyperfine shifts $E_{\rm{hf}}$ are measured. The results shown in Supplementary Fig.~\ref{Fig:SHWP}, indicate that the waveplate must be set within $\pm2^\circ$ in order to attain the highest nuclear spin polarization degree. The optimal orientations of the $\lambda/2$ and $\lambda/4$ waveplates are different for the minimum negative (triangles) and the maximum positive (squares) $E_{\rm{hf}}$. 

\begin{figure}
\includegraphics[width=0.6\linewidth]{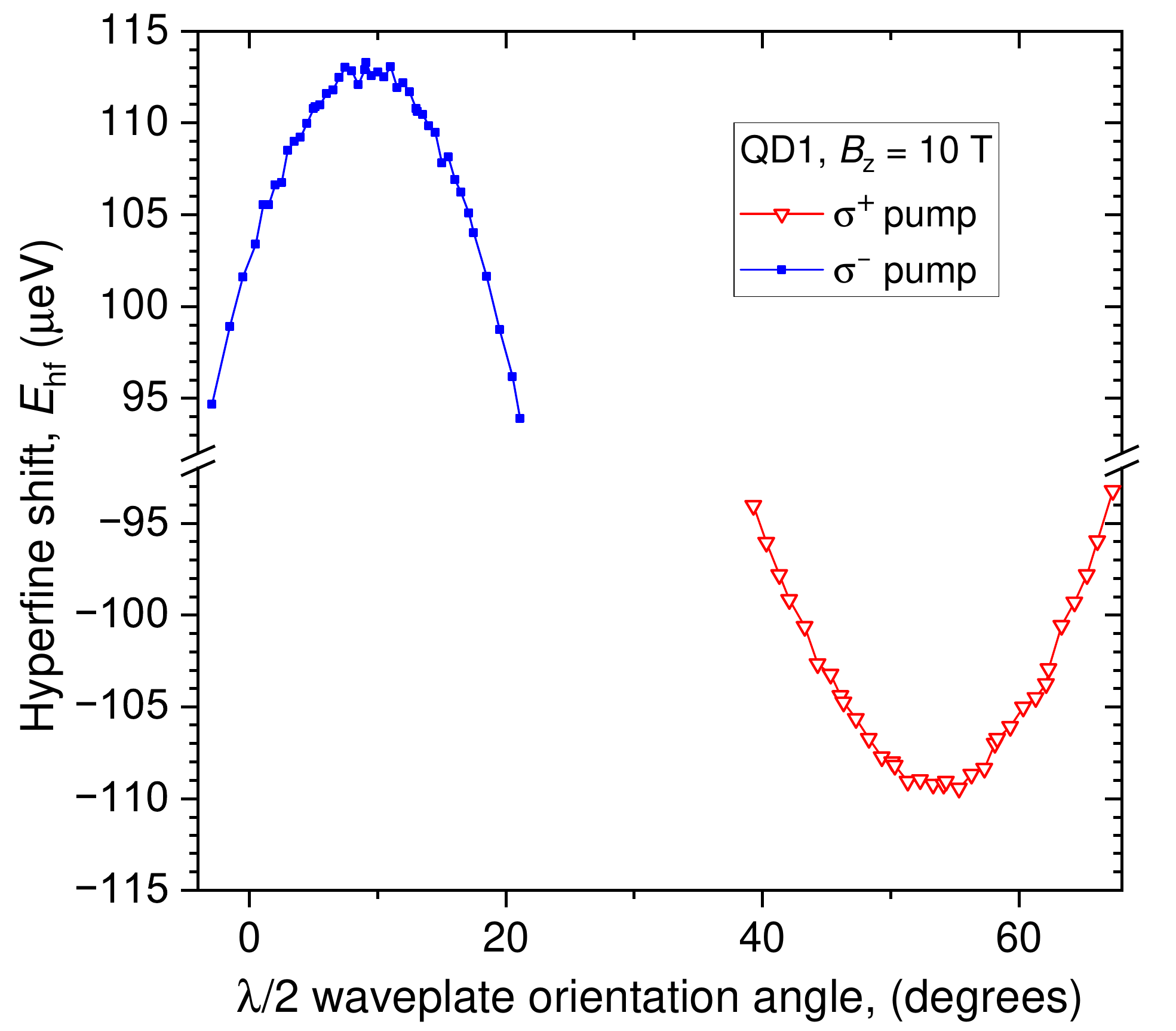}
\caption{\label{Fig:SHWP} {\bf{Calibration of the optical polarization of the pump.}} Hyperfine shift measured as a function of the orientation angle of a $\lambda/2$ waveplate installed in the pump laser beam. Other parameters, such as $\lambda/4$ waveplate orientation, pump power, wavelength and bias are optimized separately for both $\sigma^+$ (triangles) and $\sigma^-$ (squares) pumping of the QD nuclear spin polarization. Results are shown for QD1 at $B_{\rm{z}}=10$~T.}
\end{figure}

Once the orientations of the $\lambda/2$ and $\lambda/4$ waveplates are optimised, we examine the polarization state of the pump beam directed to the cryostat. To this end, we place a linear polarizer (analyzer) after the $\lambda/4$ waveplate, followed by a power meter. The linear polarizer is rotated to find the minimum ($I_{\rm{min}}$) and the maximum ($I_{\rm{max}}$) intensities of the transmitted beam. For a perfect circularly polarized beam $I_{\rm{min}}=I_{\rm{max}}$, whereas for a linearly polarized beam $I_{\rm{min}}=0$. We characterise the optimized beams using the degree of linear polarization $\rho_{\rm{lin}}=(I_{\rm{max}}-I_{\rm{min}})/I_{\rm{max}}$ and the analyzer orientation angle $\alpha_{\rm{max}}$ where the maximum intensity is achieved. These results are summarized in the polar plot of Supplementary Fig.~\ref{Fig:SLinPol}. 

\begin{figure}
\includegraphics[width=0.6\linewidth]{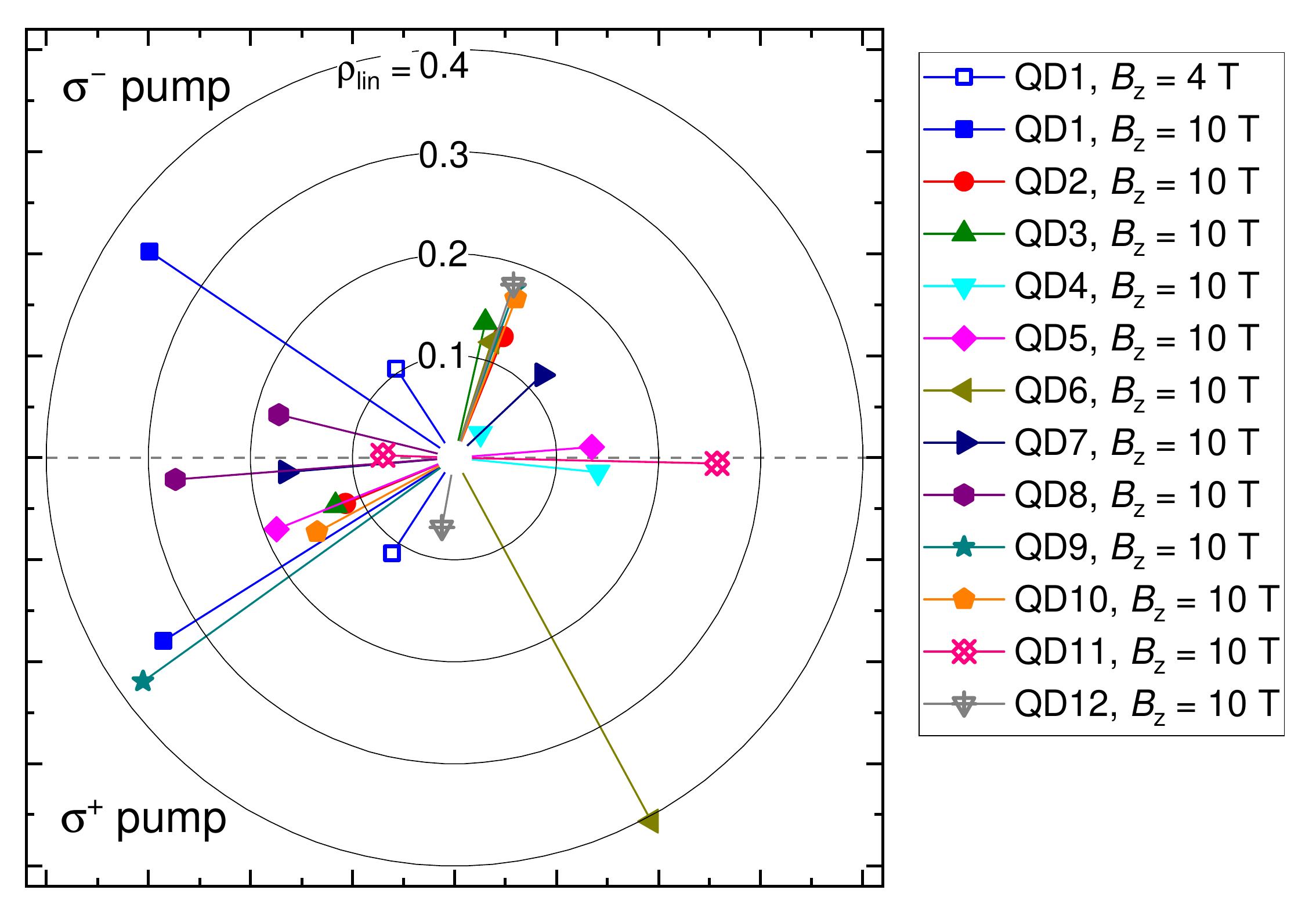}
\caption{\label{Fig:SLinPol} {\bf{Optical polarization properties of the optimal nuclear spin pumping beams.}} Vectors show the orientation of the linear polarization and the magnitude of the linear polarization degree $\rho_{\rm{lin}}$ of the pump beam that produces maximum nuclear spin polarization in individual dots QD1 - QD12. The bottom part shows the results for minimum negative hyperfine shift ($\sigma^+$ character of the circularly polarized component of the pump) while the top part shows results for maximum positive hyperfine shift ($\sigma^-$ character). The horizontal and vertical axes of the plot correspond approximately to the edges of the semiconductor sample, cleaved along the $[110]$ and $[1\bar{1}0]$ crystallographic directions.}
\end{figure}

The optimal degree  $\rho_{\rm{lin}}$ and orientation $\alpha_{\rm{max}}$ of the linearly polarized components vary between individual quantum dots. Moreover, optimal polarization parameters are different for $\sigma^+$ and $\sigma^-$ pumping and even depend on magnetic field for the same QD1. Such variability, as well as the large values of $\rho_{\rm{lin}}\lesssim0.4$ suggest that polarization imperfections in the optical elements (e.g. mechanical stress in the cryo-objective) are not the major contribution. The large deviation of the optimal optical pumping from pure circular polarization is therefore attributed to the properties of the individual quantum dots. These may include anisotropy of the QD shape and inhomogeneous microstrains that make semiconductor material around the QD act as an optical waveplate and give rise to heavy-light hole mixing \citep{Huber2019} that causes optical selection rules to depart from those of the bulk GaAs. Indeed, previous studies have shown that a sufficiently large uniaxial strain $\gtrsim0.5~\%$ can flip the valence band hole quantization axis into the sample growth plane \citep{Yuan2018}. Another measure of the QD anisotropy is the fine structure splitting (FSS) of a neutral exciton at zero magnetic field. While we have not conducted systematic correlation studies, selective measurement on QD1, where very large nuclear spin polarization was achieved, revealed a FSS of $\approx 28~\mu$eV. This is considerably larger than the few-$\mu$eV FSS observed in symmetric GaAs QDs \citep{HuoAPL2013}. This comparison suggests that QD anisotropy does not preclude large nuclear spin polarization, as long as optical selection rules permit coupling to spin-polarized conduction band electronic states. From that perspective, optimization of the $\lambda/2$ and $\lambda/4$ waveplate angles can be understood as matching of the optical pump polarization to the elliptical polarization of the QD optical transition, allowing generation of spin-polarized electrons. Further investigations (both experimental and theoretical) would be needed to elucidate which types of anisotropies are compatible with efficient nuclear spin pumping. For brevity, throughout this work we use the term ``$\sigma^+$ pumping'' (``$\sigma^-$ pumping'') to describe the optimized elliptically-polarized optical pumping with $\sigma^+$ ($\sigma^-$) character of the circularly polarized component.

\subsection{Optical probing of quantum dot nuclear spins\label{Subsec:Probe}}

\begin{figure}
\includegraphics[width=0.6\linewidth]{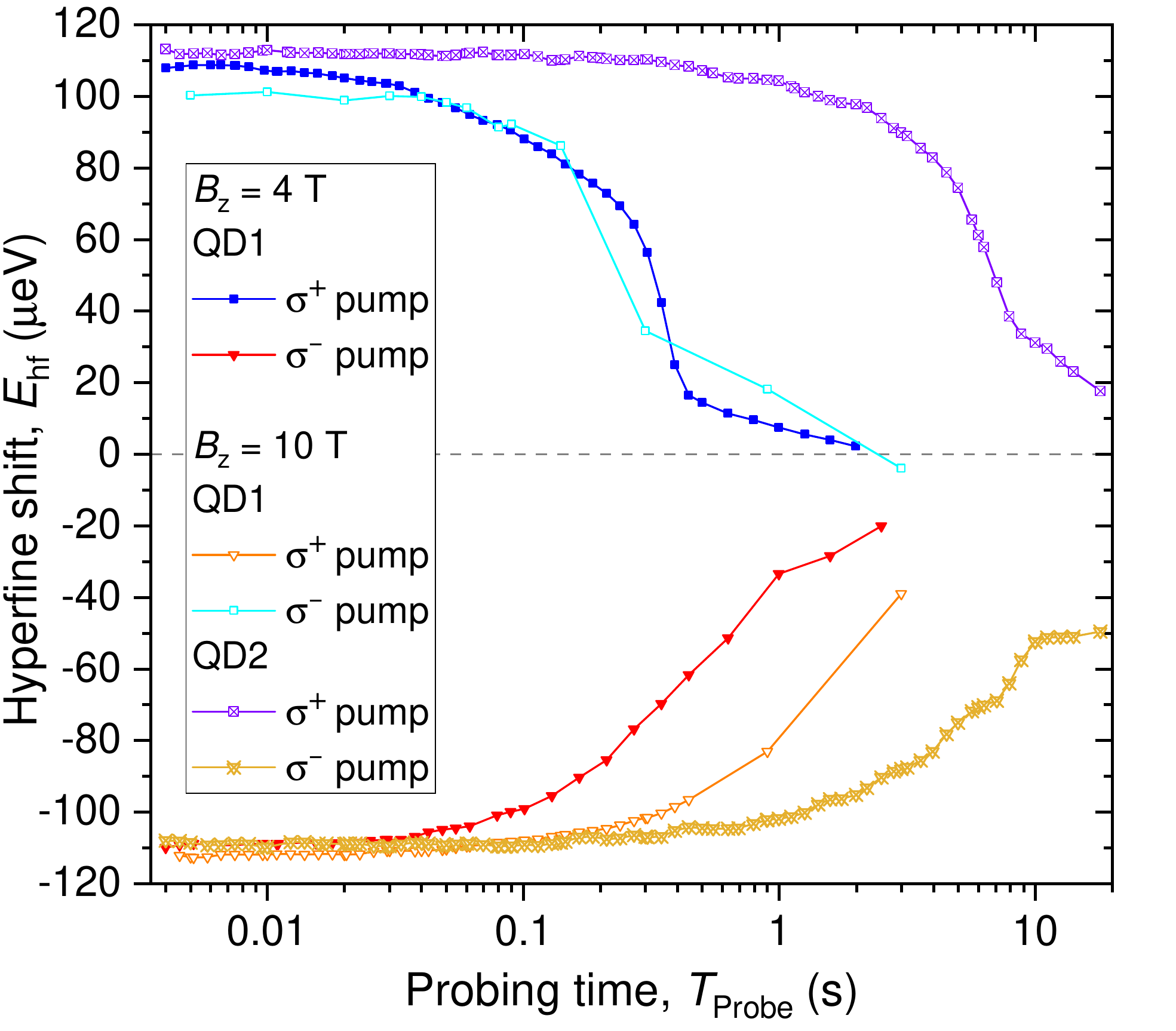}
\caption{\label{Fig:SProbe} {\bf{Calibration of the optical probing of the QD nuclear spin polarization.}} Hyperfine shift measured as a function of the probing time $T_{\rm{Probe}}$ following a $\sigma^+$ (triangles) or $\sigma^-$ (squares) pumping of the nuclear spin polarization in a QD. For QD1 results are shown for $B_{\rm{z}}=4$~T (solid symbols) and $B_{\rm{z}}=10$~T (open symbols), measured with a probe power of $P_{\rm{Probe}}=30$~nW. For QD2 results are shown for $B_{\rm{z}}=10$~T (crossed symbols), measured with a probe power of $P_{\rm{Probe}}=7$~nW.}
\end{figure}

For optical probing of the nuclear spin polarization we use a diode laser emitting at 690~nm. Sample forward bias, typically $+0.7$~V, and the probe power are chosen to maximize (nearly saturate) PL intensity of the ground state $X^-$ trion. The difference between the spectral splitting $\Delta E_{\rm{PL}}$ of the $X^-$ trion doublet and the same splitting $\Delta E_{\rm{PL,0}}$ measured for depolarized nuclei reveals the hyperfine shifts $E_{\rm{hf}}=-(\Delta E_{\rm{PL}}-\Delta E_{\rm{PL,0}})$. Illumination with a probe laser inevitably acts back on the nuclear spin polarization. In order to quantify such back-action we perform calibration measurements with examples shown in Supplementary Fig.~\ref{Fig:SProbe}. In these experiments the QD is first pumped with a $\sigma^+$ or $\sigma^-$ polarized laser with power and bias set to maximize the steady state nuclear polarization. Then the pump is switched off and the probe laser pulse is applied. The hyperfine shift $E_{\rm{hf}}$ is measured from PL spectroscopy at the end of this probe. It can be seen that the probe induces decay of the nuclear spin polarization. For QD1 we use the same probe power of $P_{\rm{Probe}}=30$~nW, but the unwanted probe-induced depolarization is faster at $B_{\rm{z}}=4$~T (solid symbols) compared to $B_{\rm{z}}=10$~T (open symbols). For selective-NMR measurements of the nuclear spin polarization (spin thermometry) in QD1 we use $T_{\rm{Probe}}=15$~ms at $B_{\rm{z}}=10$~T, so that the resulting depolarization is negligible ($<1\%$). PL intensity of the same QD1 is weaker at $B_{\rm{z}}=4$~T so we use a longer
$T_{\rm{Probe}}=24$~ms in order to obtain a sufficiently strong probe PL signal. However, this leads to a larger depolarization of $\approx4\%$ under $\sigma^-$ pumping. Depolarization itself is not an issue, since it would simply rescale all the measured $E_{\rm{hf}}$, which would not affect the differential NMR spin thermometry. In practice, the probe-induced depolarization also depends on the instantaneous $E_{\rm{hf}}$ -- such nonlinearity is what causes the distortion, resulting in larger uncertainties of the nuclear spin polarization measured at $B_{\rm{z}}=4$~T. For QD2 we use a lower probe power $P_{\rm{Probe}}=7$~nW (crossed symbols in Supplementary Fig.~\ref{Fig:SProbe}), which leads to an even slower probe-induced depolarization than for QD1. This allows to have a longer probe pulse ($T_{\rm{Probe}}=40$~ms) for QD2, while keeping parasitic depolarization small ($<1\%$).

\subsection{Radiofrequency control of nuclear spins}

The radiofrequency oscillating magnetic field $B_{\rm{x}}\perp z$ is produced by a coil placed at a distance of $\approx0.5$~mm
from the QD sample. The coil is made of 10 turns of a 0.1~mm
diameter enameled copper wire wound on a $\approx0.4$~mm diameter
spool in 5 layers, with 2 turns in each layer. Two main types of radiofrequency signals are used in this work. The first type is a frequency-swept monochromatic excitation which is used for adiabatic inversion of the nuclear spin population. The amplitude of the radiofrequency field is constant and the frequency is swept linearly in time. Radiofrequency sweeps are discussed further in~\ref{subsec:SwpCalib}.

The second type is the broadband radiofrequency excitation which is required to saturate inhomogeneously broadened quadrupolar resonances. The typical width of the resonances that needs to be saturated is tens to hundreds of kHz (further details are given in~\ref{subsec:NMROverlap}), which is significantly larger than the typical homogeneous NMR linewidth ($<1$~kHz). Therefore monochromatic radiofrequency excitation cannot provide a sufficiently uniform saturation of the entire inhomogeneously broadened resonance. This necessitates the use of a broadband radiofrequency excitation. Ideally, one wants a signal with a rectangular spectral profile, that has a constant spectral density in the required frequency interval, and a zero intensity outside that interval. In practice, when implementing the radiofrequency waveforms on a digital generator, it is convenient to approximate the required rectangular spectral band with a frequency comb. In spectral domain, the comb consists of periodically spaced monochromatic modes of constant amplitude, covering the desired interval of frequencies. The mode spacing of 120~Hz is chosen to be smaller than the homogeneous NMR linewidth. Under these conditions, by using a sufficiently small amplitude of each mode we achieve exponential depolarization (i.e. without nuclear spin Rabi oscillations) of the nuclear spin ensemble \citep{Waeber2016} with a typical time constant of $\tau\approx30$~ms. The saturation of a chosen NMR resonance is achieved by applying a frequency comb excitation for a period of $\approx5\tau$. When subject to such excitation, the nuclear spins undergo slow Rabi rotation, transitioning between the spin states parallel and antiparallel to the external magnetic field \citep{Bloch1946}. Due to the nuclear-nuclear dipole-dipole interactions each nuclear spin is subject to a local field. The randomness of these local fields results in dephasing between Rabi precessions of the individual nuclei. Consequently, the nuclear spin ensemble becomes depolarized (i.e. each nucleus is randomly polarized) after a long saturation pulse. 

For the saturation NMR spectra, shown in Fig.~3b of the main text, we use a frequency comb with a total width of $6$~kHz. The central frequency of the comb is scanned to obtain the spectra -- this frequency is the horizontal axis of the spectral plots. For the high-resolution NMR spectra, shown in Fig.~3a of the main text, we employ the ``inverse'' NMR technique \citep{Chekhovich2012} which enhances the NMR signal and allows the spectra to be measured even on those nuclear spin transitions that are depopulated at high polarization degrees. In this approach the radiofrequency excitation spectrum is a broadband frequency comb with a narrow gap. The central frequency of the gap is scanned and is used for the horizontal axis of the ``inverse'' NMR spectra. The width of the gap controls the balance between the NMR signal amplitude and the spectral resolution. For the spectra of Fig.~3a we use a 4~kHz gap to measure the satellite transitions and a 2~kHz gap to measure the narrow central transition NMR peak.

\section{Additional experimental data\label{sec:AddData}}

\subsection{Extended data from nuclear spin pumping measurements}

\begin{figure}
\includegraphics[width=0.9\linewidth]{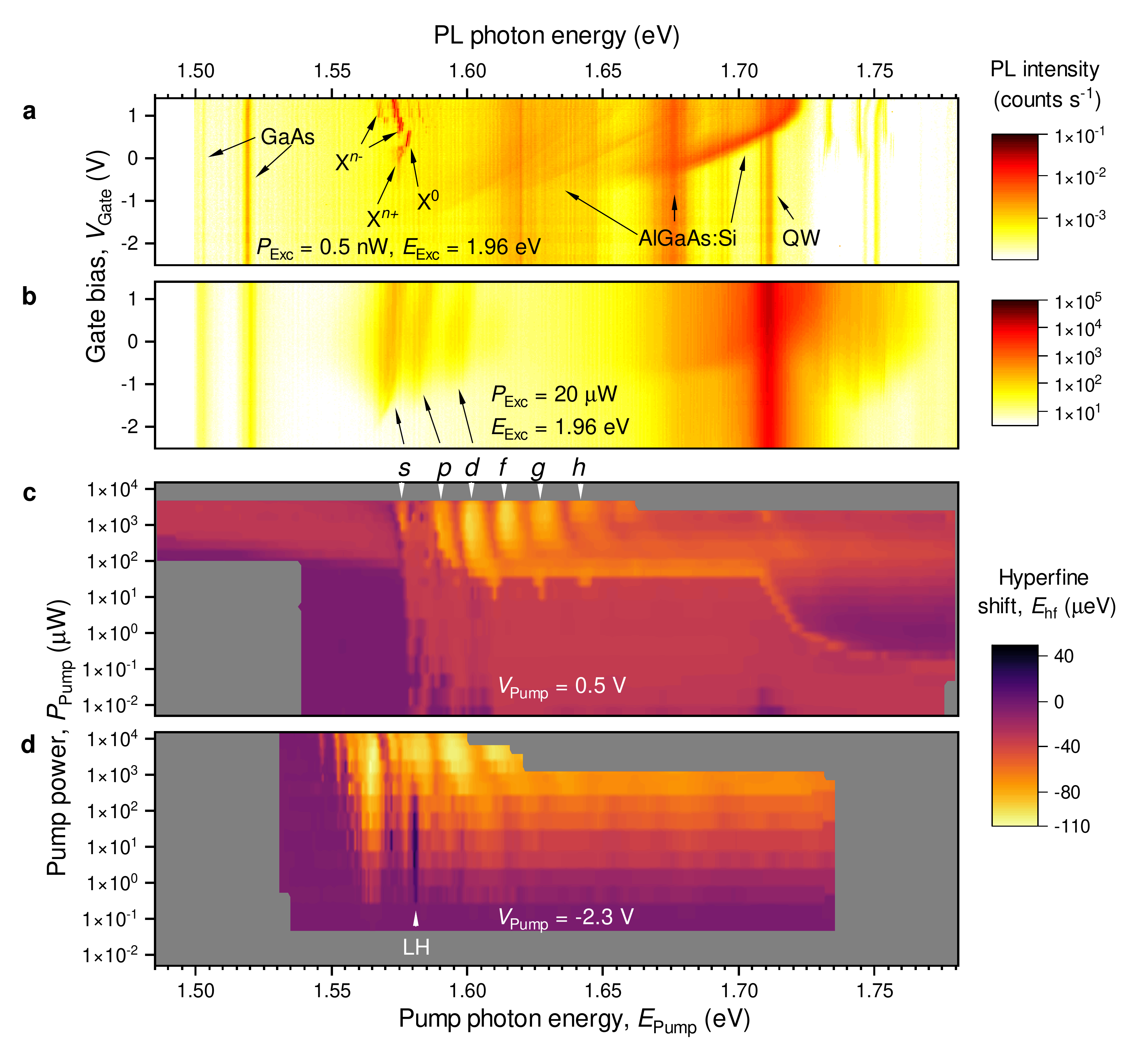}
\caption{\label{Fig:SDNPPower}{\bf{Power dependence of nuclear spin pumping.}} {\bf{a}} Bias dependent photoluminescence spectra of QD1 at $B_{\rm{z}}=10$~T measured at a low excitation power $P_{\rm{Exc}}=0.5$~nW. Optical excitation is continuous wave at a wavelength of 632.8~nm. {\bf{b}} Same photoluminescence spectra but at a high power of $P_{\rm{Exc}}=20~\mu$W. {\bf{c}} Steady-state hyperfine shift measured as a function of the pump power $P_{\rm{Pump}}$ and the pump photon energy $E_{\rm{Pump}}$. Measurement is conducted on QD1 at $B_{\rm{z}}=10$~T using $\sigma^+$ pump polarization. The sample bias is kept at $V_{\rm{Pump}}=+0.5$~V during pumping. Grey color shows the areas where no data was measured. {\bf{d}} Same as (c) but for $V_{\rm{Pump}}=-2.3$~V. ``LH'' labels the feature ascribed to the resonant absorption of a light-hole exciton.}
\end{figure}

Our approach to maximizing the nuclear spin polarization is through line-search optimization of the optical pumping parameters, such as pump photon energy $E_{\rm{Pump}}$, pump power $P_{\rm{Pump}}$, optical polarization and the sample bias $V_{\rm{Pump}}$. In order to understand the physics of the nuclear spin pumping process we also measure a systematic parametric dependence. Given the typical timescales of the nuclear spin process ($1-100$~s) it is not possible to explore the entire parameter space within a reasonable experimental time. Therefore, we measure various one- and two- dimensional sections in the multidimensional parameter space. An example is shown in Supplementary Fig.~\ref{Fig:SDNPWl}, where a one-dimensional dependence on $E_{\rm{Pump}}$ is shown. Interpretation of the nuclear spin polarization data is conducted by correlating it with photoluminescence spectroscopy.

Supplementary Fig.~\ref{Fig:SDNPPower}a is a bias-dependent photoluminescence spectroscopy map measured under non-resonant optical excitation (HeNe laser emitting at 632.8~nm). At low excitation power $P_{\rm{Exc}}=0.5$~nW we observe multiple spectral features, labelled accordingly. The emission of the bulk GaAs free exciton is observed at $\approx1.518$~eV, accompanied by a low energy band at $\approx1.50$~eV arising from doping and impurities. The emission of the Si-doped AlGaAs layer is observed as broad spectral features between $\approx1.60 - 1.70$~eV, and the peak at $\approx1.71$~eV originates from the GaAs quantum well (QW). Single-QD emission is observed between $\approx1.56 - 1.59$~eV as a series of Zeeman doublets that switch over as the gate bias $V_{\rm{Gate}}$ is changed. The higher-resolution spectra of the QD excitons are shown in Supplementary Fig.~\ref{Fig:SPL1200}. At $V_{\rm{Gate}}\approx+0.5$~V photoluminescence is dominated by the neutral exciton $X^0$, identified from its fine structure splitting at $B_{\rm{z}}=0$~T. At more negative biases the emission of positively charged excitons dominates, since electrons rapidly tunnel out of the dot, leaving excess photogenerated (non-equilibrium) holes. At more positive biases the emission of $X^0$ is superseded by the negatively charged trion $X^-$, which becomes dominant when QD confines a resident (equilibrium) electron. At even more positive biases the dot is charged with multiple resident electrons. The spectral features originating from doubly ($X^{2-}$) and triply ($X^{3-}$) charged excitons can be distinguished, while the photoluminescence peaks at even higher charge numbers tend to overlap. 

\begin{figure}
\includegraphics[width=0.9\linewidth]{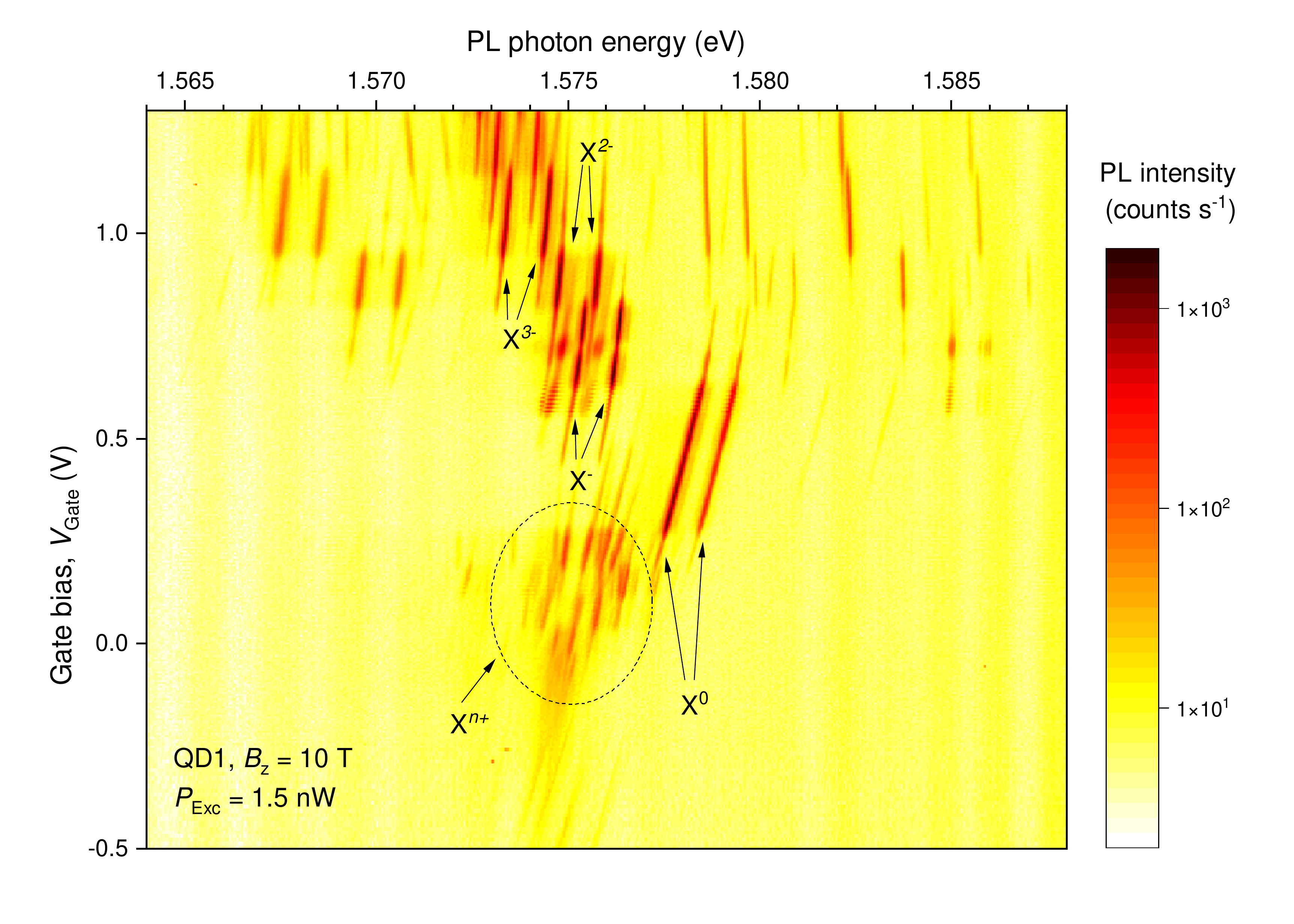}
\caption{\label{Fig:SPL1200}{\bf{Photoluminescence spectroscopy of QD excitons.}} {\bf{a}} Bias dependent photoluminescence spectra of QD1 at $B_{\rm{z}}=10$~T measured at a low excitation power $P_{\rm{Exc}}=1.5$~nW and an excitation wavelength of 632.8~nm.}
\end{figure}

When the power is increased ($P_{\rm{Exc}}=20~\mu$W in Supplementary Fig.~\ref{Fig:SDNPPower}b) the emission of the differently-charge $s$-shell QD excitons broadens into a single red-shifted band ($\approx1.572$~eV). Emission of the higher-shell excitons ($p$, $d$ and $f$) also becomes visible. By contrast, photoluminescence intensity of the QW increases with power without saturation.

Supplementary Fig.~\ref{Fig:SDNPPower}c shows the two-dimensional map of the hyperfine shift $E_{\rm{hf}}$ measured as a function of the pump power $P_{\rm{Pump}}$ and photon energy $E_{\rm{Pump}}$ at a fixed $V_{\rm{Pump}}=+0.5$~V, which roughly corresponds to a bias where the QD equilibrium state switches from 0 to 1 electron. Nuclear spin pumping evidently takes place at powers as low as $P_{\rm{Pump}}\approx10$~nW (which is close to saturation power of the $s$-shell QD excitons) as long as the pump laser is tuned above the $s$-shell exciton transition, so that the QD can absorb the pump photons. However, the resulting nuclear spin polarization degree is low, characterized by $E_{\rm{hf}}\approx-30~\mu$eV. Spin pumping efficiency increases when the pump power is increased to hundreds of $\mu$W, which is well above the ground state exciton saturation. The lowest negative $E_{\rm{hf}} \approx -90~\mu$eV is achieved at $P_{\rm{Pump}}\approx1$~mW. At this high power, a series of spectral peaks is observed. Their periodicity matches the periodicity observed in the high-power photoluminescence spectra (Supplementary Fig.~\ref{Fig:SDNPPower}b), which allows us to identify the peaks as originating from different excitonic shells (up to six visible). The mechanism of nuclear spin pumping can then be understood to arise from resonant absorption of the circularly polarized pump photons, which generate spin-polarized electrons and holes in the excited orbital states. It also follows from Supplementary Fig.~\ref{Fig:SDNPPower}c that at $V_{\rm{Pump}}=+0.5$~V the steady-state $\vert E_{\rm{hf}}\vert$ produced by pumping via the higher $d$ and $f$ shells is larger than via the $p$ and ground $s$ shells. Excitation via higher shells means that excitons can relax towards the ground state before recombination. Such energy relaxation provides a route for a simultaneous exchange of spin with the nuclei \citep{Urbaszek2007}, since it helps to absorb or supply a small amount of energy required to compensate the mismatch of the electron and nuclear Zeeman energies. Without the coupling to external energy reservoirs the electron-nuclear spin flip-flop would be energetically forbidden. At high powers $P_{\rm{Pump}}\gtrsim100~\mu$W, nuclear spins can be polarized even via optical excitation well below the ground state QD exciton transition, indicating that it's a distinct spin pumping mechanism which we further discuss below.

Supplementary Fig.~\ref{Fig:SDNPPower}d shows the same dependence of $E_{\rm{hf}}$ on $P_{\rm{Pump}}$ and $E_{\rm{Pump}}$ but with the sample gate voltage changed to a large reverse bias regime $V_{\rm{Pump}}=-2.3$~V. At this bias, no QD photoluminescence is observed, even at high optical excitation power, meaning that the excitons become ionized before they can recombine to emit a photon. Nevertheless, nuclear spin pumping is observed and is more efficient than at $V_{\rm{Pump}}=+0.5$~V. Under large reverse bias, a higher threshold power (of a few $\mu$W) is needed to induce measurable nuclear spin polarization, which can be explained by the need for the optical excitation to outpace the fast tunneling of the charges from the QD. We again observe spectral peaks that can be matched to the individual excitonic shells (at $V_{\rm{Pump}}=-2.3$~V the shells are red-shifted with respect to $V_{\rm{Pump}}=+0.5$~V because of the Stark shift). Large $\vert E_{\rm{hf}}\vert$ are observed for all four lowest exciton peaks, but from experiments on multiple individual QDs we consistently find that pumping through the ground state $s$-shell exciton under large reverse bias leads to the most efficient spin pumping (characterized by the highest $\vert E_{\rm{hf}}\vert$).

It is also worth noting that inverted $E_{\rm{hf}}$ is observed under certain pumping conditions. For example, in Supplementary Fig.~\ref{Fig:SDNPPower}d we observe $E_{\rm{hf}}>0$ around $P_{\rm{Pump}}\approx10~\mu$W and $E_{\rm{Pump}}\approx1.580$~eV, which is $\approx16$~meV above the energy of the $s$-shell nuclear spin pumping peak. This feature at $E_{\rm{Pump}}\approx1.580$~eV, labelled ``LH'', is ascribed to a light-hole exciton. Optical excitation of a heavy-hole exciton transition with a $\sigma^+$ polarized light (with photons carrying a $+1$ momentum) generates a hole with momentum projection $j_{\rm{z}}=+3/2$ and an electron with a $s_{\rm{z}}=-1/2$ spin projection. The $s_{\rm{z}}=-1/2$ electrons then lead to nuclear spin pumping with a negative $E_{\rm{hf}}<0$, as indeed observed in Supplementary Fig.~\ref{Fig:SDNPPower}d for a wide range of the pump parameters. However, when resonant with a light-hole exciton transition, the same $\sigma^+$ photon generates a hole with $j_{\rm{z}}=+1/2$ and an electron with $s_{\rm{z}}=+1/2$ (see Supplementary Fig.~\ref{Fig:SEnDiag}a), which then leads to an inverted $E_{\rm{hf}}>0$. The same argument applies to $\sigma^-$ optical excitation, and manifests in experiments as $E_{\rm{hf}}<0$, observed under resonant excitation of the light-hole exciton transition.

\begin{figure}
\includegraphics[width=0.9\linewidth]{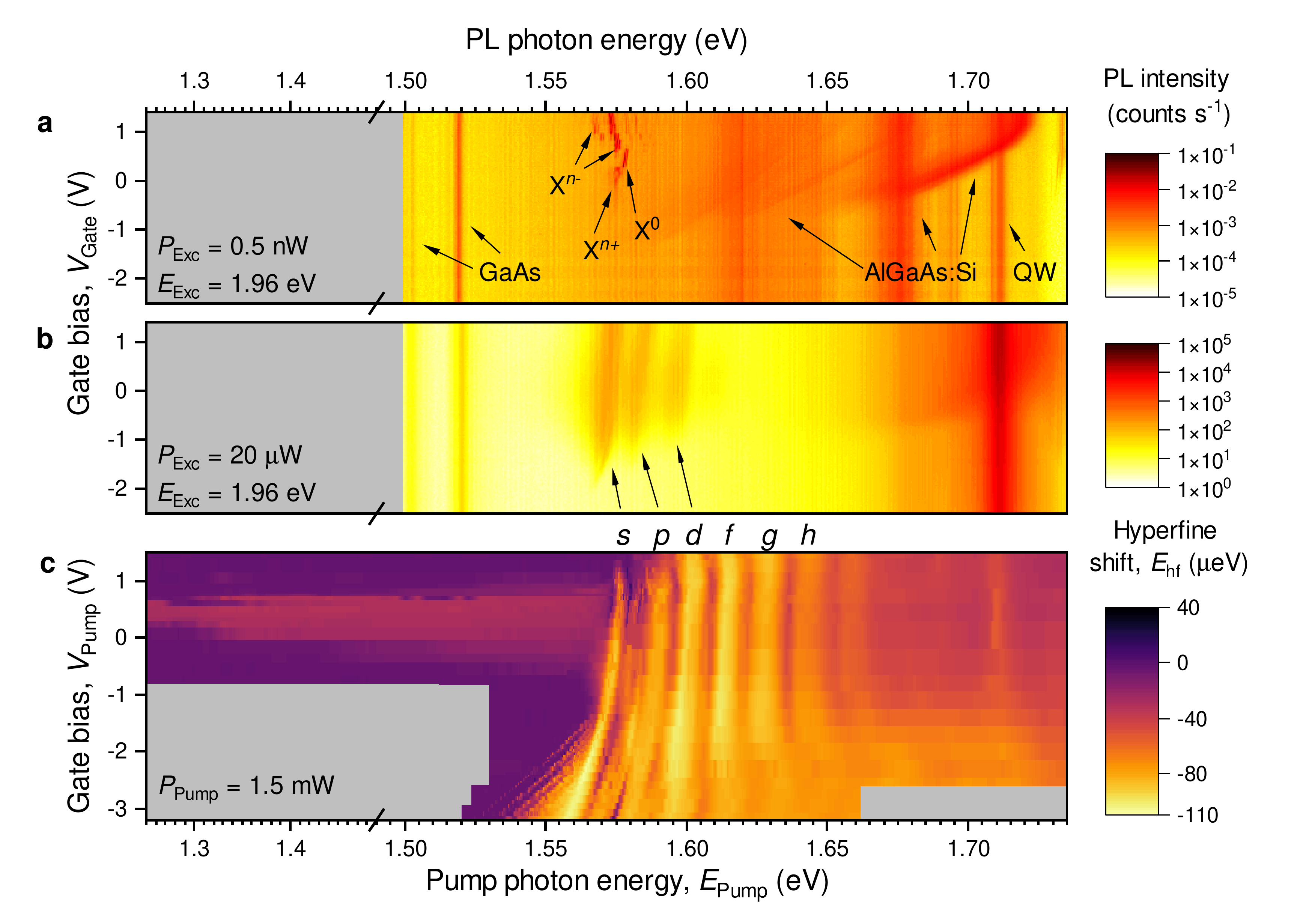}
\caption{\label{Fig:SDNPBias}{\bf{Bias dependence of nuclear spin pumping.}} {\bf{a, b}} Same bias-dependent photoluminescence spectra as in Supplementary Fig.~\ref{Fig:SDNPPower}a,b reproduced for reference. {\bf{c}} Steady-state hyperfine shift measured as a function of the pump bias $V_{\rm{Pump}}$ and the pump photon energy $E_{\rm{Pump}}$. Measurement is conducted on QD1 at $B_{\rm{z}}=10$~T. Excitation power is $P_{\rm{Exc}}=1.5$~mW with a $\sigma^+$ polarization.}
\end{figure}

In order to investigate the nuclear spin pumping mechanisms further, we fix the pump power at $P_{\rm{Exc}}=1.5$~mW and measure the dependence of $E_{\rm{hf}}$ on $E_{\rm{Pump}}$ and $V_{\rm{Pump}}$. The results are presented in Supplementary Fig.~\ref{Fig:SDNPBias}c, which shows an extension of the data from Fig.~2c of the main text.
The spectral peaks, ascribed to individual excitonic shells, are seen to Stark-shift with the applied bias (Supplementary Fig.~\ref{Fig:SDNPBias}c). The largest hyperfine shift  $\vert E_{\rm{hf}}\vert$ is again observed for the $s$-shell exciton at reverse bias, which varies between $-2.7$ and $-2.1$~V for different individual QDs. The higher-shell excitons ($p$, $d$, etc.) differ from the $s$ shell in that their excitation can be followed by relaxation into the lower energy shell(s). The slightly lower $\vert E_{\rm{hf}}\vert$ can then be ascribed to such relaxation between shells, which may involve flipping of the electron without spin transfer to the nuclei. In other words, relaxation between shells may result in a reduced electron spin polarization, which in turn leads to reduction of the maximum achievable $\vert E_{\rm{hf}}\vert$. Resonant pumping into the $s$-shell can only generate up to two electrons and two holes in the QD. This leaves only four excitonic complexes that can take part in dynamic nuclear spin polarization: neutral exciton $X^0$, neutral biexciton $XX^0$, negatively charged trion $X^{-}$ and positively charged trion $X^{+}$. We have performed the same measurements as in Supplementary Figs.~\ref{Fig:SDNPWl} and \ref{Fig:SDNPBias}c but with $\sigma^-$ pump polarization (producing positive $ E_{\rm{hf}}$). We find that the spectral positions $E_{\rm{Pump}}$ of the optimal nuclear spin pumping peaks under $\sigma^+$ and $\sigma^-$ pumping are split by $\approx900~\mu$eV at $B_{\rm{z}}=10$~T, matching excitonic spectral splitting observed in photoluminescence (Supplementary Fig.~\ref{Fig:SPL1200}). However, it is not possible to determine directly which excitonic feature is responsible for nuclear spin pumping with maximum  $\vert E_{\rm{hf}}\vert$. The biexciton $XX^0$ is unlikely to play a role -- it consists of an electron spin singlet (two electrons, one with spin projection $s_{\rm{z}}=-1/2$ and one with $s_{\rm{z}}=+1/2$) and a hole spin singlet (two holes, one with momentum projection $j_{\rm{z}}=-3/2$ and one with $j_{\rm{z}}=+3/2$) which do not couple to nuclear spins. The $X^{-}$ trion is also unlikely to cause efficient nuclear spin pumping, because the two electrons are in a singlet state. Moreover, spin-selective optical excitation of $X^{-}$ requires prior injection of another spin-flipped electron, for which there is no sufficiently fast process that could compete with rapid tunneling. The $X^{+}$ trion is more likely to contribute to nuclear spin pumping, since it contains only one (spin-polarized) electron. However,  to form a hole spin singlet, $X^{+}$ excitation would still need to be accompanied by hole spin flipping, which would create a bottleneck and slow down the cyclic nuclear spin pumping process. Therefore, we argue that resonant optical excitation of $X^0$ is the most likely route for efficient nuclear spin pumping, as it enables fast optical reexcitation upon tunneling of the previously-excited electron-hole pair out of the QD. In addition to the broad resonance that gives the most efficient nuclear spin pumping ($E_{\rm{Pump}}\approx1.564$~eV at $V_{\rm{Pump}}=-2.3$~V in Supplementary Fig.~\ref{Fig:SDNPBias}c), there are narrower and less efficient Stark-shifting resonances observed at lower $E_{\rm{Pump}}$ (intersecting around $E_{\rm{Pump}}\approx1.552$~eV and $E_{\rm{Pump}}\approx1.546$~eV for the same $V_{\rm{Pump}}=-2.3$~V). These narrow resonances may correspond to optical excitation of $X^{+}$ and $X^{-}$. This would be consistent with photoluminescence spectra (Supplementary Fig.~\ref{Fig:SDNPBias}a), which show that all charged exciton transitions appear on the low-energy side of $X^{0}$. Further investigation, both experimental and theoretical, would be needed to establish with certainty which excitonic transition is responsible for high-efficiency nuclear spin pumping in the regime of fast tunneling.


Supplementary Fig.~\ref{Fig:SDNPBias}c shows that nuclear spins can be polarized at photon energies down to $E_{\rm{Pump}}\approx1.25$~eV, which is well below the ground state QD exciton energy and bulk GaAs bandgap. This mechanism leads to negative hyperfine shifts $E_{\rm{hf}}\approx-30~\mu$eV for both $\sigma^+$ and $\sigma^-$ pumping, suggesting that optical excitation plays a different role, possibly related to activation of charge traps or Auger effect. The trapped-charge hypothesis is further supported by the bias dependence, which shows that the sub-bandgap nuclear spin pumping disappears for $V_{\rm{Pump}}<-0.5$~V and $V_{\rm{Pump}}>+0.9$~V. The buildup time of the nuclear spin polarization is found to be around $\approx 5$~s, which is approximately an order of magnitude slower than nuclear spin pumping via resonant excitation of the QD excitons (see~\ref{subsec:Buildup}). The exact mechanism of sub-bandgap optical nuclear spin pumping is currently unclear and would require a separate systematic investigation.

Focusing on the $s$-shell pumping, we plot the minimum $E_{\rm{hf}}$ (i.e. maximum $\vert E_{\rm{hf}}\vert$) as a function of bias $V_{\rm{Pump}}$ in Supplementary Fig.~\ref{Fig:SDNPTunBias}a. The corresponding photoluminescence intensity of the $s$-shell exciton under above-gap excitation is shown in Supplementary Fig.~\ref{Fig:SDNPTunBias}b. At positive $V_{\rm{Gate}}$ the electric field in the sample is small and the band structure is close to flat-band (right sketch in Supplementary Fig.~\ref{Fig:SDNPTunBias}c). As a result, optical recombination is the only way the photo-generated carriers can leave the QD. The typical radiative lifetimes for the studied type of GaAs QDs are $\approx300$~ps, which creates a bottleneck for how quickly the spin-polarized electrons can be injected into the QD, limiting in turn the rate of the nuclear spin pumping. When the sample gate is tuned towards larger reverse (negative) bias, the maximum hyperfine shift $\vert E_{\rm{hf}}\vert$ increases. At the same time, photoluminescence intensity gradually decreases when $V_{\rm{Gate}}<-1~$V, indicating that electrons and holes tunnel out of the QD (left sketch in Supplementary Fig.~\ref{Fig:SDNPTunBias}c) faster than they can recombine optically. We attribute this correlation to the key role that the tunneling plays in dynamic nuclear spin polarization. Fast tunneling allows to overcome the radiative-recombination bottleneck, so that high-power optical excitation can be used to inject spin polarized electrons at a high rate. The efficiency of nuclear spin pumping peaks at $V_{\rm{Pump}}=-2.3$~V. For even larger reverse bias (i.e. more negative $V_{\rm{Pump}}$) the maximum hyperfine shift $\vert E_{\rm{hf}}\vert$ is seen to reduce slightly. For $V_{\rm{Pump}}<-2.3$~V tunneling becomes even faster, which would require an even higher pump power $P_{\rm{Pump}}\gtrsim10$~mW to maintain steady-state occupation of the QD with spin-polarized electrons. However, when focused into a diffraction-limited spot, such high-power optical excitation causes heating of the crystal lattice, which may result in accelerated nuclear spin relaxation, explaining why the highest achievable nuclear spin polarization is reduced at very large reverse biases.

\subsection{Estimate of the electron tunnel rate}

\begin{figure}
\includegraphics[width=0.7\linewidth]{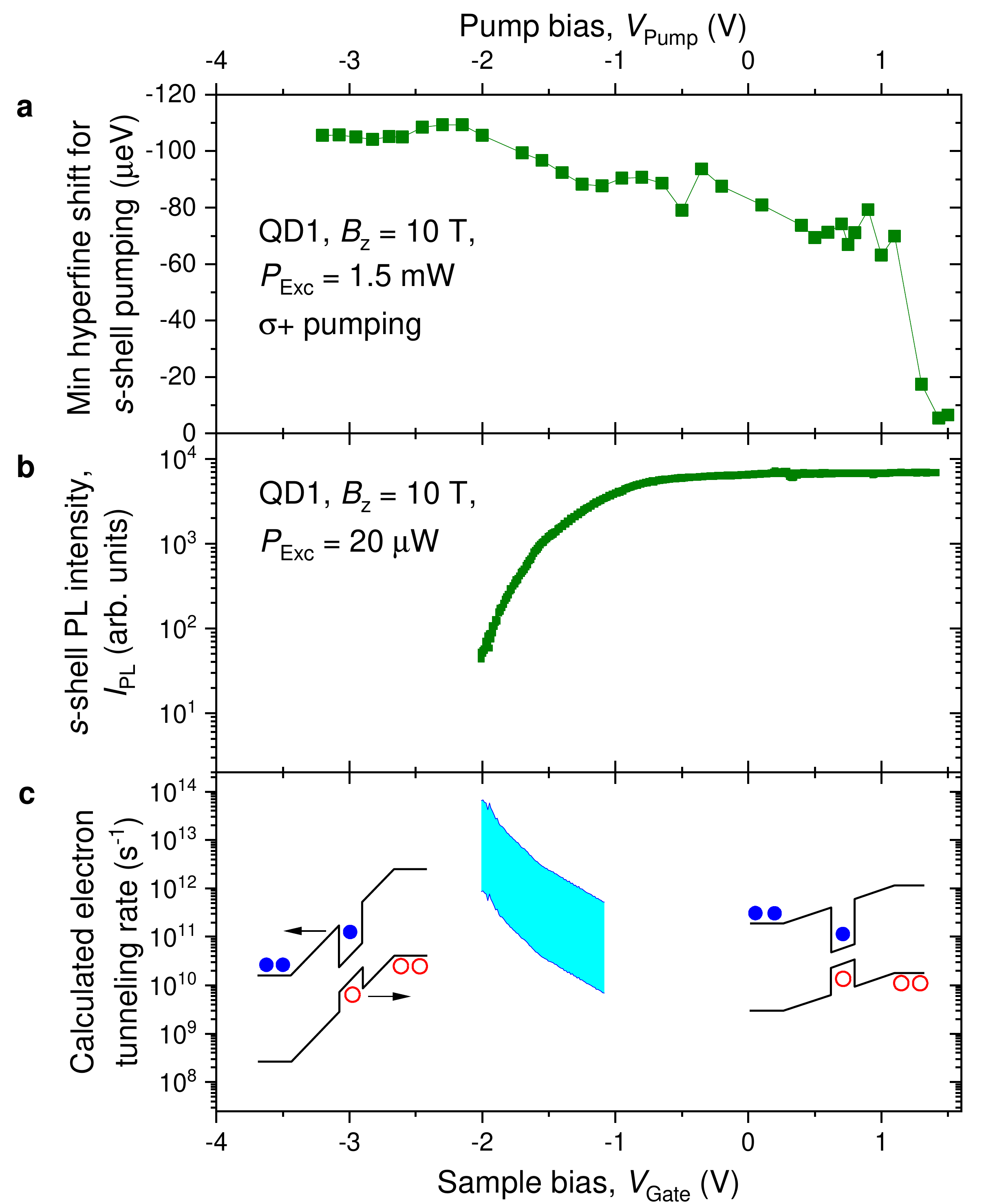}
\caption{\label{Fig:SDNPTunBias}{\bf{Bias dependence of nuclear spin pumping via $s$-shell excitation.}} {\bf{a}} Steady-state hyperfine shift measured as a function of the pump bias $V_{\rm{Pump}}$ for pump photon energy $E_{\rm{Pump}}$ tuned into resonance with the $s$-shell exciton at each bias. Measurement is conducted on QD1 at $B_{\rm{z}}=10$~T. {\bf{b}} Photoluminescence intensity of the $s$-shell exciton measured as a function of bias $V_{\rm{Gate}}$ under 632.8~nm high-power saturation. {\bf{c}} Calculated rate for the electron to tunnel out of the QD, shown as a shaded area between lower and upper bound estimates. Sketches show conduction and valence band profiles at positive and negative $V_{\rm{Gate}}$.}
\end{figure}

In order to quantify the optical nuclear spin pumping process, we estimate the electron tunneling rate using the photoluminescence intensity data. We employ a rate equation approach by considering the probability $p_{\rm{eh}}$ that the QD $s$-shell is occupied by an electron-hole pair. The steady state is defined by the balance between the rate of the optical excitation $\Gamma_{\rm{Exc}}$ and the rates of deexcitation through optical recombination ($\Gamma_{\rm{R}}$) and tunneling ionization ($\Gamma_{\rm{Tun}}$):
\begin{eqnarray}
\begin{aligned}
& (1-p_{\rm{eh}})\Gamma_{\rm{exc}}-p_{\rm{eh}}(\Gamma_{\rm{R}}+\Gamma_{\rm{Tun}})=0 \Rightarrow\\
& p_{\rm{eh}}= \Gamma_{\rm{exc}}/(\Gamma_{\rm{exc}}+\Gamma_{\rm{R}}+\Gamma_{\rm{Tun}})
\label{eq:GammaRateEq}
\end{aligned}
\end{eqnarray}
The intensity of photoluminescence $I_{\rm{PL}}$ is proportional to $p_{\rm{eh}}$. We further assume that the maximum observed PL intensity $I_{\rm{PL,max}}$ corresponds to negligible tunneling rate $\Gamma_{\rm{Tun}}\rightarrow0$. With this assumption we eliminate the unknown PL intensity that would be observed at $p_{\rm{eh}}=1$ and find for the tunneling rate at an arbitrary bias:
\begin{eqnarray}
\begin{aligned}
& \Gamma_{\rm{Tun}}=(\Gamma_{\rm{R}}+\Gamma_{\rm{exc}})\left(\frac{I_{\rm{PL,max}}}{I_{\rm{PL}}}-1\right)
\label{eq:GammaTunPL}
\end{aligned}
\end{eqnarray}
It then follows that the reduction of photoluminescence intensity under reverse bias signifies that the tunneling rate exceeds the sum $(\Gamma_{\rm{R}}+\Gamma_{\rm{exc}})$ of the radiative recombination and optical excitation rates. The radiative recombination time is $1/\Gamma_{\rm{R}}\approx 300$~ps (Ref.~\citep{Schimpf2019}) for the studied type of QDs ($\Gamma_{\rm{R}}\approx3.3\times10^9$~s$^{-1}$). Since photoluminescence intensity of the $s$ shell exciton is saturated, we assume that the excitation rate exceeds the recombination rate $\Gamma_{\rm{Exc}}> \Gamma_{\rm{R}}$. This gives the lower bound estimate for the tunneling rate. The  photoluminescence measurement shown in Supplementary Fig.~\ref{Fig:SDNPTunBias}b was conducted at high excitation power, exceeding the ground state exciton saturation power by a factor of $\approx150$. Thus we write $\Gamma_{\rm{Exc}}\lesssim 150 \Gamma_{\rm{R}}$, which gives an upper bound estimate, since some of the photo-excited electron-hole pairs can recombine from higher shells, without reaching the $s$-shell. Using Supplementary Eq.~\ref{eq:GammaTunPL} we calculate $\Gamma_{\rm{Tun}}$ taking $I_{\rm{PL}}$ and the maximum observed intensity $I_{\rm{PL,max}}$ from Supplementary Fig.~\ref{Fig:SDNPTunBias}b. The range between the lower bound ($\Gamma_{\rm{Exc}}=\Gamma_{\rm{R}}$) and the upper bound ($\Gamma_{\rm{Exc}}=150 \Gamma_{\rm{R}}$) estimates is shown by the shaded area in Supplementary Fig.~\ref{Fig:SDNPTunBias}c. Such direct evaluation of $\Gamma_{\rm{Tun}}$ is possible only when bias is high enough to detect photoluminescence, but also low enough to have a substantial reduction of $I_{\rm{PL}}$ compared to $I_{\rm{PL,max}}$. These estimates show that the tunneling time at $V_{\rm{Gate}}=-2$~V is on the order of $\approx 0.1-1$~ps.

In order to estimate $\Gamma_{\rm{Tun}}$ below $V_{\rm{Gate}}<-2$~V we consider the well-known WKB approximation of the tunneling rate through a triangular barrier (see e.g. Ref.~\citep{VillasBoas2005}). Up to a constant factor, we have:
\begin{eqnarray}
\begin{aligned}
& \Gamma_{\rm{Tun}}\propto\exp\left(-\frac{4\sqrt{2 m_{\rm{e}}}}{\hbar e F_{z}}\vert \epsilon_{\rm{e}} \vert^{3/2}\right),
\label{eq:GammaTun}
\end{aligned}
\end{eqnarray}
where $m_{\rm{e}}^*$ is the effective electron mass, $\epsilon_{\rm{e}}$ is the electron ionization energy and $F_{z}$ is the electric field in the growth direction. The total thickness of the structure between the doped layers is $\approx300$~nm, so the electric field is estimated as $F_{z}=-(V_{\rm{Gate}}-V_{\rm{Gate,0}})/300~{\rm{nm}}$, where $V_{\rm{Gate,0}}=+1.55$~V is the built-in potential of the structure, deduced as a bias where the PL Stark shift vanishes. Based on Ref.~\citep{Adachi2009} we estimate the conduction band discontinuity between GaAs and Al$_{0.33}$Ga$_{0.67}$As to be $U_{\rm{e}}\approx0.28$~eV and take $m_{\rm{e}}^*=0.067 m_{\rm{e}}$, where $m_{\rm{e}}$ is the free electron mass. The $s$-shell photoluminescence of the QD is observed $\approx0.055$~eV above the bulk GaAs. Taking that 0.63 of this offset is in the conduction band, we estimate the ionization energy $\epsilon_{\rm{e}}=U_{\rm{e}}-0.63\times(0.055\; 	{\rm eV})\approx0.25$~eV. Substituting this, we find the numerical estimate $ \Gamma_{\rm{Tun}}\propto\exp\left(-\frac{64.2}{V_{\rm{Gate}}-1.55}\right)$, where $V_{\rm{Gate}}$ is in the units of Volts. Firstly, we see that the exponent is far from saturation in the studied range of $V_{\rm{Gate}}$, so that further reduction of $V_{\rm{Gate}}$ (i.e. making the bias more negative) would result in shortened tunneling times. Secondly, going from  $V_{\rm{Gate}}=-2$~V, where photoluminescence is measurable, to $V_{\rm{Gate}}\in[-2.7,-2.3]$~V, where dynamical nuclear spin polarization is most efficient, results in an order of magnitude higher $\Gamma_{\rm{Tun}}$. Therefore we estimate $\Gamma_{\rm{Tun}}\approx10^{13} - 10^{14}$~s for optimal nuclear spin pumping (tunneling time $\approx0.01 - 0.1$~ps).

We can now independently estimate the optical excitation rate that leads to optimal nuclear spin pumping. Resonance fluorescence intensity, measured on InGaAs/GaAs QDs in the same setup and under similar experimental conditions, saturates at $P_{\rm{Exc}}\approx5$~nW. We assume that $\Gamma_{\rm{Exc}}\approx\Gamma_{\rm{R}}$ at saturation, where the radiative rate is $\Gamma_{\rm{R}}\approx10^9$~s$^{-1}$ for InGaAs QDs. Assuming that the excitation rate scales linearly with optical power, we find that the resonant pumping power of $P_{\rm{Pump}}\approx1.5$~mW, used for optimal nuclear spin pumping, corresponds to $\Gamma_{\rm{Exc}}\approx\frac{1.5~{\rm{mW}}}{5~{\rm{nW}}}10^9~\rm{s}^{-1}\approx3\times10^{14}$~s$^{-1}$. This corresponds to optical reexcitation time of $\approx0.0033$~ps. These estimates yield $\Gamma_{\rm{Exc}}$ that are comparable to or somewhat higher than the above-calculated $\Gamma_{\rm{Tun}}$, as would be expected for a steady-state condition. In other words, having $\Gamma_{\rm{Exc}}\gtrsim\Gamma_{\rm{Tun}}$ ensures that optical excitation generates as many spin-polarized electrons per unit of time as permitted by the  rate of the tunnel escape. The typical linewidths of the excitonic shell peaks in the spectra of the steady state $E_{\rm{hf}}$ at large reverse biases (Supplementary Figs.~\ref{Fig:SDNPWl}, \ref{Fig:SDNPBias}c) are on the order of $\approx10$~meV, which translates to $\approx0.4$~ps, corroborating our order-of-magnitude estimates of $\Gamma_{\rm{Tun}}$.

Apart from the fast cycling of the optically-generated electrons, we expect that fast tunneling also facilitates the nuclear spin pumping by disrupting the formation of coherent nuclear ``dark'' states, which are otherwise predicted to prevent the approach to a near-unity nuclear spin polarization \citep{ImamogluPRL2003,Christ2007}. Moreover, the short (tunneling-limited) lifetime of the electron spin can also be interpreted as spectral broadening of the electron spin levels. Such spectral broadening can facilitate nuclear spin pumping by compensating the energy mismatch of the electron and nuclear spin Zeeman energies, which otherwise inhibits the electron-nuclear spin flip-flops. The role played by tunneling is then similar to the effect that elevated lattice temperatures have on nuclear spin polarization, as studied previously in InGaAs QDs \citep{Urbaszek2007}. Future theoretical work may explain the details of the nuclear spin pumping process by treating optical excitation, tunneling and electron-nuclear spin interactions in a unified framework.

\subsection{Nuclear spin buildup dynamics\label{subsec:Buildup}}

The nuclear spin buildup dynamics under optical pumping are non-exponential (Fig.~4a of the main text). Therefore we fit the data by a sum of two stretched exponentials:
\begin{eqnarray}
\begin{aligned}
E_{\rm{hf}}= E_{\rm{hf,fast}}\left(1-\exp(-(t/\tau_{\rm{fast}})^{\beta_{\rm{fast}}})\right) + E_{\rm{hf,slow}}\left(1-\exp(-(t/\tau_{\rm{slow}})^{\beta_{\rm{slow}}})\right) . \label{eq:Buildup}
\end{aligned}
\end{eqnarray}
The best fits are shown by the solid lines in Fig.~4a of the main text. The fitting parameters for the optimal steady-state nuclear spin pumping of QD2 at $B_{\rm{z}}=10$~T are as follows:
\begin{align}
\begin{array}{l|c|c}
\textrm{Parameter} & \sigma^+\:\textrm{pumping}  &  \sigma^-\:\textrm{pumping}\\
 \hline
\tau_{\rm{fast}} &0.1215\:{\rm{s}} &0.2446\:{\rm{s}}\\ 
\tau_{\rm{slow}} &0.5174\:{\rm{s}} &1.1903\:{\rm{s}}\\ 
E_{\rm{hf,fast}} &43.03\:\mu{\rm{eV}} &-69.46\:\mu{\rm{eV}}\\ 
E_{\rm{hf,slow}} &67.51\:\mu{\rm{eV}} &-41.44\:\mu{\rm{eV}}\\ 
\beta_{\rm{fast}} &0.44 &0.88\\ 
\beta_{\rm{slow}} &0.36 &0.68\\ 
\end{array}\label{eq:BuildupFit}
\end{align}
The fit is empirical in nature, so its parameters should be treated as estimates. Nevertheless, we can establish that the fast initial buildup occurs on a $0.1 - 0.3$~s timescale, slowing down to $0.5 - 1.5$~s when the nuclear spin polarization approaches closer to its steady state. Electron-nuclear spin dynamics become nonlinear when electron spin Zeeman splitting is cancelled by the hyperfine shift. This is manifested in a kink in the nuclear spin buildup dynamics, observed at $E_{\rm{hf}}\approx+50~\mu$eV in Fig.~4a of the main text. From the zero-splitting condition $E_{\rm{hf}}\approx-\mu_{\rm{B}}g_{\rm{e,z}}B_{\rm{z}}$ we can estimate the electron $g$-factor $g_{\rm{e,z}}\approx-0.09$, in agreement with previous measurements on the same structure \citep{MillingtonHotze2022}.

It is also interesting to estimate the rate of the electron-nuclear spin flip-flops. Starting from a depolarized state, it takes $I N$ spin flips to achieve a fully polarized state of $N\approx10^5$ nuclei with spin $I$. The hyperfine shift $E_{\rm{hf}}$ corresponding to a fully polarized state is $I F_{\rm{tot}}\approx110 - 115~\mu$eV. On the other hand, from the nuclear spin buildup dynamics measurements we find that the highest rate of change in $E_{\rm{hf}}$ (the derivative at the start of pumping from a depolarized state) is $\approx600~\mu$eV/s, in agreement with $\tau_{\rm{fast}}$ derived above. Combining these parameters we estimate the electron-nuclear flip-flop rate to be $\approx 8\times10^5$~s$^{-1}$ (i.e. one nucleus flipped every $\approx1~\mu$s). Assuming that the cycling of the spin-polarized electrons is limited by the tunneling rate of $\Gamma_{\rm{Tun}}\approx10^{13}$, we estimate that only a small fraction $\approx10^{-7}$ of the injected electrons transfer their spin to the nuclei, while the rest tunnel out of the QD without polarizing the nuclear spins.


\section{Derivation of nuclear spin polarization\label{sec:Pn}}
\subsection{Nuclear magnetic resonance thermometry of spin-3/2 nuclei}

We describe the state of the nuclear spin ensemble in terms of probabilities $p_{m}$ for each nuclear spin to occupy a state with spin $z$ projections $m$. In case of the spin-3/2 nuclei $m\in\{-3/2,-1/2,+1/2,+1/2\}$. For the state induced by optical dynamical nuclear polarization we model these probabilities using the Boltzmann distribution:
\begin{eqnarray}
p_m = e^{m \beta}/\sum_{m=-I}^{+I}e^{m \beta},\label{eq:pBoltz}
\end{eqnarray}
where $\beta=h\nu_{\rm{L}}/k_\textrm{b} T_\textrm{N}$ is the dimensionless inverse temperature, expressed in terms of the nuclear spin Larmor frequency $\nu_{\rm{L}}$ and the spin temperature $T_\textrm{N}$ ($h$ is the Planck's constant and $k_\textrm{b}$
is the Boltzmann constant). In this definition we assume $\nu_{\rm{L}}>0$ and $B_{\rm{z}}>0$, so that $m=+I$ is the ground state for nuclei with $\gamma>0$, in agreement with Supplementary Eq.~\ref{Eq:HZN}. For spin $I$=1/2 where $m=\pm1/2$ any statistical distribution is described by Supplementary Eq. \ref{eq:pBoltz} with some $T_\textrm{N}$. By contrast, for $I>$1/2 Supplementary Eq. \ref{eq:pBoltz} states the non-trivial nuclear spin temperature hypothesis \cite{GoldmanBook} -- previous experimental studies on low-strain epitaxial quantum dots \citep{Chekhovich2017} have shown its validity for the state induced by optical dynamical nuclear polarization. Nuclear spin polarization degree is defined as
\begin{equation}
P_\textrm{N}=\frac{1}{I}\sum_{m=-I}^{+I}mp_{m}.\label{eq:PN}
\end{equation}
For the Boltzmann distribution of Supplementary Eq.~\ref{eq:pBoltz} the polarization degree is given by the Brillouin function:
\begin{eqnarray}
\begin{aligned}
&P_{\rm{N}}=\frac{1}{2I}\left( (2I+1)\coth[(I+1/2)\beta]-\coth[\beta/2] \right).\label{eq:PNExp}
\end{aligned}
\end{eqnarray}
It is worth noting that the polarization degree $P_{\rm{N}}$ and the dimensionless inverse temperature $\beta$ provide more relevant description of the nuclear spin state than the temperature $T_{\rm{N}}$. Indeed, when the Larmor frequency $\nu_L$ is changed (by varying the external magnetic field) $P_{\rm{N}}$ and $\beta$ are preserved, whereas $T_{\rm{N}}$ is not constant for a given optically-pumped nuclear spin state. The temperature $T_{\rm{N}}$ only gains physical meaning at low magnetic fields, comparable to the local nuclear dipolar fields. $P_{\rm{N}}$ and $\beta$ are also related to entropy \citep{Knuuttila2001,Wenckebach2008} -- the minimum in entropy is achieved only for $P_{\rm{N}}=\pm1$ ($\beta\rightarrow\pm\infty$).

The hyperfine shift experienced by the quantum dot exciton is linearly proportional to the nuclear spin polarization degree $P_{\rm{N}}$:
\begin{eqnarray}
\begin{aligned}
E_{\rm{hf}}=\sum_j F^{(j)} I_j P_{{\rm{N}},j}, \label{eq:Ehf}
\end{aligned}
\end{eqnarray}
where the sum is over individual isotope species. Although the hyperfine shift $E_{\rm{hf}}$ can be measured accurately from the photoluminescence spectra, the proportionality factor $F^{(j)}$ depends not only on the material's hyperfine constants $A^{(j)}$, but also on the leakage of the electron wavefunction into the AlGaAs barrier. Since it is difficult to estimate this leakage independently, the measurement of $E_{\rm{hf}}$ alone is not suitable for accurate derivation of $P_{\rm{N}}$. The unknown proportionality factor between $E_{\rm{hf}}$ and $P_{\rm{N}}$ can be eliminated for $I>1/2$ nuclei, provided that it is possible to address selectively the magnetic dipole transitions between states with spin projections $m$ and $m+1$. For example, if a long radiofrequency (Rf) pulse is applied to saturate the $m\leftrightarrow m+1$ NMR transition, it equalizes the
populations of these states. The resulting final population probabilities $p_{m}$, $p_{m+1}$ both equal the average $(p_{m}+p_{m+1})/2$ of their initial populations. For an ideal selective NMR Rf excitation the population probabilities of all other
nuclear spin states remain unchanged. 

One can then substitute Supplementary Eq.~\ref{eq:PN} into 
Supplementary Eq.~\ref{eq:Ehf} to calculate the change in the
optically detected hyperfine shift $\Delta E_\textrm{hf}$ resulting from selective saturation of a single NMR transition $m\leftrightarrow m+1$. For example, for $m=+1/2$ and
$m+1=+3/2$ of the $j$-th isotope, we calculate $\Delta
E_\textrm{hf}^{+1/2\leftrightarrow+3/2}=F^{(j)}\left[(+\frac{3}{2})\frac{p_{+3/2}+p_{+1/2}}{2}+(+\frac{1}{2})\frac{p_{+3/2}+p_{+1/2}}{2}\right]-F^{(j)}\left[(+\frac{3}{2})p_{+3/2}+(+\frac{1}{2})p_{+1/2}\right]=-F^{(j)}
(p_{+3/2}-p_{+1/2})/2$. This result has a simple interpretation that the hyperfine shift variation $\Delta
E_\textrm{hf}$ depends only on the difference in the initial
populations of the states that are selectively saturated with Rf.

In the same manner, simultaneous selective saturation of the NMR transitions
$m \leftrightarrow m+1$ and $m+1 \leftrightarrow m+2$ leads to complete averaging of the populations of the three involved spin states. Their final population probabilities become $p_{m}, p_{m+1},
p_{m+2}\rightarrow (p_{m}+p_{m+1}+p_{m+2})/3$. Saturation of all three
NMR transitions of spin $I$=3/2 nuclei leads to complete depolarization and equal populations of all four spin states $p_{-3/2}=
p_{-1/2}= p_{+1/2}= p_{+3/2} = 1/4$. Using
Supplementary Eqs.~\ref{eq:PN} and \ref{eq:Ehf} we evaluate the changes in the
hyperfine shift $\Delta E_{{\rm{hf}},j}$ arising from the $j$-th isotope, to arrive to the following results, derived previously in Ref.~\citep{Chekhovich2017}:
\begin{eqnarray}
\begin{aligned}
&\Delta E_{{\rm{hf}},j}^{m\leftrightarrow m+1}=-F^{(j)}
(p_{m+1,j}-p_{m,j})/2=-F^{(j)} \frac{e^{(m+1)\beta_j}-e^{m\beta_j}}{4\cosh(\beta_j/2)+4\cosh(3\beta_j/2)},\\
&\Delta E_{{\rm{hf}},j}^{m\leftrightarrow m+2}=-F^{(j)} (p_{m+2,j}-p_{m,j})=-F^{(j)} e^{(m+1)\beta_j} \sinh(\beta_j/2)/\cosh(\beta_j),\\
&\Delta E_{{\rm{hf}},j}^{-I\leftrightarrow +I}=-F^{(j)} P_{{\rm{N}},j}I_j=-F^{(j)}[3/2+1/\cosh(\beta_j)]\tanh(\beta_j/2).\label{eq:DEhf}
\end{aligned}
\end{eqnarray}
The last expression in each of these equations is obtained by substituting the Boltzmann distribution (Supplementary Eq.~\ref{eq:pBoltz}) for spin $I=3/2$.

\subsection{Corrections for the nuclei with small or inverted quadrupolar shifts\label{subsec:NMROverlap}}

For a fully resolved NMR triplet, Supplementary Eq.~\ref{eq:DEhf} is sufficient to extract the inverse temperatures $\beta_j$ and derive the polarization degree of the spin-3/2 nuclei. In a real semiconductor system the separation of the quadrupolar NMR components is not perfect. Here we examine the role that the nuclei with small or inverted quadrupolar shift $\nu_{\rm{Q}}$ have on the derivation of nuclear spin polarization from experimental data. The case of an experiment where a radiofrequency comb is used to saturate two out of three NMR transitions is considered in Supplementary Fig.~\ref{Fig:SContribComb}. For unstrained GaAs, nuclear quadrupolar effects are absent ($\nu_{\rm{Q}}=0$) and all NMR transitions of the spin-3/2 nucleus appear at the same Larmor frequency $\nu_{\rm{L}}$. Strain induces quadrupolar effects which are characterized to first order by the shift $\nu_{\rm{Q}}$. In all our experiments $\nu_{\rm{Q}}\ll \nu_{\rm{L}}$, so that first-order approximation is valid. The central transition $-1/2 \leftrightarrow +1/2$ between nuclear states with spin projections $m=\pm1/2$ is unaffected by quadrupolar shifts in the first order, hence its NMR frequency is  $\nu_{\rm{L}}$ for all nuclei (vertical solid line in Supplementary Fig.~\ref{Fig:SContribComb}). The satellite transitions are affected by quadrupolar shifts: the NMR frequency of the $-3/2 \leftrightarrow -1/2$ transition is $\nu_{\rm{L}}+\nu_{\rm{Q}}$, whereas the NMR frequency of the $+1/2 \leftrightarrow +3/2$ transition is $\nu_{\rm{L}}-\nu_{\rm{Q}}$ (solid lines in Supplementary Fig.~\ref{Fig:SContribComb} with slopes $+1$ and $-1$, respectively). 

\begin{figure}
\includegraphics[width=0.98\linewidth]{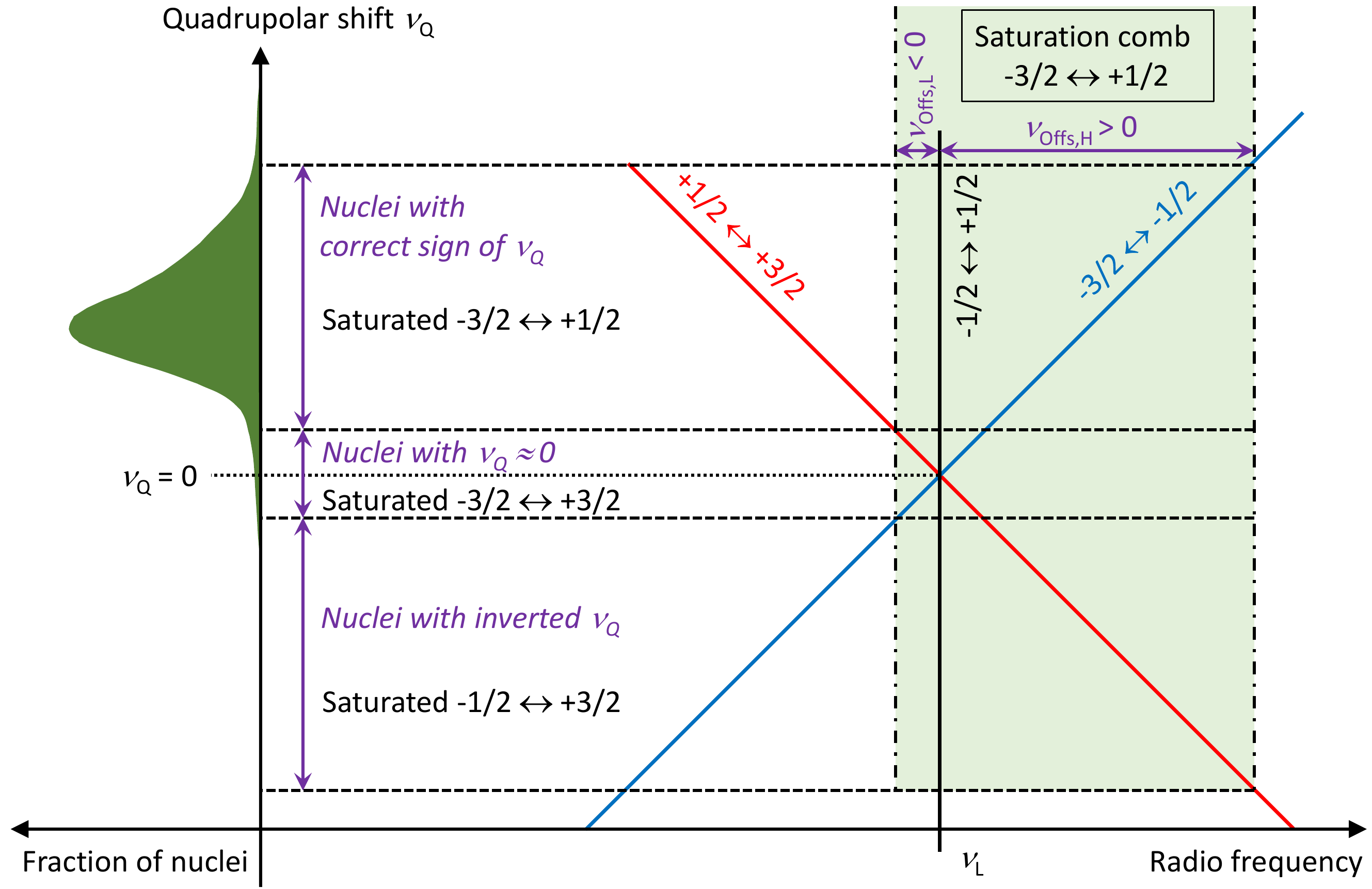}
\caption{\label{Fig:SContribComb}{\bf{Effect of the nuclei with different quadrupolar shifts in a two-transition comb saturation measurement.}}}
\end{figure}

The strain varies within the QD volume, so there is a statistical distribution of $\nu_{\rm{Q}}$ values within the ensemble of the nuclei (sketched in the left part of Supplementary Fig.~\ref{Fig:SContribComb}). The majority of the $^{75}$As nuclei have a positive quadrupolar shift $\nu_{\rm{Q}}>0$ (for Ga nuclei the shift is predominantly negative $\nu_{\rm{Q}}<0$). Therefore, if we want to saturate simultaneously the two NMR transitions $-3/2 \leftrightarrow -1/2$ and $-1/2 \leftrightarrow +1/2$ (labelled $-3/2 \leftrightarrow +1/2$ for brevity) we choose a radiofrequency comb band sketched by the shaded area in Supplementary Fig.~\ref{Fig:SContribComb}. The low-offset edge of the band at frequency $\nu_{\rm{L}}+\nu_{\rm{Offs,L}}$ (with negative $\nu_{\rm{Offs,L}}<0$) is tuned just below the Larmor frequency, in order to saturate the narrow $-1/2 \leftrightarrow +1/2$ transition. The high-offset edge of the band  $\nu_{\rm{L}}+\nu_{\rm{Offs,H}}$ (with positive $\nu_{\rm{Offs,H}}>0$) is chosen to be far enough from the Larmor frequency to cover the $-3/2 \leftrightarrow -1/2$ satellite transition for most nuclei. The typical values in two-transition comb saturation experiments are $\nu_{\rm{Offs,L}}=-5$~kHz, $\nu_{\rm{Offs,L}}=+178$~kHz for $^{75}$As and $\nu_{\rm{Offs,L}}=+5$~kHz, $\nu_{\rm{Offs,L}}=-80$~kHz for $^{69}$Ga. For the $-1/2 \leftrightarrow +3/2$ two-transition saturation the values of $\nu_{\rm{Offs,L}}$ and $\nu_{\rm{Offs,H}}$ are inverted.

As can be seen in Supplementary Fig.~\ref{Fig:SContribComb} the nominal $-3/2 \leftrightarrow +1/2$ comb saturates the desired transitions for the majority of nuclei, which have $-\nu_{\rm{Offs,L}}<\nu_{\rm{Q}}<\nu_{\rm{Offs,H}}$ (note that the lower bound $-\nu_{\rm{Offs,L}}$ is positive for a negative $\nu_{\rm{Offs,L}}<0$). These nuclei give a correct contribution to the Rf-induced hyperfine shifts. But there are also several cases, where nuclei give contributions that differ from those intended. For a fraction of nuclei with small quadrupolar shifts $\nu_{\rm{Offs,L}}<\nu_{\rm{Q}}<-\nu_{\rm{Offs,L}}$ all three NMR transitions are excited by the comb, resulting in full depolarization of such nuclei. Furthermore, for those $^{75}$As nuclei where quadrupolar shift is inverted $-\nu_{\rm{Offs,H}}<\nu_{\rm{Q}}<\nu_{\rm{Offs,L}}$, the $-3/2 \leftrightarrow -1/2$ transition will be out of resonance with the Rf band, while the $+1/2 \leftrightarrow +3/2$ transition will be saturated. Such nuclei will produce hyperfine shifts that would correspond to the $-1/2 \leftrightarrow +3/2$ two-transition saturation, rather than the intended $-3/2 \leftrightarrow +1/2$ saturation. Finally, for a small fraction of nuclei with very large absolute quadrupolar shifts $\vert\nu_{\rm{Q}}\vert > \vert\nu_{\rm{Offs,H}}\vert$, only the central $-1/2 \leftrightarrow +1/2$ transition will be saturated. Thus, introducing the empirical coefficients $c$, the observed hyperfine shift $\Delta E_\textrm{hf,Obs}^{-3/2\leftrightarrow +1/2}$ in the $-3/2 \leftrightarrow +1/2$ two-transition saturation experiment can be written as:
\begin{eqnarray}
\begin{aligned}
&\Delta E_\textrm{hf,Obs}^{-3/2\leftrightarrow +1/2}=c_{\rm{Sat,Ideal}} \Delta E_\textrm{hf}^{-3/2\leftrightarrow +1/2}+\\
&+c_{\rm{Sat,Full}} \Delta E_\textrm{hf}^{-3/2\leftrightarrow +3/2}+c_{\rm{Sat,Inv}} \Delta E_\textrm{hf}^{-1/2\leftrightarrow +3/2}+c_{\rm{Sat,CT}} \Delta E_\textrm{hf}^{-1/2\leftrightarrow +1/2},
\label{eq:EhfObsSat}
\end{aligned}
\end{eqnarray}
where we have dropped the isotope index. If the quadrupolar NMR triplet is fully resolved, then $c_{\rm{Sat,Ideal}}=1$ with all other $c$ coefficient equal to zero. In a real quantum dot $c_{\rm{Sat,Ideal}}<1$ and the remaining coefficients are non-zero, so that the observed hyperfine shift $\Delta E_\textrm{hf,Obs}^{-3/2\leftrightarrow +1/2}$ deviates from the ideal $\Delta E_\textrm{hf}^{-3/2\leftrightarrow +1/2}$. For example, if all nuclei are in a $m=+3/2$ state (that is $P_{\rm{N}}=+1$), the expected ideal $\Delta E_\textrm{hf}^{-3/2\leftrightarrow +1/2}=0$. In reality, due to the nonzero contributions of the fully-saturated nuclei ($c_{\rm{Sat,Full}}>0$) and the nuclei with an inverted quadrupolar shift ($c_{\rm{Sat,Inv}}>0$) the observed hyperfine shift $\Delta E_\textrm{hf,Obs}^{-3/2\leftrightarrow +1/2}$ is nonzero even if nuclei are fully polarized.

The expression for the hyperfine shift in the other two-transition experiment $\Delta E_\textrm{hf,Obs}^{-1/2\leftrightarrow +3/2}$ can be obtained from Supplementary Eq.~\ref{eq:EhfObsSat} by changing the sings of all the $m$ indices. It is worth noting that in the experiment with intentional saturation of all three-transitions ($-3/2\leftrightarrow +3/2$) all nuclei get fully depolarized as long as the satellite transitions fit within the Rf band. In the three-transition experiments we use frequency combs with total widths of 347~kHz ($^{69}$Ga) and 578~kHz ($^{75}$As), centred at the Larmor frequency $\nu_{\rm{L}}$. These widths are sufficient for complete depolarization of essentially all the nuclei of the quantum dot and the surrounding barriers. Therefore, the three-transition non-selective saturation measurement is more robust than the selective two-transition Rf depolarization. 

\begin{figure}
\includegraphics[width=0.98\linewidth]{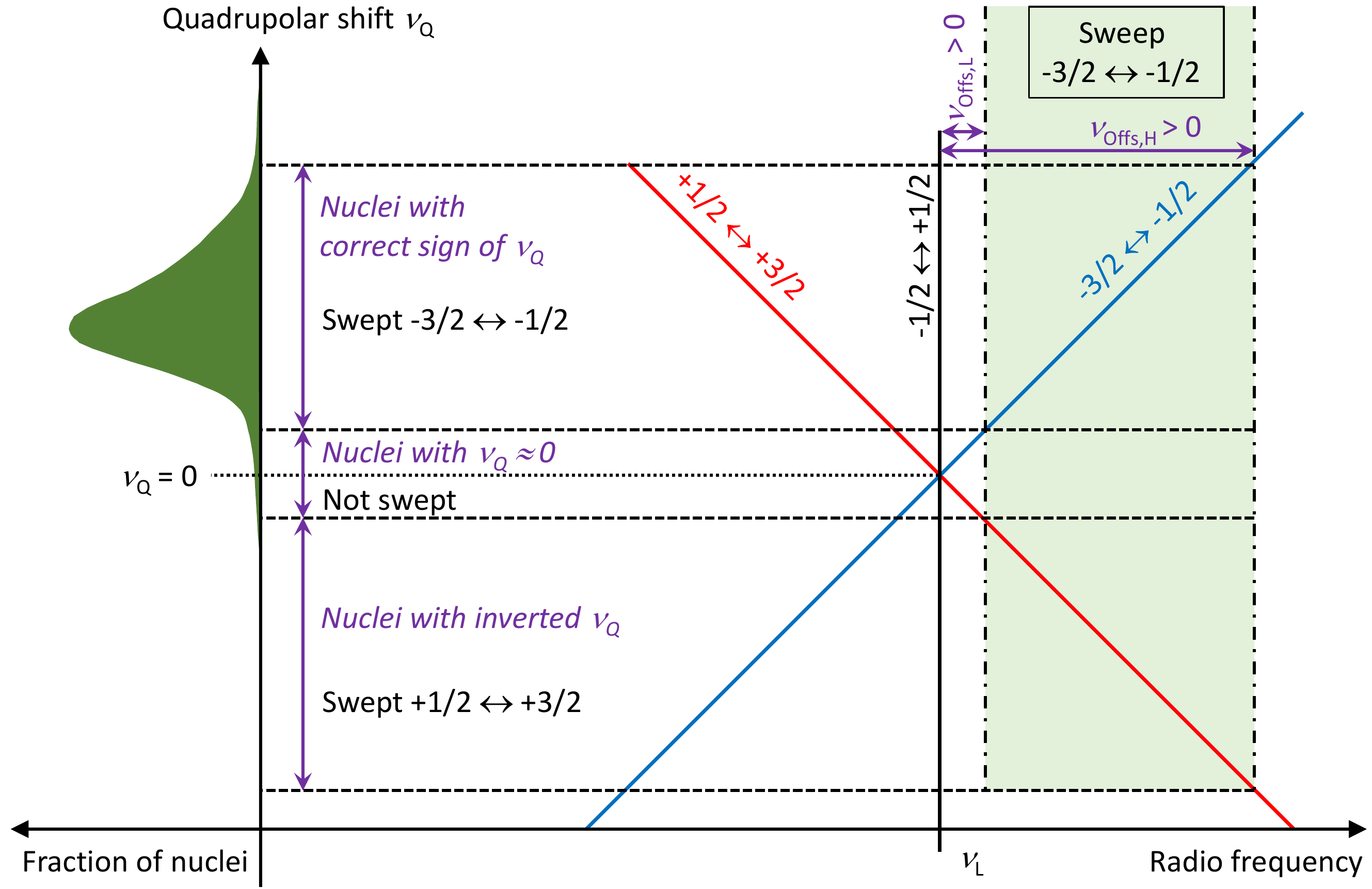}
\caption{\label{Fig:SContribSwp}{\bf{Effect of the nuclei with different quadrupolar shifts in a single-transition adiabatic inversion measurement.}}}
\end{figure}

Similar analysis applies to selective Rf excitation of a single NMR transition (Supplementary Fig.~\ref{Fig:SContribSwp}. Here, rather than saturating the NMR resonance we perform a radiofrequency sweep, which adiabatically inverts the populations of the chosen pair of spin states. Adiabatic sweep has the advantage of doubling the hyperfine shift compared to saturation -- this simple relation holds only if a single NMR transition is excited. By contrast, for an adiabatic sweep over multiple quadrupolar NMR transitions the result is more complicated, making comb saturation preferable for two-transition and three-transition NMR excitation. The radiofrequency is always swept in the direction away from the central NMR transition, starting at $\nu_{\rm{L}}+\nu_{\rm{Offs,L}}$ and ending at $\nu_{\rm{L}}+\nu_{\rm{Offs,H}}$. The amplitude and the sweep rate are derived from calibration measurements discussed in \ref{subsec:SwpCalib}. For adiabatic sweeping of the $-3/2\leftrightarrow -1/2$ transition we use $\nu_{\rm{Offs,L}}=+8$~kHz, $\nu_{\rm{Offs,L}}=+180$~kHz for $^{75}$As and $\nu_{\rm{Offs,L}}=-5$~kHz, $\nu_{\rm{Offs,L}}=-50$~kHz for $^{69}$Ga. For the sweep over the $+1/2 \leftrightarrow +3/2$ transition the values of $\nu_{\rm{Offs,L}}$ and $\nu_{\rm{Offs,H}}$ are inverted. The  $-3/2\leftrightarrow -1/2$ sweep works as designed for the majority of nuclei with $\nu_{\rm{Offs,L}}<\nu_{\rm{Q}}<\nu_{\rm{Offs,H}}$ (note that both $\nu_{\rm{Offs,L}}$ and $\nu_{\rm{Offs,H}}$ are positive). For the small number of nuclei with $-\nu_{\rm{Offs,H}}<\nu_{\rm{Q}}<-\nu_{\rm{Offs,L}}$ the nominal $-3/2\leftrightarrow -1/2$ sweep results in an adiabatic inversion of the $+1/2\leftrightarrow +3/2$ transition instead. The remaining nuclei with $\vert\nu_{\rm{Q}}\vert<\vert\nu_{\rm{Offs,L}}\vert$ or $\vert\nu_{\rm{Q}}\vert > \vert\nu_{\rm{Offs,H}}\vert$ are not affected by the Rf sweep. Thus the observed hyperfine shift $\Delta E_\textrm{hf,Obs}^{-3/2\leftrightarrow -1/2}$ in the $-3/2 \leftrightarrow -1/2$ single-transition sweep experiment can be written as:
\begin{eqnarray}
\begin{aligned}
&\Delta E_\textrm{hf,Obs}^{-3/2\leftrightarrow -1/2}=2\left(c_{\rm{Swp,Ideal}} \Delta E_\textrm{hf}^{-3/2\leftrightarrow -1/2}+c_{\rm{Swp,Inv}} \Delta E_\textrm{hf}^{+1/2\leftrightarrow +3/2}\right)
\label{eq:EhfObsSwp}
\end{aligned}
\end{eqnarray}
The main difference of the single-transition selective excitation is that the nuclei with small quadrupolar shifts $\nu_{\rm{Q}}$ are eliminated from the measured hyperfine shifts. This is preferred over the two-transition saturation measurement, where such nuclei are fully depolarized, resulting in a parasitic hyperfine shift characterised by the  $c_{\rm{Sat,Full}}$ coefficient in Supplementary Eq.~\ref{eq:EhfObsSat}.

\subsection{NMR spectra of the QD nuclei \label{subsec:NMRSpec}}

\begin{figure}
\includegraphics[width=0.98\linewidth]{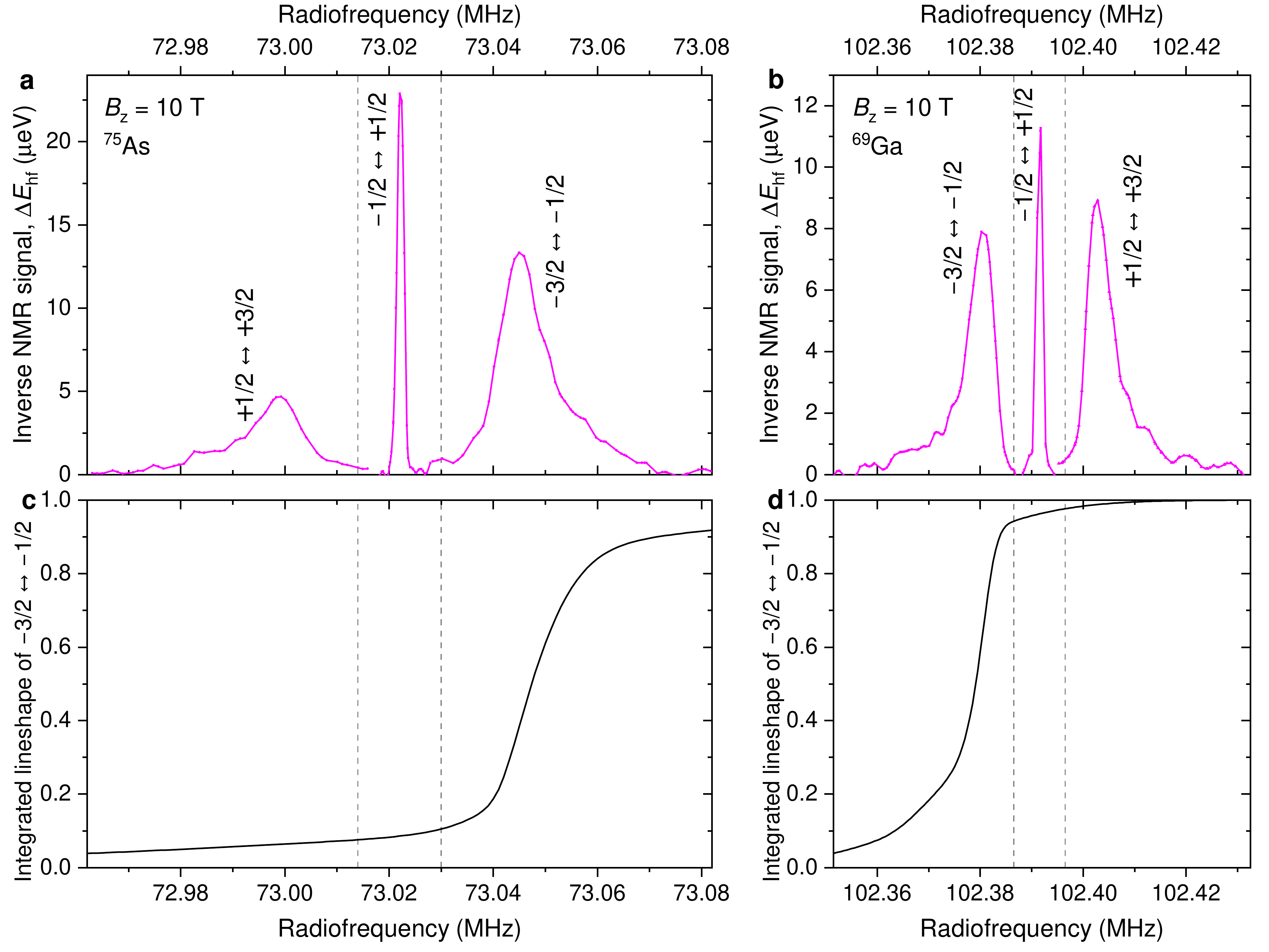}
\caption{\label{Fig:SInvNMR}{\bf{Nuclear magnetic resonance spectra of a single quantum dot.}} {\bf{a}} Nuclear magnetic resonance spectra of the $^{75}$As nuclei measured in QD1 at $B_{\rm{z}}=10$~T using the inverse NMR technique for signal enhancement \citep{Chekhovich2012}. The $-1/2 \leftrightarrow +1/2$ central transition is measured with a 2~kHz resolution, while the satellites $-3/2 \leftrightarrow -1/2$ and $+1/2 \leftrightarrow +3/2$ are measured with a 6~kHz resolution. Vertical dashed lines are offset from the central transition by $\pm8$~kHz and indicate the starting points of the frequency sweeps over the satellite peaks.  {\bf{b}} Same as (a) but for $^{69}$Ga nuclei. The $-1/2 \leftrightarrow +1/2$ central transition is measured with a 2~kHz resolution, while the satellites are measured with a 4~kHz resolution. Vertical dashed lines are offset from the central transition by $\pm5$~kHz. {\bf{c}} Integrated lineshape of the $-3/2 \leftrightarrow -1/2$ transition of $^{75}$As, derived from experiments on a stressed piece of the same semiconductor QD sample. {\bf{d}} Same as (c) but for the $-3/2 \leftrightarrow -1/2$ transition of $^{69}$Ga.}
\end{figure}

Supplementary Fig.~\ref{Fig:SInvNMR}a,b shows typical nuclear magnetic resonance spectra measured on $^{75}$As and $^{69}$Ga nuclei in QD1. The three magnetic dipole transitions of each of the spin-3/2 isotopes are split due to the natural elastic strain within the quantum dot volume, arising most likely from the residual lattice mismatch between the GaAs QD and the AlGaAs barriers. Although the NMR  triplet is well resolved, there is a few-percent overlap between the spectral components. When quantifying nuclear spin polarization degrees close to unity, such overlap must be taken into consideration. In order to quantify the spectral overlap, we study a piece of the same QD sample but subject to a uniaxial stress along the $[110]$ crystallographic direction (i.e. the strain is applied perpendicular to the sample growth direction). Nuclear quadrupolar shifts induced by the external stress significantly exceed the intrinsic quadrupolar shifts. As a result the NMR triplet is fully resolved, as can be seen in the inverse NMR spectra of Supplementary Figs.~\ref{Fig:SNMRStress}a,b where we focus on the $-1/2 \leftrightarrow +1/2$ and $-3/2 \leftrightarrow -1/2$ transitions. 

The spectral shapes of the $-3/2 \leftrightarrow -1/2$ satellites measured in a stressed QD sample are similar to those in the unstressed sample (Supplementary Fig.~\ref{Fig:SInvNMR}). The satellite lineshape consists mainly of an asymmetric peak, but also shows evidence of spectral wings that are broad enough to overlap with the $-1/2 \leftrightarrow +1/2$ central transition in an unstressed sample. In principle, the overlap can be derived by integrating the relevant part of the satellite lineshape, measured with inverse NMR and shown in Supplementary Figs.~\ref{Fig:SNMRStress}a,b. However, this approach is vulnerable to noise -- it is more efficient to incorporate integration into the NMR spectroscopy method \citep{Ragunathan2019Thesis,Zaporski2022}. Such an integral NMR measurement is performed by selectively saturating all nuclear spin transitions within a certain spectral band. The high-frequency edge of the saturating band (implemented as a frequency comb) is kept fixed. The low-frequency edge is scanned and the resulting change in the hyperfine shift $\Delta E_\textrm{hf}$ is measured. This dependence of $\Delta E_\textrm{hf}$  reveals the fraction of the nuclei covered by the saturating Rf band, and therefore provides a scaled definite integral of the NMR lineshape. For $^{75}$As nuclei the fixed-frequency edge of the Rf band is detuned by $+500$~kHz from the central transition to ensure that the entire $-3/2 \leftrightarrow -1/2$ transition can be covered. For $^{69}$Ga nuclei the fixed-frequency edge of the Rf band is detuned by $+230$~kHz from the central transition, so that the entire $-1/2 \leftrightarrow +1/2$ and $+1/2 \leftrightarrow +3/2$ transitions are also included in the band, in order to amplify the integral NMR signal of the $-3/2 \leftrightarrow -1/2$ satellite. 

\begin{figure}
\includegraphics[width=0.98\linewidth]{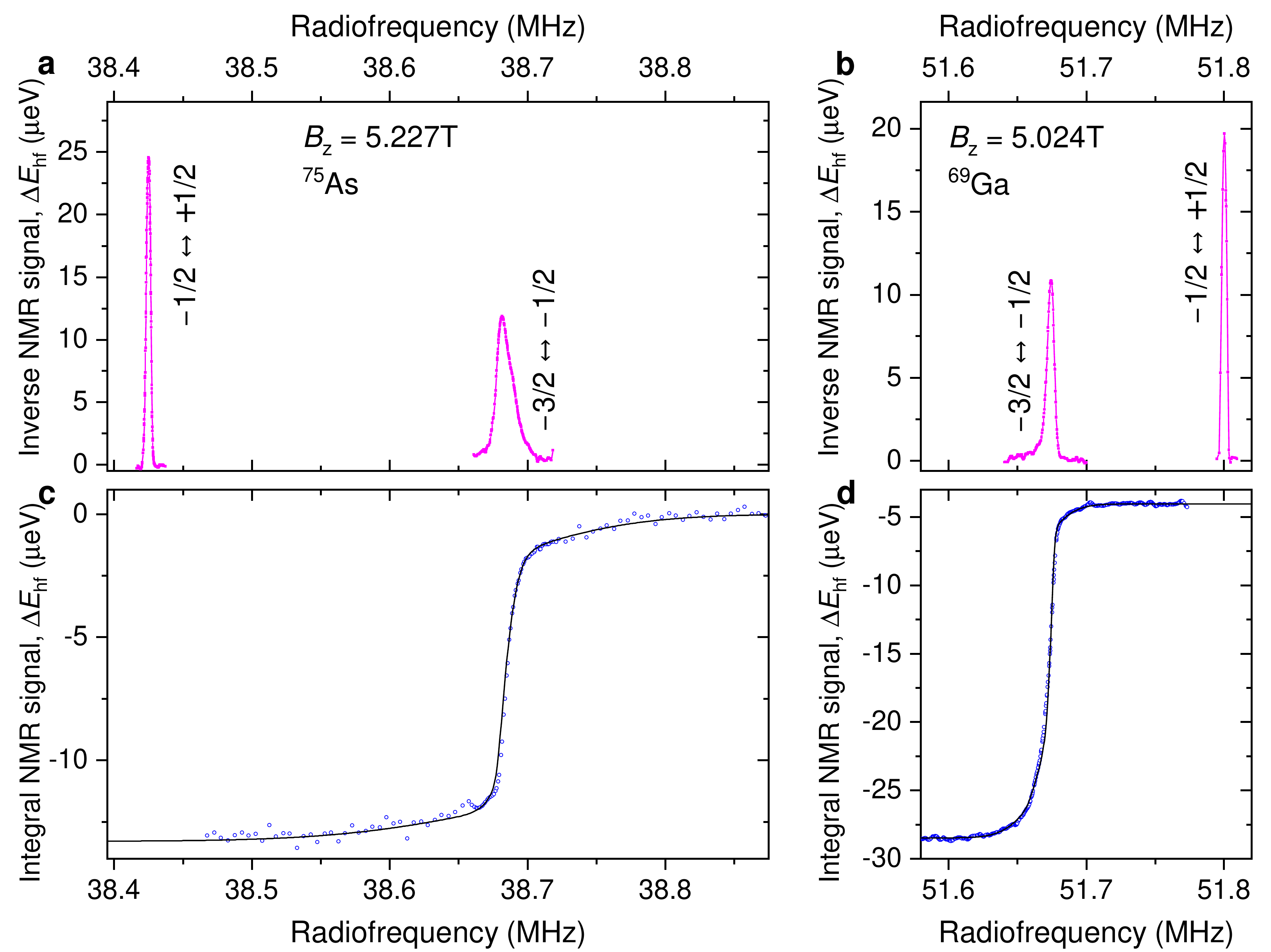}
\caption{\label{Fig:SNMRStress}{\bf{Nuclear magnetic resonance spectra of a quantum dot under external uniaxial stress.}} {\bf{a}} Inverse NMR spectrum of $^{75}$As at $B_{\rm{z}}=5.227$~T measured with a 4~kHz resolution. {\bf{b}} Inverse NMR spectrum of $^{69}$Ga at $B_{\rm{z}}=5.024$~T measured with a 4~kHz resolution. {\bf{c}} Integral saturation NMR spectrum of $^{75}$As measured (symbols) under the same condition as inverse NMR in (a). Line shows a smoothed fitted profile. {\bf{d}} Integral saturation NMR spectrum of $^{69}$Ga measured under the same condition as inverse NMR in (b).}
\end{figure}

Supplementary Fig.~\ref{Fig:SNMRStress}c shows the integral NMR spectrum of the $-3/2 \leftrightarrow -1/2$ transition of $^{75}$As nuclei in a stressed sample. The steep rise in the integral signal matches the position of the sharp peak in the inverse NMR spectrum of  Supplementary Fig.~\ref{Fig:SNMRStress}a. However, we also observe the slopes that stretch as far as $\approx\pm100$~kHz from the satellite peak maximum, indicating the contribution of a broad NMR signal. Broad spectral features are also observed in Supplementary Fig.~\ref{Fig:SNMRStress}d for $^{69}$Ga nuclei, though in a narrower spectral range and with an overall smaller contribution. This broad background can be ascribed to the NMR signal from AlGaAs barriers or any Al atoms diffusing into the GaAs QD layer \citep{Zaporski2022}. The Al atoms that randomly replace the Ga atoms distort the tetrahedral symmetry of the four nearest neighbours surrounding each As atom. The resulting unit-cell-scale strain results in pronounced quadrupolar shifts. By contrast, all Ga atoms have four identical As atoms as nearest neighbours. Therefore, Ga atoms are affected by Al/Ga random alloying only through next-nearest neighbours, explaining why the broad nuclear quadrupolar wings are smaller than for As nuclei.

Integral NMR spectra are processed in order to derive the correction coefficients. The experimental data is first smoothed (lines in Supplementary Fig.~\ref{Fig:SNMRStress}c,d) by fitting with a sum of three skew normal distribution peaks. The integral lineshapes are then normalized and shifted along the frequency scale to have the $-3/2 \leftrightarrow -1/2$ satellite NMR peaks in the stressed sample (Supplementary Fig.~\ref{Fig:SNMRStress}a,b) match the peak positions in the unstressed sample (Supplementary Fig.~\ref{Fig:SInvNMR}a,b). The resulting integrals of the $-3/2 \leftrightarrow -1/2$ lineshapes are shown by the solid lines in Supplementary Fig.~\ref{Fig:SInvNMR}c,d and are used to derive the $c$ coefficients in Supplementary Eqs.~\ref{eq:EhfObsSat},~\ref{eq:EhfObsSwp} . For example, the vertical dashed lines indicate the starting points of the frequency sweeps over the satellite peaks. The integral value at the lower starting point for $^{75}$As is $\approx0.07$ and approximately corresponds to the fraction $c_{\rm{Swp,Inv}}$ of the nuclei where the $+1/2 \leftrightarrow +3/2$ satellite is swept instead of the intended $-3/2 \leftrightarrow -1/2$. The difference of the integral at the higher and lower sweep starting points gives approximately the fraction of nuclei ($\approx0.03$) that are not swept at all. The summary of all the coefficients derived from the integrated lineshapes can be found in the following table:
\begin{align}
\begin{array}{l|c|c}
\textrm{Coefficient} & ^{75}{\rm{As}} & ^{69}{\rm{Ga}}   \\
 \hline
c_{\rm{Sat,Ideal}} & 0.8979 & 0.9424 \\
c_{\rm{Sat,Full}} & 0.0170 & 0.0341 \\
c_{\rm{Sat,Inv}}& 0.0750 & 0.0234 \\
c_{\rm{Sat,CT}} & 0.0100 & 0.0003 \\
c_{\rm{Swp,Ideal}} & 0.8893 & 0.9272 \\
c_{\rm{Swp,Inv}} & 0.0721 & 0.0234 \\
\end{array}\label{eq:cCoefs}
\end{align}
It can be seen that the contributions of the ideal signals are higher for the $^{69}$Ga nuclei due to their smaller quadrupolar broadening. As a result, nuclear spin polarization measurements are more accurate for $^{69}$Ga than for $^{75}$As. It is worth noting that the nuclear spin thermometry data measured on QDs in an unstrained sample is corrected with the $c$ coefficients measured on a different individual QD (in a stressed sample). However, measurements conducted on several individual QDs from the same sample reveal NMR spectra very similar to those shown Supplementary Fig.~\ref{Fig:SInvNMR}a,b. Thus, while there is always some uncertainty arising from dot-to-dot variation, its effect is expected to be smaller than the actual correction introduced through the $c$ coefficients.

\subsection{Calibration of the adiabatic radiofrequency sweeps\label{subsec:SwpCalib}}

\begin{figure}
\includegraphics[width=0.98\linewidth]{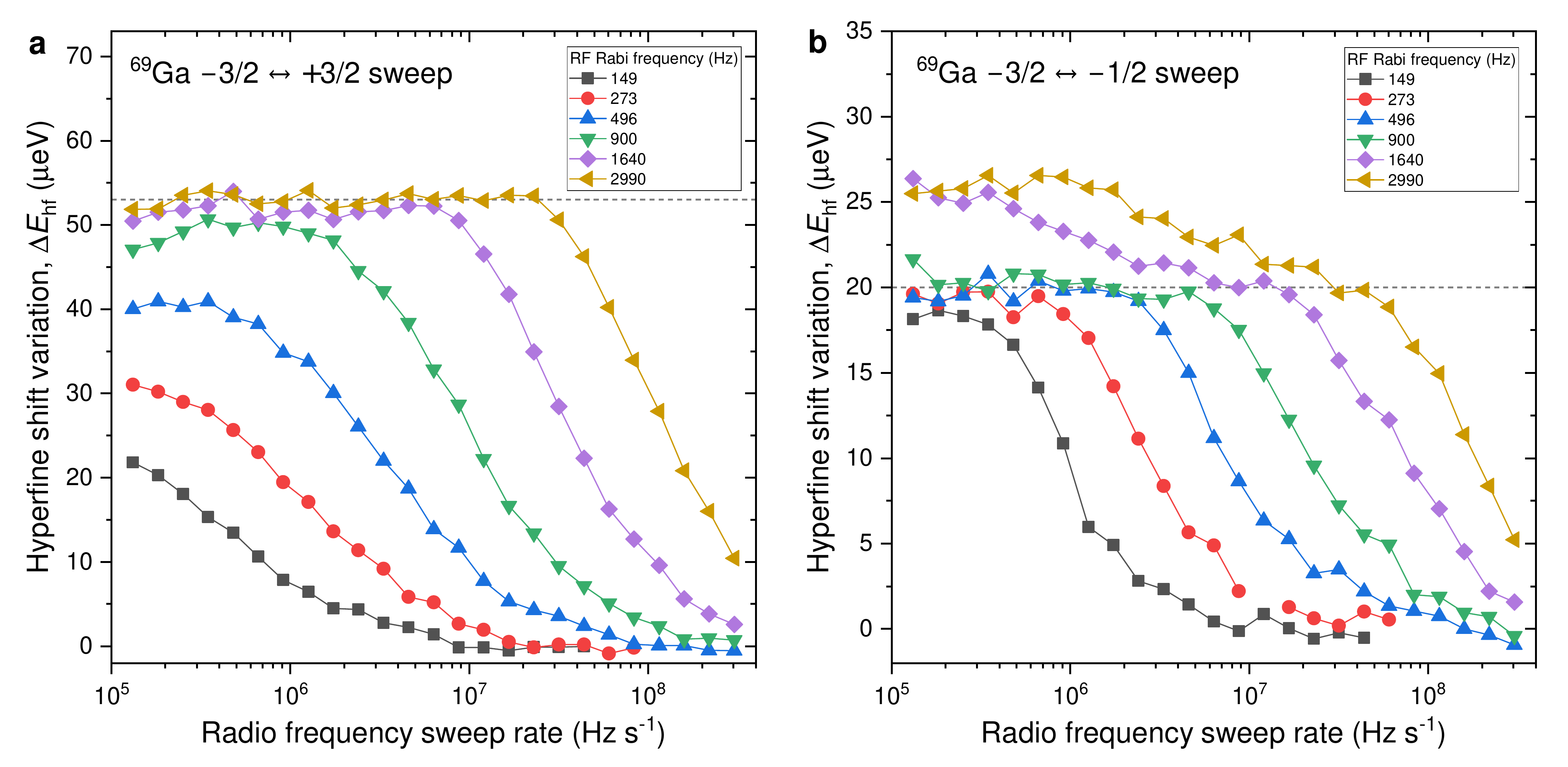}
\caption{\label{Fig:SwpCalib}{\bf{Calibration of the radiofrequency adiabatic sweeps.}} {\bf{a}} In these experiments $\sigma^+$ optical pumping is first used to produce negative nuclear spin polarization. Optical pumping is followed by a frequency-swept Rf burst and the resulting hyperfine shift variation is plotted as a function of the frequency sweep rate. The radiofrequency field is swept from $-50$~kHz to $+50$~kHz with respect to the Larmor frequency of the $^{69}$Ga nuclear spins. This range covers all three quadrupolar-split NMR transitions. Results are shown for several amplitudes of the radiofrequency field, expressed in terms of the Rabi frequency $\nu_1$ that such a field produces when tuned in resonance with the satellite NMR transition $-3/2\leftrightarrow -1/2$. Dashed horizontal line shows the hyperfine shift variation under adiabatic conditions. {\bf{b}} Same as (a), but the frequency range of the sweep is from  $-5$~kHz to $-50$~kHz, covering only the satellite NMR transition $-3/2\leftrightarrow -1/2$.}
\end{figure}

Supplementary Fig.~\ref{Fig:SwpCalib} shows the dependence of the Rf-induced hyperfine shift on the frequency sweep rate. The amplitude of the Rf field is expressed in terms of the corresponding Rabi frequency $\nu_1$. Supplementary Fig.~\ref{Fig:SwpCalib}a shows the results of an experiment where radiofrequency is swept from $-50$~kHz to $+50$~kHz around the $^{69}$Ga Larmor frequency. This sweep range covers nearly the entire $^{69}$Ga quadrupolar triplet. For a sufficiently large Rf amplitude $\nu_1\gtrsim1.64$~kHz and a sufficiently low rate the sweep is adiabatic, resulting in population transfer from the optically-populated $m=-3/2$ states into the $m=+3/2$ states. The variation of the hyperfine shift under adiabatic conditions is $\Delta E_\textrm{hf}\approx53~\mu$eV (dashed horizontal line). As expected \citep{Janzen1973}, the sweep rate that produces adiabatic transfer increases quadratically with the Rf amplitude $\nu_1$. When $\nu_1$ is reduced below $\approx1$~kHz, the magnitude of the hyperfine shift $\Delta E_\textrm{hf}$ in the slow-sweep limit decreases, indicating that population transfer becomes non-adiabatic. This non-adiabaticity is a result of demagnetization in the rotating frame, where Zeeman energy is transferred into the nuclear dipole-dipole interaction reservoir \citep{Janzen1973,Goldman1975,SlichterBook}. For all $\nu_1$ the sweep also becomes non-adiabatic in the large-rate limit.

Supplementary Fig.~\ref{Fig:SwpCalib}b shows sweep rate dependence for the range starting from $-5$~kHz to $-50$~kHz, which selectively covers the $-3/2\leftrightarrow -1/2$ satellite NMR transition (the starting points of the sweeps are shown by the dashed lines in Supplementary Fig.~\ref{Fig:SInvNMR}b). The adiabatic inversion of the subspace spanned by the $m=-1/2$ and $m=-3/2$ states results in a hyperfine shift of $\Delta E_\textrm{hf}\approx20~\mu$eV. Unlike for  $-1/2\leftrightarrow +1/2$, adiabaticity is achieved at a lower Rf amplitudes $\nu_1\gtrsim0.27$~kHz. This is explained by the difference in the inhomogeneous broadening of the satellite transitions and the central transition  $-1/2\leftrightarrow +1/2$. In case of the $-1/2\leftrightarrow +1/2$ transition (that is driven when the frequency is swept over the entire quadrupolar triplet) the inhomogeneous broadening is due to the second order quadrupolar shifts which are small compared to dipolar nuclear-nuclear interactions. By contrast, the first-order inhomogeneous quadrupolar broadening of the $-3/2\leftrightarrow -1/2$ satellite ($\gtrsim10$~kHz) is much larger than the dipole-dipole interaction. As a result the Zeeman and the dipolar energy reservoirs remain isolated during the sweep over the satellite, inhibiting the demagnetization. In other words, the Rf field sweeping over the broadened $-3/2\leftrightarrow -1/2$ satellite excites only a small fraction of the nuclei at any given frequency, while the majority of the nuclear spins remain out of resonance and therefore cannot participate in the exchange between the Zeeman and dipolar reservoirs. When the Rf amplitude is increased ($\nu_1\gtrsim1.64$~kHz), the magnitude $\vert\Delta E_\textrm{hf}\vert$ of the hyperfine shift increases further beyond the adiabatic-inversion level. This can be explained by the parasitic driving of the $-1/2\leftrightarrow +1/2$ transition, which occurs when $\nu_1$ becomes non-negligible compared to the minimal offset ($-5$~kHz from the $-1/2\leftrightarrow +1/2$ frequency) during the sweep over the $-3/2\leftrightarrow -1/2$ transition. Based on these calibrations, we use adiabatic frequency sweeps only on the $\pm3/2\leftrightarrow \pm1/2$ satellites, avoiding any sweeps that involve the $-1/2\leftrightarrow +1/2$ central transition. For the spin temperature measurements on $^{69}$Ga we use $\nu_1\approx0.496$~kHz and a sweep rate of 1~MHz~s$^{-1}$. From the data of Supplementary Fig.~\ref{Fig:SwpCalib}b this combination of parameters is seen to provide good adiabatic inversion of the satellite transitions without any noticeable parasitic excitation of the central transition. Similar results were obtained from calibrations on the $^{75}$As satellite transition resonance -- the optimal Rf amplitude was found to be $\nu_1\approx0.560$~kHz with a sweep rate of 0.8~MHz~s$^{-1}$.

\subsection{Model fitting for derivation of the nuclear spin polarization degree\label{subsec:PnFit}}

\begin{figure}
\includegraphics[width=0.5\linewidth]{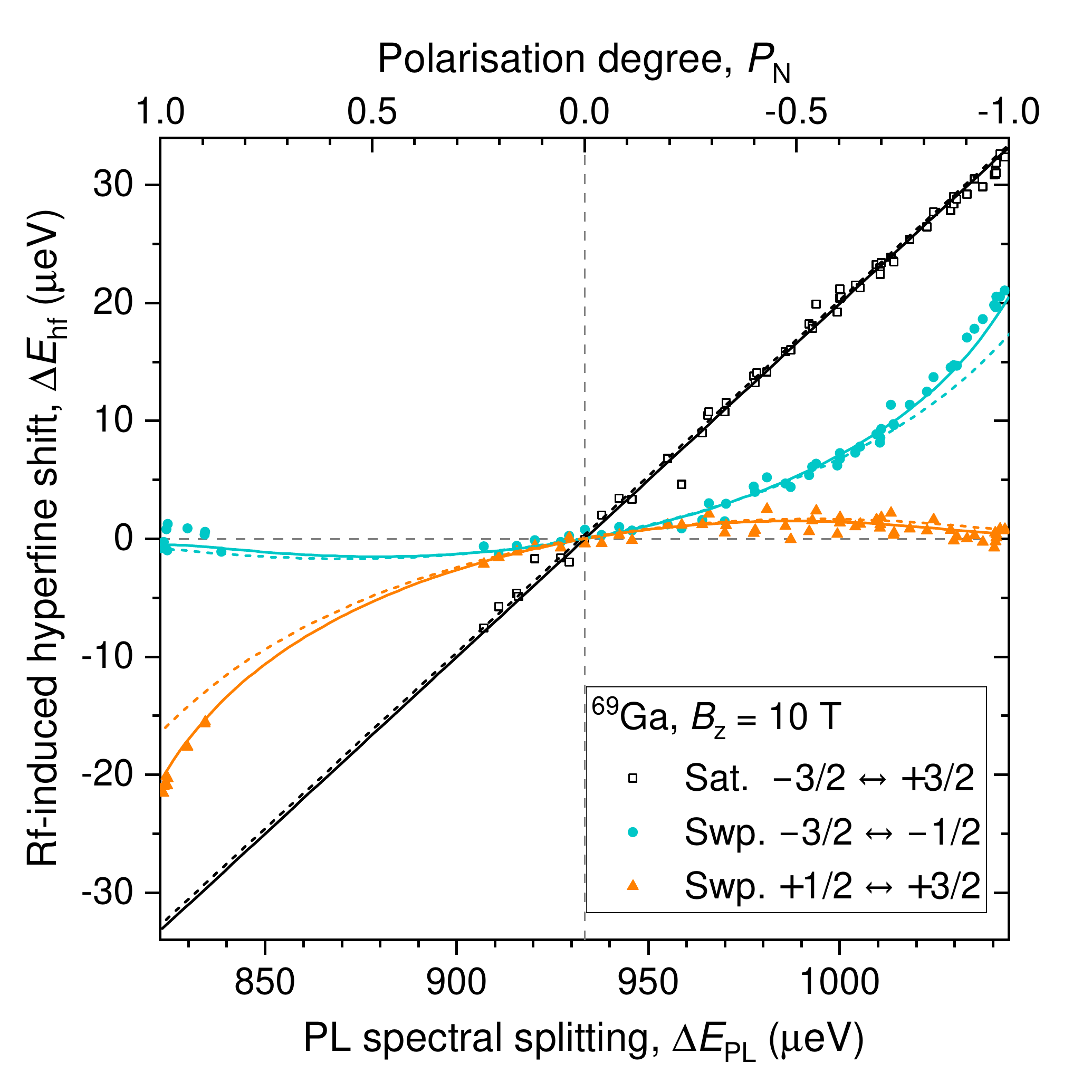}
\caption{\label{Fig:TempFitFullRange}{\bf{Derivation of nuclear spin polarization from selective NMR excitation.}} Spectral splitting $\Delta E_{\rm{PL}}$ of a QD negatively charge trion measured without Rf excitation is plotted on the bottom horizontal axis. Rf-induced hyperfine shift variation $\Delta E_{\rm{hf}}$ is plotted on the vertical axis. Each point is obtained by changing the initial degree of the optically induced nuclear spin polarization and conducting two measurements: with selective Rf excitation and without Rf excitation. $\Delta E_{\rm{hf}}$ is the difference of these two measurements. Several types of $^{69}$Ga Rf excitation are employed: non-selective saturation of the entire NMR triplet (squares), adiabatic frequency sweep over the  $-3/2\leftrightarrow -1/2$ satellite (circles) and adiabatic frequency sweep over the  $+1/2\leftrightarrow +3/2$ satellite (triangles). Solid lines show the best fit, whereas dashed lines show a fit constrained by the $\vert P_{\rm{N}}\vert<0.9$ hypothesis. The top horizontal scale shows the nuclear spin polarization degree $P_{\rm{N}}$ evaluated from the best fit. Experiments are conducted on an individual QD1 at $B_{\rm{z}}=10$~T.}
\end{figure}

\begin{figure}[tp]
\includegraphics[width=0.65\linewidth]{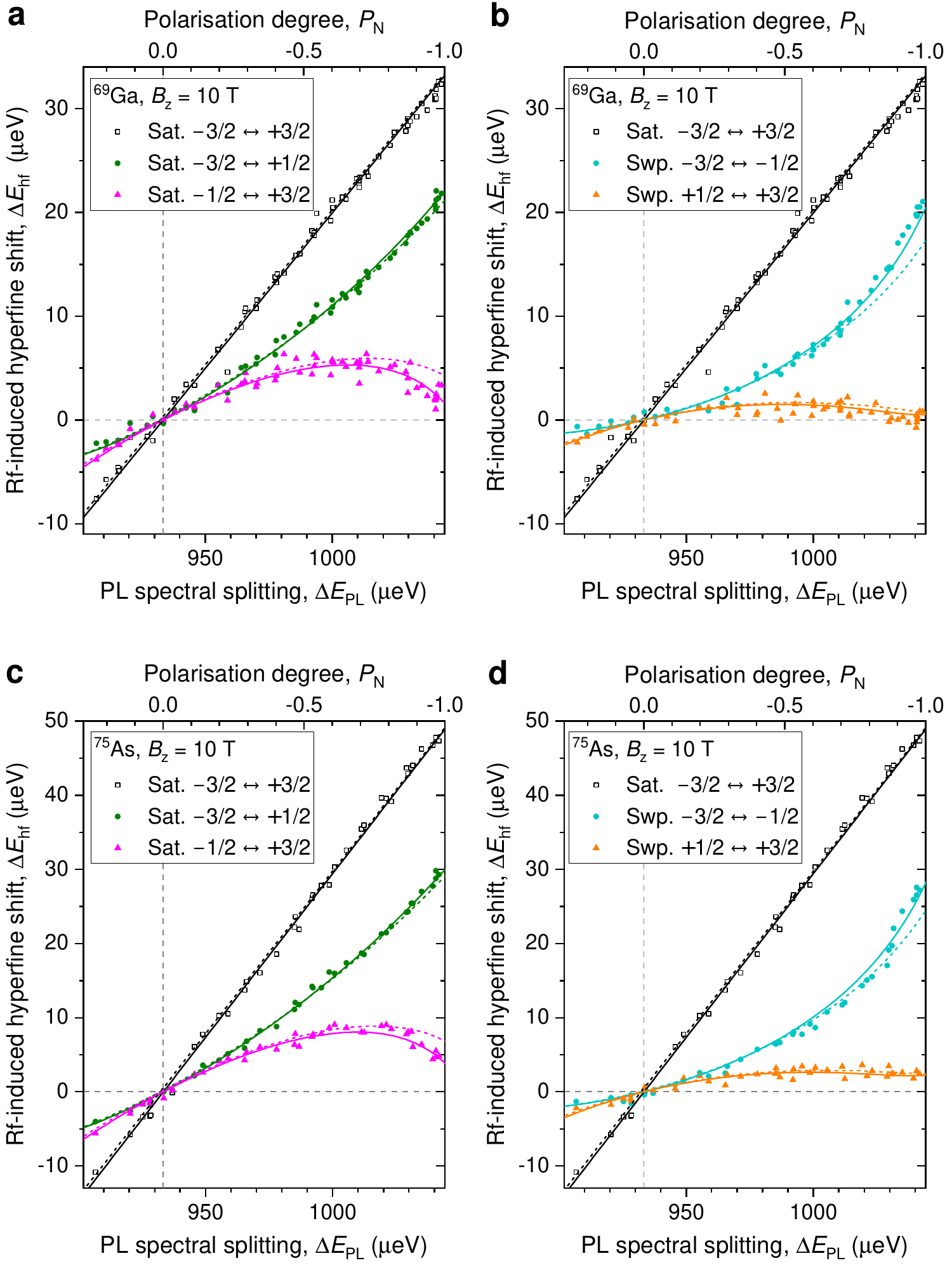}
\caption{\label{Fig:TempFitAll}{\bf{Derivation of nuclear spin polarization from selective NMR measurements.}} Same as Supplementary Fig.~\ref{Fig:TempFitFullRange}, but focusing on the range of negative nuclear spin polarizations. {\bf{a}} Data and fits for $^{69}$Ga nuclei. Hyperfine shift variations are shown for saturation of the entire NMR triplet (squares) and selective two-transition saturation (circles for $-3/2\leftrightarrow +1/2$, triangles for $-1/2\leftrightarrow +3/2$). {\bf{b}} Data for the non-selective saturation (same data as in (a), squares) and selective adiabatic frequency sweeps over NMR satellites (circles for $-3/2\leftrightarrow -1/2$, triangles for $+1/2\leftrightarrow +3/2$). {\bf{c,d}} Same as (a,b) but for $^{75}$As nuclei.}
\end{figure}

The experimental measurement of the nuclear spin polarization (spin thermometry) uses the pump-rf-probe cycle shown in Supplementary Fig.~\ref{Fig:STDiagr}a. The variable parameter is the initial degree of nuclear spin polarization produced by the optical pumping. The steady-state nuclear spin polarization is changed either by detuning the pump laser wavelength away from the optimum or by reducing the degree of circular polarisation of the pump. For any given initial nuclear polarization pump-rf-probe measurements are carried out with different types of Rf excitation or with no Rf pulse. The spectral splitting of the trion $\Delta E_{\rm{PL}}$ detected in the probe pulse is then used as the horizontal axis for the data plots in Supplementary Fig.~\ref{Fig:TempFitFullRange}. On the vertical axis we plot the difference between the trion spectral splitting measured with the Rf pulse (final state) and without the Rf pulse (initial state). This difference yields the change in the hyperfine shift resulting from selective Rf excitation of a certain NMR transition for a chosen isotope, whereas hyperfine shifts arising from other transitions and isotopes remain unaffected. In the experiment we avoid a certain range of positive initial nuclear spin polarizations (characterised by $850~\mu{\rm{eV}}<\Delta E_{\rm{PL}}<900~\mu$eV for QD1 at $B_{\rm{z}}=10$~T) where electron spin energy splitting is close to zero due to the hyperfine shift and the Zeeman effect cancelling each other. Such cancellation is characterised by accelerated nuclear spin dynamics (see Fig.~4a of the main text), making it difficult to perform non-perturbing optical probing. Therefore, in order to discuss the spin thermometry fitting, we focus on the negative nuclear polarizations, where most of the datapoints are collected (Supplementary Fig.~\ref{Fig:TempFitAll}).

The splitting in the photoluminescence spectrum of a negatively charged trion $X^{-}$ (see Supplementary Fig.~\ref{Fig:SEnDiag}b) can be written as (see Supplementary Eq.~\ref{eq:DEPLhf}):
\begin{eqnarray}
\begin{aligned}
\Delta E_{\rm{PL}}=\Delta E_{\rm{PL,0}} - \sum_j F^{(j)} I_j P_{{\rm{N}},j}. \label{eq:EhfIso}
\end{aligned}
\end{eqnarray}
where $\Delta E_{\rm{PL,0}}$ is the trion splitting corresponding to depolarized nuclei and the summation goes over all isotopes with their individual polarization degrees $P_{{\rm{N}},j}$. The individual proportionality constants can be written as $F^{(j)}=k_j(A^{(j)}-C^{(j)})$, where both the electron ($A^{(j)}$) and the hole ($C^{(j)}$) hyperfine material constants are included since the photoluminescence of the trion is only observed for recombination of an electron and a hole with the opposite spin $z$ projections (see \ref{sec:eNSystem}). The factors $0<k_j\leq1$ account for the Ga nuclei atoms replaced by Al, resulting in a reduced hyperfine shift experienced by the electron spin. If all $I=3/2$ isotopes have the same polarization degree, Supplementary Eq.~\ref{eq:EhfIso} simplifies to
\begin{eqnarray}
\begin{aligned}
\Delta E_{\rm{PL}}=\Delta E_{\rm{PL,0}} - I P_{{\rm{N}}} \sum_j F^{(j)} . \label{eq:EhfIso2}
\end{aligned}
\end{eqnarray}
Resolving this for $P_{{\rm{N}}}$ and substituting into the last of Supplementary Eq.~\ref{eq:DEhf}, we find that the change in the hyperfine shift arising from non-selective saturation of all 3 NMR transitions of the $j$-th isotope is a linear function of the trion spectral splitting:
\begin{eqnarray}
\begin{aligned}
&\Delta E_{{\rm{hf}},j}^{-I\leftrightarrow +I}=(\Delta E_{\rm{PL}}-\Delta E_{\rm{PL,0}}) \frac{F^{(j)}}{F_{\rm{tot}}}=w_j(\Delta E_{\rm{PL}}-\Delta E_{\rm{PL,0}}),\\
&F_{\rm{tot}}=\sum_j F^{(j)}, \label{eq:EhfFullSat}
\end{aligned}
\end{eqnarray}
where $w_j$ is the weight coefficient of the $j$-th isotope in the total hyperfine shift $E_{{\rm{hf}}}$, and $F_{\rm{tot}}$ is the total proportionality factor. The measured $\Delta E_{{\rm{hf}},j}^{-I\leftrightarrow +I}$ are shown by the squares in Supplementary Fig.~\ref{Fig:TempFitAll}a,c for $^{75}$As and $^{69}$Ga, respectively. The dependence on $\Delta E_{\rm{PL}}$ is indeed seen to be linear. For precise modelling we take the squared differences between the measured $\Delta E_{{\rm{hf}},j}^{-I\leftrightarrow +I}$ and $\Delta E_{{\rm{hf}},j}^{-I\leftrightarrow +I}$ computed from Supplementary Eq.~\ref{eq:EhfFullSat} with spectral splitting $\Delta E_{\rm{PL}}$ measured under the same optical pumping but without radiofrequency depolarization.

The same approach is applied to the hyperfine shift variations $\Delta E_{{\rm{hf}},j}^{m\leftrightarrow m+1}$ and $\Delta E_{{\rm{hf}},j}^{m\leftrightarrow m+2}$ arising from selective saturation (or adiabatic inversion) of one or two NMR transitions, respectively. Here, Supplementary Eq.~\ref{eq:EhfIso2} is first resolved to find $P_{{\rm{N}}}$ as a function of $\Delta E_{\rm{PL}}$, and $P_{{\rm{N}}}$ is then substituted into Supplementary Eq.~\ref{eq:PNExp} to find $\beta$. Since there is no explicit form for the inverse of the Brillouin function, this relation is kept in an exact analytical form using the Root[~] function in Wolfram Mathematica 12.3 software. The inverse temperature $\beta$ is then inserted into Supplementary Eqs.~\ref{eq:DEhf}. Finally, the ideal $\Delta E_{{\rm{hf}},j}^{m\leftrightarrow m+1}$ and $\Delta E_{{\rm{hf}},j}^{m\leftrightarrow m+2}$ calculated in this way, are inserted into Supplementary Eqs.~\ref{eq:EhfObsSat}, \ref{eq:EhfObsSwp} using the $c$ coefficients from Supplementary Eq.~\ref{eq:cCoefs} to account for the small spectral overlaps between the individual components of the quadrupolar NMR triplet. Taking these model hyperfine shifts at the experimentally measured $\Delta E_{\rm{PL}}$, we calculate the squared differences with respect to the measured Rf-induced hyperfine shifts (triangles and circles in Supplementary Figs.~\ref{Fig:TempFitAll}a,c for two-transition saturation and in Supplementary Figs.~\ref{Fig:TempFitAll}b,d for single-transition sweeps). We then sum up the squared differences for all individual Rf types and all isotopes to form the total $\chi^2$ functional. 

As a last step, we include in our model the possibility that different isotopes have different polarization degrees $P_{{\rm{N}},j}$. For arbitrary $P_{{\rm{N}},j}$ it is not possible to resolve the trion PL splitting $\Delta E_{\rm{PL}}$ as a function of $P_{{\rm{N}},j}$, requiring some explicit assumptions. As a simplest approximation, we assume that polarization degrees of the three abundant spin-3/2 isotopes ( $^{75}$As, $^{69}$Ga and $^{71}$Ga) are linearly interdependent. Mathematically, this is equivalent to allowing the total scaling factor $F_{\rm{tot}}$ to deviate for the different measured isotopes. The introduction of $F_{\rm{tot}}$ and $w_j$ as model parameters is also convenient in that the data does not have to be collected on all isotopes, in particular on $^{71}$Ga, which has not be studied in this work.

The $\chi^2$ is a function of only five fitting parameters: the zero-polarization trion splitting $\Delta E_{\rm{PL,0}}$, the total scaling factors $F_{\rm{tot}}$ of $^{75}$As and $^{69}$Ga as well as the weight coefficients $w_j$ of $^{75}$As and $^{69}$Ga. In case of QD1, where the data was measured at two different magnetic fields, we fit these datasets independently in order to account for the different degree to which the probe laser pulse introduces parasitic depolarization in the optically measured hyperfine shifts. The best fits obtained by minimizing $\chi^2$ functional are plotted by the solid lines in Supplementary Fig.~\ref{Fig:TempFitAll} and show a good match to the measured data. The best fit parameters are listed below together with the total number of experimental datapoints $N_{\rm{data}}$ and the root-mean-square (RMS) residual $R_{\rm{min}}=\sqrt{\chi_{\rm{min}}^2/N_{\rm{data}}}$ derived from the minimized functional value $\chi_{\rm{min}}^2$. In addition, we quote the residual obtained from a separate linear fit where only the full-saturation hyperfine shift $\Delta E_{{\rm{hf}},j}^{-I\leftrightarrow +I}$ is considered:
\begin{align}
\begin{array}{l|c|c|c|c|c|c|c|c}
\textrm{QD} & B_{\rm{z}} & w(^{75}{\rm{As}}) & w(^{69}{\rm{Ga}}) & I F_{\rm{tot}}(^{75}{\rm{As}}) & I F_{\rm{tot}}(^{69}{\rm{Ga}}) & N_{\rm{data}} & R_{\rm{min}} & R_{\rm{min}}\:{\rm{of}}\:\Delta E_{{\rm{hf}},j}^{-I\leftrightarrow +I}\\
 \hline
 {\rm{QD1}} &4\:{\rm{T}} & 0.416 & 0.292 & 112.0\:\mu{\rm{eV}} & 104.5\:\mu{\rm{eV}} & 346 & 0.875\:\mu{\rm{eV}} & 0.803\:\mu{\rm{eV}}\\
{\rm{QD1}} &10\:{\rm{T}} & 0.442 & 0.300 & 112.3\:\mu{\rm{eV}} & 110.7\:\mu{\rm{eV}} & 510 & 0.749\:\mu{\rm{eV}} & 0.778\:\mu{\rm{eV}}\\
{\rm{QD2}} &10\:{\rm{T}} & 0.429 & 0.294 & 112.3\:\mu{\rm{eV}} &  110.3\:\mu{\rm{eV}} & 336 & 1.108\:\mu{\rm{eV}} & 0.946\:\mu{\rm{eV}}\\
{\rm{QD3}} &10\:{\rm{T}} & 0.424 & 0.294 & 113.9\:\mu{\rm{eV}} & 112.8\:\mu{\rm{eV}}  & 170 & 0.987\:\mu{\rm{eV}} & 0.968\:\mu{\rm{eV}}\\	
\end{array}\label{eq:BestFit}
\end{align}

As discussed above, the full-saturation experiment is the most robust against the errors arising from NMR spectral overlaps. Therefore, the RMS residual $R_{\rm{min}}$ obtained from linear fitting of $\Delta E_{{\rm{hf}},j}^{-I\leftrightarrow +I}$ alone characterizes the random measurement errors. These errors in the optically-detected hyperfine shifts originate mainly from the noise of the CCD detector used to collect the optical photoluminescence spectra of a single quantum dot. Any excess of $R_{\rm{min}}$ obtained from the nonlinear fit of the selective-NMR data is an indicator of systematic deviation from the fitting model. According to Supplementary Eq.~\ref{eq:BestFit} such excess is small, confirming the validity of the Boltzmann distribution model (Supplementary Eq.~\ref{eq:pBoltz}). The spread in the isotope-specific weights $w_i$, and the scaling factor $F_{\rm{tot}}$ is on the order of 1\% for the $B_{\rm{z}}=10$~T data collected from three individual quantum dots, affirming the systematic nature of these results. The fit of the $B_{\rm{z}}=4$~T data shows the most deviation, which is explained by the need for shorter probe pulses $T_{\rm{Probe}}$, resulting in more noisy photoluminescence spectra as well as larger systematic deviations arising from the probe-induced nuclear spin depolarization.

In order to derive the nuclear spin polarization degree, we use the highest and the lowest trion spectral splitting $\Delta E_{\rm{PL}}$ detected without any radiofrequency manipulation. For this measurement we use the timing diagram of Supplementary Fig.~\ref{Fig:STDiagr}d, where we allow the nuclear spin polarization to build up over $T_{\rm{Buildup}}>100$~s, giving a closer approach to the steady state than what can be achieved in the NMR thermometry measurements (Supplementary Fig.~\ref{Fig:TempFitAll}), where the pumping time $T_{\rm{Pump}}\lesssim30$~s is limited by the maximum duration of the CCD detector exposure. It is worth noting that this approach of using the separately measured steady-state $\Delta E_{\rm{PL}}$ is the reason why we build our fitting model on relating the polarization degree $P_{\rm{N}}$ to spectral splitting $\Delta E_{\rm{PL}}$ via Supplementary Eq.~\ref{eq:EhfIso}. Otherwise, $\Delta E_{\rm{PL}}$ can be eliminated and polarization degree can be derived purely from the Rf-induced hyperfine shifts of Supplementary Eq.~\ref{eq:DEhf}. The best fit value of $\Delta E_{\rm{PL,0}}$ is subtracted from the steady state $\Delta E_{\rm{PL}}$ to derive the lowest negative and the highest positive hyperfine shifts. For QD1 from at $B_{\rm{z}}=10$~T we find $E_{\rm{hf}}=-109.6~\mu$eV and $E_{\rm{hf}}=+112.3~\mu$eV. The latter number exceeds by $\approx1$\% the best-fit product $I F_{\rm{tot}}$($^{69}$Ga). By definition, the $I F_{\rm{tot}}$ product is the maximum $\vert E_{\rm{hf}}\vert$ corresponding to full polarization $P_{\rm{N}}=\pm1$. The discrepancy with the measured $E_{\rm{hf}}$ reveals the scale of errors in the derived polarization degrees $P_{\rm{N}}$, both due to the random noise in the raw data and any systematic inaccuracy of the fitting model. 

\subsection{Error analysis in model fitting of the nuclear spin polarization data}

In order to systematically analyze the fitting errors we construct a multidimensional confidence region (Chapter 9 in Ref.\citep{CowanBook}) defined as a collection of all points in the fitting parameter space for which
\begin{eqnarray}
\begin{aligned}
&\chi^2<(1+Q(\gamma,n)/N_{\rm{data}})\chi_{\rm{min}}^2, \label{eq:Chi2}
\end{aligned}
\end{eqnarray}
where we have approximated the standard error in the experimental data by the RMS fit residual $R_{\rm{min}}=\sqrt{\chi_{\rm{min}}^2/N_{\rm{data}}}$. (Note that here we define $\chi^2$ and $\chi_{\rm{min}}$ as sums that are not normalized by the standard error.) We define $Q(\gamma,n)$ as a quantile of the $\chi^2$-distribution with $n$ parameters corresponding to the confidence level $1-\gamma$. We use $1-\gamma=0.95$ where the relevant quantile is $Q(1-0.95,5)\approx11.07$. We implement a Monte-Carlo calculation, where the $\chi^2$ sum is computed for a large number of random sets of the fitting parameters around the best-fit point.  For each trial point that satisfies Supplementary Eq.~\ref{eq:Chi2} we calculate the polarization degrees $P_{\rm{N}}$ from the maximum and minimum steady-state spectral splitting $\Delta E_{\rm{PL}}$ measured with long $T_{\rm{Buildup}}>100$~s. Finally, the confidence intervals are derived separately for the maximum positive and the minimum negative polarization degree as maximum and minimum $P_{\rm{N}}$ values from the random Monte-Carlo set, coerced to satisfy the condition $-1\leq P_{\rm{N}}\leq1$. The confidence intervals are shown by the symbols in Fig.~3e of the main text and are tabulated below:
\begin{align}
\begin{array}{l|c|c|c|c|c}
\textrm{QD} & B_{\rm{z}} & {\rm{max}}\:^{75}{\rm{As}}\: P_{\rm{N}}& {\rm{min}}\:^{75}{\rm{As}}\: P_{\rm{N}}& {\rm{max}}\:^{69}{\rm{Ga}}\: P_{\rm{N}}& {\rm{min}}\:^{69}{\rm{Ga}}\: P_{\rm{N}}\\
 \hline
{\rm{QD1}} &4\:{\rm{T}} & [0.890,0.974] & [-0.989,-0.903] & [0.982,1.0] & [-1,-0.998]\\ 
{\rm{QD1}} &10\:{\rm{T}} & [0.969,1.0] & [-0.994,-0.954]  & [1.0,1.0] & [-1,-0.984]\\
{\rm{QD2}} &10\:{\rm{T}} & [0.956,1.0] & [-1.0,-0.962] & [0.989,1.0] & [-1,-0.998]\\
{\rm{QD3}} &10\:{\rm{T}} & [0.954,1.0] & [-1,-0.951] & [0.964,1.0] & [-1,-0.959]\\
\end{array}\label{eq:PNCIs}
\end{align}
This systematic evaluation agrees with the rough estimates above, confirming that the accuracy of our $P_{\rm{N}}$ estimates is on the other of a few percent. The fit returns similar values for polarization degrees of $^{75}$As and $^{69}$Ga -- this is expected for a spin pumping mechanism \citep{Chekhovich2017} where the inverse temperature $\beta$ of each individual nucleus is independently equilibrated with the $\beta$ of a spin-polarized electron. The somewhat broader confidence intervals of $^{75}$As could be simply due to the larger overlaps of the NMR spectral components, making the fit less sensitive to $P_{\rm{N}}$ and more dependent on the accuracy of the $c$ coefficients tabulated in Supplementary Eq.~\ref{eq:cCoefs}.

In order to further evaluate the error estimates we approach the problem of data modelling from the opposite direction. Namely, we start with a hypothesis that the maximum absolute polarization degree $\vert P_{\rm{N}}\vert$ is no more than a certain value $<1$, and then evaluate how well our experimental data can be matched to this hypothesis. Constraining the fit to $\vert P_{\rm{N}}\vert<0.9$ we search for the fitting parameter combination that minimizes the $\chi^2$ functional for the same model as the one used to derive the best fit. The resulting constrained best-fit is shown for QD1 by the dashed lines in Supplementary Fig.~\ref{Fig:TempFitAll}. There are visible systematic deviations from the measured data, already suggesting that the $\vert P_{\rm{N}}\vert<0.9$ hypothesis is inadequate, and the actual absolute polarization degree is well above $0.9$. Quantitatively, the RMS residual from the fit of the $B_{\rm{z}}=10$~T datasets constrained to $\vert P_{\rm{N}}\vert<0.9$ is $\approx1.73~\mu$eV for QD1, $\approx2.91~\mu$eV for QD2 and $\approx2.17~\mu$eV for QD3. These residuals are a factor of $\gtrsim2$ higher than the best-fit values tabulated in Supplementary Eq.~\ref{eq:BestFit} -- statistically, such deviations are improbable for our datasets containing hundreds of datapoints.

\begin{figure}
\includegraphics[width=0.55\linewidth]{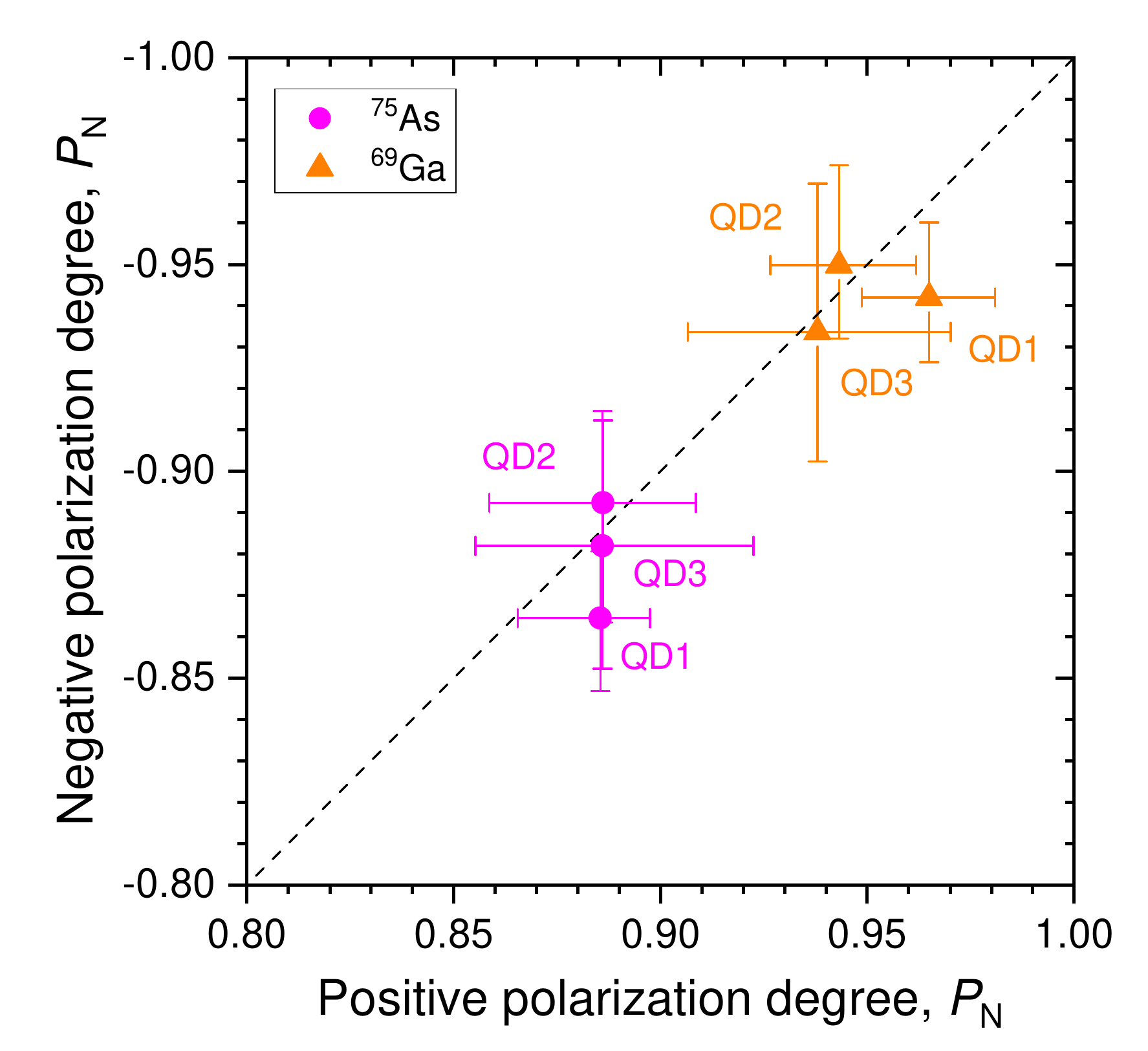}
\caption{\label{Fig:TempFitUnCor}{\bf{Nuclear spin polarization derived from an uncorrected fit.}} Maximum positive (horizontal axis) and minimum negative (vertical axis) nuclear spin polarization degree $P_{\rm{N}}$ derived from the measurements on $^{75}$As (circles) and $^{69}$Ga (triangles) nuclei in individual dots QD1 - QD3. The $P_{\rm{N}}$ values in this plot are derived from a fit that ignores the small spectral overlaps between the NMR triplet components. Error bars are 95\% confidence intervals.}
\end{figure}

Next we fit the same experimental data but without correcting for the overlaps of the NMR spectral components. This is equivalent to setting $c_{\rm{Sat,Ideal}} = c_{\rm{Swp,Ideal}}=1$ with the remaining $c$ coefficients set to 0. Such a fit can be seen as a lower bound estimate for the absolute polarization degree $\vert P_{\rm{N}}\vert$. Without correction, the RMS fit residual slightly increases from 0.749~$\mu$eV to 0.819~$\mu$eV. The resulting uncorrected $P_{\rm{N}}$ are shown in Supplementary Fig.~\ref{Fig:TempFitUnCor}. The uncorrected polarization degrees for $^{75}$As ($P_{\rm{N}}\approx0.88$) are lower than for $^{69}$Ga ($P_{\rm{N}}\approx0.94$), contradicting the expectation of equal $\beta$ across different isotopes. Moreover, the maximum uncorrected $\vert P_{\rm{N}}\vert$ are very close to the corresponding $c_{\rm{Sat,Ideal}}$ coefficients. Additional computations confirm that this is to be expected -- a naive uncorrected fit of the data affected by NMR spectral overlap returns for fully polarized nuclei a reduced polarization $\vert P_{\rm{N}}\vert$ which roughly equals the fraction of the ``ideal'' nuclei that are not affected by the overlap. Nevertheless, even without the corrections, a high polarization degree is derived for $^{69}$Ga nuclei, since they are less prone to NMR spectral overlaps than $^{75}$As.

We now consider the different possible sources of systematic errors. Since electron localization in a GaAs quantum dot is not infinitely strong, the electron wavefunction leaks into the AlGaAs barriers where it gradually decays with the increasing distance from the dot. This means that both the nuclear spin pumping efficiency and the sensitivity of the electron hyperfine shift to nuclear spin polarization are spatially inhomogeneous. The resulting steady-state nuclear spin polarization is also spatially inhomogeneous, if only because the QD layer is sandwiched between the two doped semiconductor layers, where free charge carriers result in $P_{\rm{N}}\approx0$. On the other hand, when considering spins as a quantum resource, a significant role is played only by the nuclei within the QD electron wavefunction.  Recent studies of nuclear spin relaxation in the same sample have shown that spin diffusion is the dominant mechanism of nuclear spin decay in a QD \citep{MillingtonHotze2022}. The nuclear spins at the center of the QD are quickly polarized by the optically-pumped electron spin and then transfer their polarization to more distant nuclei via nuclear spin flip-flops. For a sufficiently long pumping the nuclei in the AlGaAs barriers around the dot become gradually polarized. This manifests in a slow-down of the subsequent relaxation without the pump. The relaxation times of the nuclear spins are on the order of hundreds of seconds, much longer than the nuclear spin buildup times, which are less than one second.  Such a large ratio of the timescales suggests that competition between spin pumping and polarization leakage would not be a limiting factor for achieving  $\vert P_{\rm{N}}\vert$ up to $\approx0.99$. Moreover, relaxation much slower than pumping means that spin diffusion creates a smooth spatial profile of the nuclear spin polarization -- the extent of the polarized volume is larger than the volume of the electron wavefunction. Therefore, we expect that the electron probes a volume with a nearly uniform nuclear spin polarization degree $P_{\rm{N}}$. In other words, the existence of spin diffusion combined with long nuclear spin pumping means that there is no realistic mechanism that would result in abrupt spatial variations of $P_{\rm{N}}$. Under these conditions the selective-NMR thermometry measurements return the average of nuclear polarization, weighted by the electron envelope wavefunction density. Observation of a near-unity average polarization itself implies that polarization is very homogeneous for all nuclei within the electron wavefunction volume. For example, if we take the typical leakage of the electron wavefunction into the AlGaAs barriers at $\approx0.1$ (estimated previously in Ref.~\citep{Chekhovich2017}) and assume full polarization within the GaAs layer ($\vert P_{\rm{N}}\vert=1$), then observation of a weighted average of $\vert P_{\rm{N}}\vert=0.99$ implies that polarization within the AlGaAs barriers cannot be much smaller than $\vert P_{\rm{N}}\vert=0.9$. To summarize, although our present technique is not capable of revealing the spatial profile of the nuclear spin polarization, the most plausible hypothesis is that nuclear spin polarization achieved under steady-state optical pumping is nearly uniform within the volume of the QD electron wavefunction. In practice, this implies the ability to polarize nearly all nuclei whose coupling to the electron is strong enough to have any relevance to electron-nuclear coherent spin dynamics \citep{Taylor2003}. This also justifies our model, which assumes $P_{\rm{N}}$ to be constant within the quantum dot volume and its surrounding.


Although $^{75}$As, $^{69}$Ga and $^{71}$Ga are the three abundant isotopes, the inevitable penetration of the electron into the AlGaAs barriers implies some hyperfine interaction with the spin-5/2 $^{27}$Al nuclei. And yet it turns out that $^{27}$Al hyperfine shift is to small to be studied quantitatively. As it has been shown previously \citep{Chekhovich2017}, this is a combined effect of several factors. The small fraction of the wavefunction overlapping with AlGaAs ($\approx0.1$), the small fraction of Al atoms (0.33 in our sample) and the small hyperfine constant ($\approx0.3$ of that of As and Ga) mean that the $^{27}$Al relative contribution to the total electron hyperfine shift is within $\approx1\%$. In addition to that, the lack of Al at the center of the QD, where the overlap with the electron is the strongest, suggests inhibition of the pumping-through-diffusion mechanism discussed above. The $^{27}$Al spins can only be polarized through direct (and weak) contact with the spin-polarized electron, meaning that aluminium polarization can be reduced. In the context of the present work where we focus on polarization of As and Ga, these observations mean that any systematic errors arising from $^{27}$Al are small (within $\approx1\%$). Investigation of $^{27}$Al spin polarization would be an interesting subject for future work -- this would require more sensitive experimental techniques, such as trigger detection via abundant isotopes \cite{GoldmanBook}.

Summarising this analysis, we see that there is a handful of potential error sources, both random and systematic, but all on the order of 1\%. Taking a conservative approach we conclude with confidence that nuclear spin polarization degrees well above 0.95 are achieved. In reality, the polarization is likely to be higher, with rigorous confidence-interval analysis returning polarization degrees as high as $\vert P_{\rm{N}}\vert\approx0.99$ (for $^{69}$Ga at high magnetic field $B_{\rm{z}}=10$~T in all three selected individual quantum dots). Achieving even higher polarizations would depend critically on development of more sensitive thermometry techniques. One possibility is to use the dephasing dynamics of the electron spin qubit, since this would gain sensitivity at high polarizations as $\propto 1/\sqrt{1-P_{\rm{N}}^2}$ (Ref.~\citep{Kloeffel2013}).



\begin{thebibliography}{37}%


\makeatletter
\providecommand \@ifxundefined [1]{%
 \@ifx{#1\undefined}
}%
\providecommand \@ifnum [1]{%
 \ifnum #1\expandafter \@firstoftwo
 \else \expandafter \@secondoftwo
 \fi
}%
\providecommand \@ifx [1]{%
 \ifx #1\expandafter \@firstoftwo
 \else \expandafter \@secondoftwo
 \fi
}%
\providecommand \natexlab [1]{#1}%
\providecommand \enquote  [1]{``#1''}%
\providecommand \bibnamefont  [1]{#1}%
\providecommand \bibfnamefont [1]{#1}%
\providecommand \citenamefont [1]{#1}%
\providecommand \href@noop [0]{\@secondoftwo}%
\providecommand \href [0]{\begingroup \@sanitize@url \@href}%
\providecommand \@href[1]{\@@startlink{#1}\@@href}%
\providecommand \@@href[1]{\endgroup#1\@@endlink}%
\providecommand \@sanitize@url [0]{\catcode `\\12\catcode `\$12\catcode
  `\&12\catcode `\#12\catcode `\^12\catcode `\_12\catcode `\%12\relax}%
\providecommand \@@startlink[1]{}%
\providecommand \@@endlink[0]{}%
\providecommand \url  [0]{\begingroup\@sanitize@url \@url }%
\providecommand \@url [1]{\endgroup\@href {#1}{\urlprefix }}%
\providecommand \urlprefix  [0]{URL }%
\providecommand \Eprint [0]{\href }%
\providecommand \doibase [0]{https://doi.org/}%
\providecommand \selectlanguage [0]{\@gobble}%
\providecommand \bibinfo  [0]{\@secondoftwo}%
\providecommand \bibfield  [0]{\@secondoftwo}%
\providecommand \translation [1]{[#1]}%
\providecommand \BibitemOpen [0]{}%
\providecommand \bibitemStop [0]{}%
\providecommand \bibitemNoStop [0]{.\EOS\space}%
\providecommand \EOS [0]{\spacefactor3000\relax}%
\providecommand \BibitemShut  [1]{\csname bibitem#1\endcsname}%
\let\auto@bib@innerbib\@empty
\bibitem [{\citenamefont {Jacques}\ \emph {et~al.}(2009)\citenamefont
  {Jacques}, \citenamefont {Neumann}, \citenamefont {Beck}, \citenamefont
  {Markham}, \citenamefont {Twitchen}, \citenamefont {Meijer}, \citenamefont
  {Kaiser}, \citenamefont {Balasubramanian}, \citenamefont {Jelezko},\ and\
  \citenamefont {Wrachtrup}}]{Jacques2009}%
  \BibitemOpen
  \bibfield  {author} {\bibinfo {author} {\bibfnamefont {V.}~\bibnamefont
  {Jacques}}, \bibinfo {author} {\bibfnamefont {P.}~\bibnamefont {Neumann}},
  \bibinfo {author} {\bibfnamefont {J.}~\bibnamefont {Beck}}, \bibinfo {author}
  {\bibfnamefont {M.}~\bibnamefont {Markham}}, \bibinfo {author} {\bibfnamefont
  {D.}~\bibnamefont {Twitchen}}, \bibinfo {author} {\bibfnamefont
  {J.}~\bibnamefont {Meijer}}, \bibinfo {author} {\bibfnamefont
  {F.}~\bibnamefont {Kaiser}}, \bibinfo {author} {\bibfnamefont
  {G.}~\bibnamefont {Balasubramanian}}, \bibinfo {author} {\bibfnamefont
  {F.}~\bibnamefont {Jelezko}},\ and\ \bibinfo {author} {\bibfnamefont
  {J.}~\bibnamefont {Wrachtrup}},\ }\bibfield  {title} {\bibinfo {title}
  {Dynamic polarization of single nuclear spins by optical pumping of
  nitrogen-vacancy color centers in diamond at room temperature},\ }\href
  {https://doi.org/10.1103/PhysRevLett.102.057403} {\bibfield  {journal}
  {\bibinfo  {journal} {Phys. Rev. Lett.}\ }\textbf {\bibinfo {volume} {102}},\
  \bibinfo {pages} {057403} (\bibinfo {year} {2009})}\BibitemShut {NoStop}%
\bibitem [{\citenamefont {Falk}\ \emph {et~al.}(2015)\citenamefont {Falk},
  \citenamefont {Klimov}, \citenamefont {Iv\'ady}, \citenamefont {Sz\'asz},
  \citenamefont {Christle}, \citenamefont {Koehl}, \citenamefont {Gali},\ and\
  \citenamefont {Awschalom}}]{Falk2015}%
  \BibitemOpen
  \bibfield  {author} {\bibinfo {author} {\bibfnamefont {A.~L.}\ \bibnamefont
  {Falk}}, \bibinfo {author} {\bibfnamefont {P.~V.}\ \bibnamefont {Klimov}},
  \bibinfo {author} {\bibfnamefont {V.}~\bibnamefont {Iv\'ady}}, \bibinfo
  {author} {\bibfnamefont {K.}~\bibnamefont {Sz\'asz}}, \bibinfo {author}
  {\bibfnamefont {D.~J.}\ \bibnamefont {Christle}}, \bibinfo {author}
  {\bibfnamefont {W.~F.}\ \bibnamefont {Koehl}}, \bibinfo {author}
  {\bibfnamefont {A.}~\bibnamefont {Gali}},\ and\ \bibinfo {author}
  {\bibfnamefont {D.~D.}\ \bibnamefont {Awschalom}},\ }\bibfield  {title}
  {\bibinfo {title} {Optical polarization of nuclear spins in silicon
  carbide},\ }\href {https://doi.org/10.1103/PhysRevLett.114.247603} {\bibfield
   {journal} {\bibinfo  {journal} {Phys. Rev. Lett.}\ }\textbf {\bibinfo
  {volume} {114}},\ \bibinfo {pages} {247603} (\bibinfo {year}
  {2015})}\BibitemShut {NoStop}%
\bibitem [{\citenamefont {Reichertz}\ \emph {et~al.}(1994)\citenamefont
  {Reichertz}, \citenamefont {Dutz}, \citenamefont {Goertz}, \citenamefont
  {Kramer}, \citenamefont {Meyer}, \citenamefont {Reicherz}, \citenamefont
  {Thiel},\ and\ \citenamefont {Thomas}}]{Reichertz1994}%
  \BibitemOpen
  \bibfield  {author} {\bibinfo {author} {\bibfnamefont {L.}~\bibnamefont
  {Reichertz}}, \bibinfo {author} {\bibfnamefont {H.}~\bibnamefont {Dutz}},
  \bibinfo {author} {\bibfnamefont {S.}~\bibnamefont {Goertz}}, \bibinfo
  {author} {\bibfnamefont {D.}~\bibnamefont {Kramer}}, \bibinfo {author}
  {\bibfnamefont {W.}~\bibnamefont {Meyer}}, \bibinfo {author} {\bibfnamefont
  {G.}~\bibnamefont {Reicherz}}, \bibinfo {author} {\bibfnamefont
  {W.}~\bibnamefont {Thiel}},\ and\ \bibinfo {author} {\bibfnamefont
  {A.}~\bibnamefont {Thomas}},\ }\bibfield  {title} {\bibinfo {title}
  {Polarization reversal of proton spins in a solid-state target by
  superradiance},\ }\href
  {https://doi.org/https://doi.org/10.1016/0168-9002(94)90103-1} {\bibfield
  {journal} {\bibinfo  {journal} {Nuclear Instruments and Methods in Physics
  Research Section A: Accelerators, Spectrometers, Detectors and Associated
  Equipment}\ }\textbf {\bibinfo {volume} {340}},\ \bibinfo {pages} {278}
  (\bibinfo {year} {1994})}\BibitemShut {NoStop}%
\bibitem [{\citenamefont {Knuuttila}\ \emph {et~al.}(2001)\citenamefont
  {Knuuttila}, \citenamefont {Tuoriniemi}, \citenamefont {Lefmann},
  \citenamefont {Juntunen}, \citenamefont {Rasmussen},\ and\ \citenamefont
  {Nummila}}]{Knuuttila2001}%
  \BibitemOpen
  \bibfield  {author} {\bibinfo {author} {\bibfnamefont {T.~A.}\ \bibnamefont
  {Knuuttila}}, \bibinfo {author} {\bibfnamefont {J.~T.}\ \bibnamefont
  {Tuoriniemi}}, \bibinfo {author} {\bibfnamefont {K.}~\bibnamefont {Lefmann}},
  \bibinfo {author} {\bibfnamefont {K.~I.}\ \bibnamefont {Juntunen}}, \bibinfo
  {author} {\bibfnamefont {F.~B.}\ \bibnamefont {Rasmussen}},\ and\ \bibinfo
  {author} {\bibfnamefont {K.~K.}\ \bibnamefont {Nummila}},\ }\bibfield
  {title} {\bibinfo {title} {Polarized nuclei in normal and superconducting
  rhodium},\ }\href {https://doi.org/10.1023/A:1017545531677} {\bibfield
  {journal} {\bibinfo  {journal} {Journal of Low Temperature Physics}\ }\textbf
  {\bibinfo {volume} {123}},\ \bibinfo {pages} {65} (\bibinfo {year}
  {2001})}\BibitemShut {NoStop}%
\bibitem [{\citenamefont {Jacquinot}\ \emph {et~al.}(1974)\citenamefont
  {Jacquinot}, \citenamefont {Wenckebach}, \citenamefont {Goldman},\ and\
  \citenamefont {Abragam}}]{Jacquinot1974}%
  \BibitemOpen
  \bibfield  {author} {\bibinfo {author} {\bibfnamefont {J.~F.}\ \bibnamefont
  {Jacquinot}}, \bibinfo {author} {\bibfnamefont {W.~T.}\ \bibnamefont
  {Wenckebach}}, \bibinfo {author} {\bibfnamefont {M.}~\bibnamefont
  {Goldman}},\ and\ \bibinfo {author} {\bibfnamefont {A.}~\bibnamefont
  {Abragam}},\ }\bibfield  {title} {\bibinfo {title} {{Polarization and NMR
  Observation of $^{43}\mathrm{Ca}$ Nuclei in Ca${\mathrm{F}}_{2}$}},\ }\href
  {https://doi.org/10.1103/PhysRevLett.32.1096} {\bibfield  {journal} {\bibinfo
   {journal} {Phys. Rev. Lett.}\ }\textbf {\bibinfo {volume} {32}},\ \bibinfo
  {pages} {1096} (\bibinfo {year} {1974})}\BibitemShut {NoStop}%
\bibitem [{\citenamefont {Goldman}\ and\ \citenamefont
  {Jacquinot}(1976)}]{Goldman1976}%
  \BibitemOpen
  \bibfield  {author} {\bibinfo {author} {\bibfnamefont {M.}~\bibnamefont
  {Goldman}}\ and\ \bibinfo {author} {\bibfnamefont {J.~F.}\ \bibnamefont
  {Jacquinot}},\ }\bibfield  {title} {\bibinfo {title} {{Measurement of
  $^{43}\mathrm{Ca}$-$^{19}\mathrm{F}$ Dipolar Energy in Antiferromagnetic
  Ca${\mathrm{F}}_{2}$}},\ }\href {https://doi.org/10.1103/PhysRevLett.36.330}
  {\bibfield  {journal} {\bibinfo  {journal} {Phys. Rev. Lett.}\ }\textbf
  {\bibinfo {volume} {36}},\ \bibinfo {pages} {330} (\bibinfo {year}
  {1976})}\BibitemShut {NoStop}%
\bibitem [{\citenamefont {Petersen}\ \emph {et~al.}(2013)\citenamefont
  {Petersen}, \citenamefont {Hoffmann}, \citenamefont {Schuh}, \citenamefont
  {Wegscheider}, \citenamefont {Giedke},\ and\ \citenamefont
  {Ludwig}}]{Petersen2013}%
  \BibitemOpen
  \bibfield  {author} {\bibinfo {author} {\bibfnamefont {G.}~\bibnamefont
  {Petersen}}, \bibinfo {author} {\bibfnamefont {E.~A.}\ \bibnamefont
  {Hoffmann}}, \bibinfo {author} {\bibfnamefont {D.}~\bibnamefont {Schuh}},
  \bibinfo {author} {\bibfnamefont {W.}~\bibnamefont {Wegscheider}}, \bibinfo
  {author} {\bibfnamefont {G.}~\bibnamefont {Giedke}},\ and\ \bibinfo {author}
  {\bibfnamefont {S.}~\bibnamefont {Ludwig}},\ }\bibfield  {title} {\bibinfo
  {title} {Large nuclear spin polarization in gate-defined quantum dots using a
  single-domain nanomagnet},\ }\href
  {https://doi.org/10.1103/PhysRevLett.110.177602} {\bibfield  {journal}
  {\bibinfo  {journal} {Phys. Rev. Lett.}\ }\textbf {\bibinfo {volume} {110}},\
  \bibinfo {pages} {177602} (\bibinfo {year} {2013})}\BibitemShut {NoStop}%
\bibitem [{\citenamefont {Chekhovich}\ \emph {et~al.}(2017)\citenamefont
  {Chekhovich}, \citenamefont {Ulhaq}, \citenamefont {Zallo}, \citenamefont
  {Ding}, \citenamefont {Schmidt},\ and\ \citenamefont
  {Skolnick}}]{Chekhovich2017}%
  \BibitemOpen
  \bibfield  {author} {\bibinfo {author} {\bibfnamefont {E.~A.}\ \bibnamefont
  {Chekhovich}}, \bibinfo {author} {\bibfnamefont {A.}~\bibnamefont {Ulhaq}},
  \bibinfo {author} {\bibfnamefont {E.}~\bibnamefont {Zallo}}, \bibinfo
  {author} {\bibfnamefont {F.}~\bibnamefont {Ding}}, \bibinfo {author}
  {\bibfnamefont {O.~G.}\ \bibnamefont {Schmidt}},\ and\ \bibinfo {author}
  {\bibfnamefont {M.~S.}\ \bibnamefont {Skolnick}},\ }\bibfield  {title}
  {\bibinfo {title} {Measurement of the spin temperature of optically cooled
  nuclei and {{GaAs}} hyperfine constants in {{GaAs/AlGaAs}} quantum dots},\
  }\href {https://doi.org/10.1038/nmat4959} {\bibfield  {journal} {\bibinfo
  {journal} {Nature Mater.}\ }\textbf {\bibinfo {volume} {16}},\ \bibinfo
  {pages} {982} (\bibinfo {year} {2017})}\BibitemShut {NoStop}%
\bibitem [{\citenamefont {Kloeffel}\ and\ \citenamefont
  {Loss}(2013)}]{Kloeffel2013}%
  \BibitemOpen
  \bibfield  {author} {\bibinfo {author} {\bibfnamefont {C.}~\bibnamefont
  {Kloeffel}}\ and\ \bibinfo {author} {\bibfnamefont {D.}~\bibnamefont
  {Loss}},\ }\bibfield  {title} {\bibinfo {title} {Prospects for spin-based
  quantum computing in quantum dots},\ }\href
  {https://doi.org/10.1146/annurev-conmatphys-030212-184248} {\bibfield
  {journal} {\bibinfo  {journal} {Annual Review of Cond. Matt. Phys.}\ }\textbf
  {\bibinfo {volume} {4}},\ \bibinfo {pages} {51} (\bibinfo {year}
  {2013})}\BibitemShut {NoStop}%
\bibitem [{\citenamefont {Imamo\ifmmode~\bar{g}\else \={g}\fi{}lu}\ \emph
  {et~al.}(2003)\citenamefont {Imamo\ifmmode~\bar{g}\else \={g}\fi{}lu},
  \citenamefont {Knill}, \citenamefont {Tian},\ and\ \citenamefont
  {Zoller}}]{ImamogluPRL2003}%
  \BibitemOpen
  \bibfield  {author} {\bibinfo {author} {\bibfnamefont {A.}~\bibnamefont
  {Imamo\ifmmode~\bar{g}\else \={g}\fi{}lu}}, \bibinfo {author} {\bibfnamefont
  {E.}~\bibnamefont {Knill}}, \bibinfo {author} {\bibfnamefont
  {L.}~\bibnamefont {Tian}},\ and\ \bibinfo {author} {\bibfnamefont
  {P.}~\bibnamefont {Zoller}},\ }\bibfield  {title} {\bibinfo {title} {Optical
  pumping of quantum-dot nuclear spins},\ }\href
  {https://doi.org/10.1103/PhysRevLett.91.017402} {\bibfield  {journal}
  {\bibinfo  {journal} {Phys. Rev. Lett.}\ }\textbf {\bibinfo {volume} {91}},\
  \bibinfo {pages} {017402} (\bibinfo {year} {2003})}\BibitemShut {NoStop}%
\bibitem [{\citenamefont {Christ}\ \emph {et~al.}(2007)\citenamefont {Christ},
  \citenamefont {Cirac},\ and\ \citenamefont {Giedke}}]{Christ2007}%
  \BibitemOpen
  \bibfield  {author} {\bibinfo {author} {\bibfnamefont {H.}~\bibnamefont
  {Christ}}, \bibinfo {author} {\bibfnamefont {J.~I.}\ \bibnamefont {Cirac}},\
  and\ \bibinfo {author} {\bibfnamefont {G.}~\bibnamefont {Giedke}},\
  }\bibfield  {title} {\bibinfo {title} {Quantum description of nuclear spin
  cooling in a quantum dot},\ }\href
  {https://doi.org/10.1103/PhysRevB.75.155324} {\bibfield  {journal} {\bibinfo
  {journal} {Phys. Rev. B}\ }\textbf {\bibinfo {volume} {75}},\ \bibinfo
  {pages} {155324} (\bibinfo {year} {2007})}\BibitemShut {NoStop}%
\bibitem [{\citenamefont {Khaetskii}\ \emph {et~al.}(2002)\citenamefont
  {Khaetskii}, \citenamefont {Loss},\ and\ \citenamefont
  {Glazman}}]{Khaetskii2002}%
  \BibitemOpen
  \bibfield  {author} {\bibinfo {author} {\bibfnamefont {A.~V.}\ \bibnamefont
  {Khaetskii}}, \bibinfo {author} {\bibfnamefont {D.}~\bibnamefont {Loss}},\
  and\ \bibinfo {author} {\bibfnamefont {L.}~\bibnamefont {Glazman}},\
  }\bibfield  {title} {\bibinfo {title} {{Electron Spin Decoherence in Quantum
  Dots due to Interaction with Nuclei}},\ }\href
  {https://doi.org/10.1103/PhysRevLett.88.186802} {\bibfield  {journal}
  {\bibinfo  {journal} {Phys. Rev. Lett.}\ }\textbf {\bibinfo {volume} {88}},\
  \bibinfo {pages} {186802} (\bibinfo {year} {2002})}\BibitemShut {NoStop}%
\bibitem [{\citenamefont {Deng}\ and\ \citenamefont {Hu}(2008)}]{Deng2008}%
  \BibitemOpen
  \bibfield  {author} {\bibinfo {author} {\bibfnamefont {C.}~\bibnamefont
  {Deng}}\ and\ \bibinfo {author} {\bibfnamefont {X.}~\bibnamefont {Hu}},\
  }\bibfield  {title} {\bibinfo {title} {Electron-spin dephasing via hyperfine
  interaction in a quantum dot: An equation-of-motion calculation of
  electron-spin correlation functions},\ }\href
  {https://doi.org/10.1103/PhysRevB.78.245301} {\bibfield  {journal} {\bibinfo
  {journal} {Phys. Rev. B}\ }\textbf {\bibinfo {volume} {78}},\ \bibinfo
  {pages} {245301} (\bibinfo {year} {2008})}\BibitemShut {NoStop}%
\bibitem [{\citenamefont {Giedke}\ \emph {et~al.}(2006)\citenamefont {Giedke},
  \citenamefont {Taylor}, \citenamefont {D'Alessandro}, \citenamefont {Lukin},\
  and\ \citenamefont {Imamo\ifmmode~\breve{g}\else \u{g}\fi{}lu}}]{Giedke2006}%
  \BibitemOpen
  \bibfield  {author} {\bibinfo {author} {\bibfnamefont {G.}~\bibnamefont
  {Giedke}}, \bibinfo {author} {\bibfnamefont {J.~M.}\ \bibnamefont {Taylor}},
  \bibinfo {author} {\bibfnamefont {D.}~\bibnamefont {D'Alessandro}}, \bibinfo
  {author} {\bibfnamefont {M.~D.}\ \bibnamefont {Lukin}},\ and\ \bibinfo
  {author} {\bibfnamefont {A.}~\bibnamefont {Imamo\ifmmode~\breve{g}\else
  \u{g}\fi{}lu}},\ }\bibfield  {title} {\bibinfo {title} {Quantum measurement
  of a mesoscopic spin ensemble},\ }\href
  {https://doi.org/10.1103/PhysRevA.74.032316} {\bibfield  {journal} {\bibinfo
  {journal} {Phys. Rev. A}\ }\textbf {\bibinfo {volume} {74}},\ \bibinfo
  {pages} {032316} (\bibinfo {year} {2006})}\BibitemShut {NoStop}%
\bibitem [{\citenamefont {Kessler}\ \emph {et~al.}(2010)\citenamefont
  {Kessler}, \citenamefont {Yelin}, \citenamefont {Lukin}, \citenamefont
  {Cirac},\ and\ \citenamefont {Giedke}}]{Kessler2010}%
  \BibitemOpen
  \bibfield  {author} {\bibinfo {author} {\bibfnamefont {E.~M.}\ \bibnamefont
  {Kessler}}, \bibinfo {author} {\bibfnamefont {S.}~\bibnamefont {Yelin}},
  \bibinfo {author} {\bibfnamefont {M.~D.}\ \bibnamefont {Lukin}}, \bibinfo
  {author} {\bibfnamefont {J.~I.}\ \bibnamefont {Cirac}},\ and\ \bibinfo
  {author} {\bibfnamefont {G.}~\bibnamefont {Giedke}},\ }\bibfield  {title}
  {\bibinfo {title} {Optical superradiance from nuclear spin environment of
  single-photon emitters},\ }\href
  {https://doi.org/10.1103/PhysRevLett.104.143601} {\bibfield  {journal}
  {\bibinfo  {journal} {Phys. Rev. Lett.}\ }\textbf {\bibinfo {volume} {104}},\
  \bibinfo {pages} {143601} (\bibinfo {year} {2010})}\BibitemShut {NoStop}%
\bibitem [{\citenamefont {Schuetz}\ \emph {et~al.}(2012)\citenamefont
  {Schuetz}, \citenamefont {Kessler}, \citenamefont {Cirac},\ and\
  \citenamefont {Giedke}}]{Schuetz2012}%
  \BibitemOpen
  \bibfield  {author} {\bibinfo {author} {\bibfnamefont {M.~J.~A.}\
  \bibnamefont {Schuetz}}, \bibinfo {author} {\bibfnamefont {E.~M.}\
  \bibnamefont {Kessler}}, \bibinfo {author} {\bibfnamefont {J.~I.}\
  \bibnamefont {Cirac}},\ and\ \bibinfo {author} {\bibfnamefont
  {G.}~\bibnamefont {Giedke}},\ }\bibfield  {title} {\bibinfo {title}
  {Superradiance-like electron transport through a quantum dot},\ }\href
  {https://doi.org/10.1103/PhysRevB.86.085322} {\bibfield  {journal} {\bibinfo
  {journal} {Phys. Rev. B}\ }\textbf {\bibinfo {volume} {86}},\ \bibinfo
  {pages} {085322} (\bibinfo {year} {2012})}\BibitemShut {NoStop}%
\bibitem [{\citenamefont {Oja}\ and\ \citenamefont
  {Lounasmaa}(1997)}]{Oja1997}%
  \BibitemOpen
  \bibfield  {author} {\bibinfo {author} {\bibfnamefont {A.~S.}\ \bibnamefont
  {Oja}}\ and\ \bibinfo {author} {\bibfnamefont {O.~V.}\ \bibnamefont
  {Lounasmaa}},\ }\bibfield  {title} {\bibinfo {title} {Nuclear magnetic
  ordering in simple metals at positive and negative nanokelvin temperatures},\
  }\href {https://doi.org/10.1103/RevModPhys.69.1} {\bibfield  {journal}
  {\bibinfo  {journal} {Rev. Mod. Phys.}\ }\textbf {\bibinfo {volume} {69}},\
  \bibinfo {pages} {1} (\bibinfo {year} {1997})}\BibitemShut {NoStop}%
\bibitem [{\citenamefont {Kotur}\ \emph {et~al.}(2021)\citenamefont {Kotur},
  \citenamefont {Tolmachev}, \citenamefont {Litvyak}, \citenamefont {Kavokin},
  \citenamefont {Suter}, \citenamefont {Yakovlev},\ and\ \citenamefont
  {Bayer}}]{Kotur2021}%
  \BibitemOpen
  \bibfield  {author} {\bibinfo {author} {\bibfnamefont {M.}~\bibnamefont
  {Kotur}}, \bibinfo {author} {\bibfnamefont {D.~O.}\ \bibnamefont
  {Tolmachev}}, \bibinfo {author} {\bibfnamefont {V.~M.}\ \bibnamefont
  {Litvyak}}, \bibinfo {author} {\bibfnamefont {K.~V.}\ \bibnamefont
  {Kavokin}}, \bibinfo {author} {\bibfnamefont {D.}~\bibnamefont {Suter}},
  \bibinfo {author} {\bibfnamefont {D.~R.}\ \bibnamefont {Yakovlev}},\ and\
  \bibinfo {author} {\bibfnamefont {M.}~\bibnamefont {Bayer}},\ }\bibfield
  {title} {\bibinfo {title} {{Ultra-deep optical cooling of coupled nuclear
  spin-spin and quadrupole reservoirs in a GaAs/(Al,Ga)As quantum well}},\
  }\href {https://doi.org/10.1038/s42005-021-00681-6} {\bibfield  {journal}
  {\bibinfo  {journal} {Communications Physics}\ }\textbf {\bibinfo {volume}
  {4}},\ \bibinfo {pages} {193} (\bibinfo {year} {2021})}\BibitemShut {NoStop}%
\bibitem [{\citenamefont {Miller}\ \emph {et~al.}(1997)\citenamefont {Miller},
  \citenamefont {Hansen}, \citenamefont {Manus}, \citenamefont {Luyken},
  \citenamefont {Lorke}, \citenamefont {Kotthaus}, \citenamefont {Huant},
  \citenamefont {Medeiros-Ribeiro},\ and\ \citenamefont
  {Petroff}}]{Miller1997}%
  \BibitemOpen
  \bibfield  {author} {\bibinfo {author} {\bibfnamefont {B.~T.}\ \bibnamefont
  {Miller}}, \bibinfo {author} {\bibfnamefont {W.}~\bibnamefont {Hansen}},
  \bibinfo {author} {\bibfnamefont {S.}~\bibnamefont {Manus}}, \bibinfo
  {author} {\bibfnamefont {R.~J.}\ \bibnamefont {Luyken}}, \bibinfo {author}
  {\bibfnamefont {A.}~\bibnamefont {Lorke}}, \bibinfo {author} {\bibfnamefont
  {J.~P.}\ \bibnamefont {Kotthaus}}, \bibinfo {author} {\bibfnamefont
  {S.}~\bibnamefont {Huant}}, \bibinfo {author} {\bibfnamefont
  {G.}~\bibnamefont {Medeiros-Ribeiro}},\ and\ \bibinfo {author} {\bibfnamefont
  {P.~M.}\ \bibnamefont {Petroff}},\ }\bibfield  {title} {\bibinfo {title}
  {Few-electron ground states of charge-tunable self-assembled quantum dots},\
  }\href {https://doi.org/10.1103/PhysRevB.56.6764} {\bibfield  {journal}
  {\bibinfo  {journal} {Phys. Rev. B}\ }\textbf {\bibinfo {volume} {56}},\
  \bibinfo {pages} {6764} (\bibinfo {year} {1997})}\BibitemShut {NoStop}%
\bibitem [{\citenamefont {Warburton}\ \emph {et~al.}(2000)\citenamefont
  {Warburton}, \citenamefont {Sch{\"a}flein}, \citenamefont {Haft},
  \citenamefont {Bickel}, \citenamefont {Lorke}, \citenamefont {Karrai},
  \citenamefont {Garcia}, \citenamefont {Schoenfeld},\ and\ \citenamefont
  {Petroff}}]{Warburton2000}%
  \BibitemOpen
  \bibfield  {author} {\bibinfo {author} {\bibfnamefont {R.~J.}\ \bibnamefont
  {Warburton}}, \bibinfo {author} {\bibfnamefont {C.}~\bibnamefont
  {Sch{\"a}flein}}, \bibinfo {author} {\bibfnamefont {D.}~\bibnamefont {Haft}},
  \bibinfo {author} {\bibfnamefont {F.}~\bibnamefont {Bickel}}, \bibinfo
  {author} {\bibfnamefont {A.}~\bibnamefont {Lorke}}, \bibinfo {author}
  {\bibfnamefont {K.}~\bibnamefont {Karrai}}, \bibinfo {author} {\bibfnamefont
  {J.~M.}\ \bibnamefont {Garcia}}, \bibinfo {author} {\bibfnamefont
  {W.}~\bibnamefont {Schoenfeld}},\ and\ \bibinfo {author} {\bibfnamefont
  {P.~M.}\ \bibnamefont {Petroff}},\ }\bibfield  {title} {\bibinfo {title}
  {Optical emission from a charge-tunable quantum ring},\ }\href
  {https://doi.org/10.1038/35016030} {\bibfield  {journal} {\bibinfo  {journal}
  {Nature}\ }\textbf {\bibinfo {volume} {405}},\ \bibinfo {pages} {926}
  (\bibinfo {year} {2000})}\BibitemShut {NoStop}%
\bibitem [{\citenamefont {Urbaszek}\ \emph {et~al.}(2013)\citenamefont
  {Urbaszek}, \citenamefont {Marie}, \citenamefont {Amand}, \citenamefont
  {Krebs}, \citenamefont {Voisin}, \citenamefont {Maletinsky}, \citenamefont
  {H\"ogele},\ and\ \citenamefont {Imamo\ifmmode~\breve{g}\else
  \u{g}\fi{}lu}}]{Urbaszek2013}%
  \BibitemOpen
  \bibfield  {author} {\bibinfo {author} {\bibfnamefont {B.}~\bibnamefont
  {Urbaszek}}, \bibinfo {author} {\bibfnamefont {X.}~\bibnamefont {Marie}},
  \bibinfo {author} {\bibfnamefont {T.}~\bibnamefont {Amand}}, \bibinfo
  {author} {\bibfnamefont {O.}~\bibnamefont {Krebs}}, \bibinfo {author}
  {\bibfnamefont {P.}~\bibnamefont {Voisin}}, \bibinfo {author} {\bibfnamefont
  {P.}~\bibnamefont {Maletinsky}}, \bibinfo {author} {\bibfnamefont
  {A.}~\bibnamefont {H\"ogele}},\ and\ \bibinfo {author} {\bibfnamefont
  {A.}~\bibnamefont {Imamo\ifmmode~\breve{g}\else \u{g}\fi{}lu}},\ }\bibfield
  {title} {\bibinfo {title} {Nuclear spin physics in quantum dots: An optical
  investigation},\ }\href {https://doi.org/10.1103/RevModPhys.85.79} {\bibfield
   {journal} {\bibinfo  {journal} {Rev. Mod. Phys.}\ }\textbf {\bibinfo
  {volume} {85}},\ \bibinfo {pages} {79} (\bibinfo {year} {2013})}\BibitemShut
  {NoStop}%
\bibitem [{\citenamefont {Raymond}\ \emph {et~al.}(2004)\citenamefont
  {Raymond}, \citenamefont {Studenikin}, \citenamefont {Sachrajda},
  \citenamefont {Wasilewski}, \citenamefont {Cheng}, \citenamefont {Sheng},
  \citenamefont {Hawrylak}, \citenamefont {Babinski}, \citenamefont {Potemski},
  \citenamefont {Ortner},\ and\ \citenamefont {Bayer}}]{Raymond2004}%
  \BibitemOpen
  \bibfield  {author} {\bibinfo {author} {\bibfnamefont {S.}~\bibnamefont
  {Raymond}}, \bibinfo {author} {\bibfnamefont {S.}~\bibnamefont {Studenikin}},
  \bibinfo {author} {\bibfnamefont {A.}~\bibnamefont {Sachrajda}}, \bibinfo
  {author} {\bibfnamefont {Z.}~\bibnamefont {Wasilewski}}, \bibinfo {author}
  {\bibfnamefont {S.~J.}\ \bibnamefont {Cheng}}, \bibinfo {author}
  {\bibfnamefont {W.}~\bibnamefont {Sheng}}, \bibinfo {author} {\bibfnamefont
  {P.}~\bibnamefont {Hawrylak}}, \bibinfo {author} {\bibfnamefont
  {A.}~\bibnamefont {Babinski}}, \bibinfo {author} {\bibfnamefont
  {M.}~\bibnamefont {Potemski}}, \bibinfo {author} {\bibfnamefont
  {G.}~\bibnamefont {Ortner}},\ and\ \bibinfo {author} {\bibfnamefont
  {M.}~\bibnamefont {Bayer}},\ }\bibfield  {title} {\bibinfo {title}
  {{Excitonic Energy Shell Structure of Self-Assembled InGaAs/GaAs Quantum
  Dots}},\ }\href {https://doi.org/10.1103/PhysRevLett.92.187402} {\bibfield
  {journal} {\bibinfo  {journal} {Phys. Rev. Lett.}\ }\textbf {\bibinfo
  {volume} {92}},\ \bibinfo {pages} {187402} (\bibinfo {year}
  {2004})}\BibitemShut {NoStop}%
\bibitem [{\citenamefont {Babinski}\ \emph {et~al.}(2006)\citenamefont
  {Babinski}, \citenamefont {Potemski}, \citenamefont {Raymond}, \citenamefont
  {Lapointe},\ and\ \citenamefont {Wasilewski}}]{Babinski2006}%
  \BibitemOpen
  \bibfield  {author} {\bibinfo {author} {\bibfnamefont {A.}~\bibnamefont
  {Babinski}}, \bibinfo {author} {\bibfnamefont {M.}~\bibnamefont {Potemski}},
  \bibinfo {author} {\bibfnamefont {S.}~\bibnamefont {Raymond}}, \bibinfo
  {author} {\bibfnamefont {J.}~\bibnamefont {Lapointe}},\ and\ \bibinfo
  {author} {\bibfnamefont {Z.~R.}\ \bibnamefont {Wasilewski}},\ }\bibfield
  {title} {\bibinfo {title} {Emission from a highly excited single
  $\mathrm{InAs}\text{\ensuremath{-}}\mathrm{GaAs}$ quantum dot in magnetic
  fields: An excitonic {Fock-Darwin} diagram},\ }\href
  {https://doi.org/10.1103/PhysRevB.74.155301} {\bibfield  {journal} {\bibinfo
  {journal} {Phys. Rev. B}\ }\textbf {\bibinfo {volume} {74}},\ \bibinfo
  {pages} {155301} (\bibinfo {year} {2006})}\BibitemShut {NoStop}%
\bibitem [{\citenamefont {Schimpf}\ \emph {et~al.}(2019)\citenamefont
  {Schimpf}, \citenamefont {Reindl}, \citenamefont {Klenovsk\'{y}},
  \citenamefont {Fromherz}, \citenamefont {Covre Da~Silva}, \citenamefont
  {Hofer}, \citenamefont {Schneider}, \citenamefont {H\"{o}fling},
  \citenamefont {Trotta},\ and\ \citenamefont {Rastelli}}]{Schimpf2019}%
  \BibitemOpen
  \bibfield  {author} {\bibinfo {author} {\bibfnamefont {C.}~\bibnamefont
  {Schimpf}}, \bibinfo {author} {\bibfnamefont {M.}~\bibnamefont {Reindl}},
  \bibinfo {author} {\bibfnamefont {P.}~\bibnamefont {Klenovsk\'{y}}}, \bibinfo
  {author} {\bibfnamefont {T.}~\bibnamefont {Fromherz}}, \bibinfo {author}
  {\bibfnamefont {S.~F.}\ \bibnamefont {Covre Da~Silva}}, \bibinfo {author}
  {\bibfnamefont {J.}~\bibnamefont {Hofer}}, \bibinfo {author} {\bibfnamefont
  {C.}~\bibnamefont {Schneider}}, \bibinfo {author} {\bibfnamefont
  {S.}~\bibnamefont {H\"{o}fling}}, \bibinfo {author} {\bibfnamefont
  {R.}~\bibnamefont {Trotta}},\ and\ \bibinfo {author} {\bibfnamefont
  {A.}~\bibnamefont {Rastelli}},\ }\bibfield  {title} {\bibinfo {title}
  {Resolving the temporal evolution of line broadening in single quantum
  emitters},\ }\href {https://doi.org/10.1364/OE.27.035290} {\bibfield
  {journal} {\bibinfo  {journal} {Opt. Express}\ }\textbf {\bibinfo {volume}
  {27}},\ \bibinfo {pages} {35290} (\bibinfo {year} {2019})}\BibitemShut
  {NoStop}%
\bibitem [{\citenamefont {Taylor}\ \emph {et~al.}(2003)\citenamefont {Taylor},
  \citenamefont {Marcus},\ and\ \citenamefont {Lukin}}]{Taylor2003}%
  \BibitemOpen
  \bibfield  {author} {\bibinfo {author} {\bibfnamefont {J.~M.}\ \bibnamefont
  {Taylor}}, \bibinfo {author} {\bibfnamefont {C.~M.}\ \bibnamefont {Marcus}},\
  and\ \bibinfo {author} {\bibfnamefont {M.~D.}\ \bibnamefont {Lukin}},\
  }\bibfield  {title} {\bibinfo {title} {Long-lived memory for mesoscopic
  quantum bits},\ }\href {https://doi.org/10.1103/PhysRevLett.90.206803}
  {\bibfield  {journal} {\bibinfo  {journal} {Phys. Rev. Lett.}\ }\textbf
  {\bibinfo {volume} {90}},\ \bibinfo {pages} {206803} (\bibinfo {year}
  {2003})}\BibitemShut {NoStop}%
\bibitem [{\citenamefont {Goldman}(1970)}]{GoldmanBook}%
  \BibitemOpen
  \bibfield  {author} {\bibinfo {author} {\bibfnamefont {M.}~\bibnamefont
  {Goldman}},\ }\href@noop {} {\emph {\bibinfo {title} {Spin temperature and
  nuclear magnetic resonance in solids}}}\ (\bibinfo  {publisher} {Oxford
  University Press},\ \bibinfo {address} {Oxford},\ \bibinfo {year}
  {1970})\BibitemShut {NoStop}%
\bibitem [{\citenamefont {Chekhovich}\ \emph {et~al.}(2012)\citenamefont
  {Chekhovich}, \citenamefont {Kavokin}, \citenamefont {Puebla}, \citenamefont
  {Krysa}, \citenamefont {Hopkinson}, \citenamefont {Andreev}, \citenamefont
  {Sanchez}, \citenamefont {Beanland}, \citenamefont {Skolnick},\ and\
  \citenamefont {Tartakovskii}}]{Chekhovich2012}%
  \BibitemOpen
  \bibfield  {author} {\bibinfo {author} {\bibfnamefont {E.~A.}\ \bibnamefont
  {Chekhovich}}, \bibinfo {author} {\bibfnamefont {K.~V.}\ \bibnamefont
  {Kavokin}}, \bibinfo {author} {\bibfnamefont {J.}~\bibnamefont {Puebla}},
  \bibinfo {author} {\bibfnamefont {A.~B.}\ \bibnamefont {Krysa}}, \bibinfo
  {author} {\bibfnamefont {M.}~\bibnamefont {Hopkinson}}, \bibinfo {author}
  {\bibfnamefont {A.~D.}\ \bibnamefont {Andreev}}, \bibinfo {author}
  {\bibfnamefont {A.~M.}\ \bibnamefont {Sanchez}}, \bibinfo {author}
  {\bibfnamefont {R.}~\bibnamefont {Beanland}}, \bibinfo {author}
  {\bibfnamefont {M.~S.}\ \bibnamefont {Skolnick}},\ and\ \bibinfo {author}
  {\bibfnamefont {A.~I.}\ \bibnamefont {Tartakovskii}},\ }\bibfield  {title}
  {\bibinfo {title} {{Structural analysis of strained quantum dots using
  nuclear magnetic resonance}},\ }\href
  {https://doi.org/10.1038/nnano.2012.142} {\bibfield  {journal} {\bibinfo
  {journal} {Nature Nanotech.}\ }\textbf {\bibinfo {volume} {7}},\ \bibinfo
  {pages} {646} (\bibinfo {year} {2012})}\BibitemShut {NoStop}%
\bibitem [{\citenamefont {Bloch}(1946)}]{Bloch1946}%
  \BibitemOpen
  \bibfield  {author} {\bibinfo {author} {\bibfnamefont {F.}~\bibnamefont
  {Bloch}},\ }\bibfield  {title} {\bibinfo {title} {Nuclear induction},\ }\href
  {https://doi.org/10.1103/PhysRev.70.460} {\bibfield  {journal} {\bibinfo
  {journal} {Phys. Rev.}\ }\textbf {\bibinfo {volume} {70}},\ \bibinfo {pages}
  {460} (\bibinfo {year} {1946})}\BibitemShut {NoStop}%
\bibitem [{\citenamefont {Huo}\ \emph {et~al.}(2014)\citenamefont {Huo},
  \citenamefont {Witek}, \citenamefont {Kumar}, \citenamefont {Cardenas},
  \citenamefont {Zhang}, \citenamefont {Akopian}, \citenamefont {Singh},
  \citenamefont {Zallo}, \citenamefont {Grifone}, \citenamefont {Kriegner},
  \citenamefont {Trotta}, \citenamefont {Ding}, \citenamefont {Stangl},
  \citenamefont {Zwiller}, \citenamefont {Bester}, \citenamefont {Rastelli},\
  and\ \citenamefont {Schmidt}}]{Huo2014}%
  \BibitemOpen
  \bibfield  {author} {\bibinfo {author} {\bibfnamefont {Y.~H.}\ \bibnamefont
  {Huo}}, \bibinfo {author} {\bibfnamefont {B.~J.}\ \bibnamefont {Witek}},
  \bibinfo {author} {\bibfnamefont {S.}~\bibnamefont {Kumar}}, \bibinfo
  {author} {\bibfnamefont {J.~R.}\ \bibnamefont {Cardenas}}, \bibinfo {author}
  {\bibfnamefont {J.~X.}\ \bibnamefont {Zhang}}, \bibinfo {author}
  {\bibfnamefont {N.}~\bibnamefont {Akopian}}, \bibinfo {author} {\bibfnamefont
  {R.}~\bibnamefont {Singh}}, \bibinfo {author} {\bibfnamefont
  {E.}~\bibnamefont {Zallo}}, \bibinfo {author} {\bibfnamefont
  {R.}~\bibnamefont {Grifone}}, \bibinfo {author} {\bibfnamefont
  {D.}~\bibnamefont {Kriegner}}, \bibinfo {author} {\bibfnamefont
  {R.}~\bibnamefont {Trotta}}, \bibinfo {author} {\bibfnamefont
  {F.}~\bibnamefont {Ding}}, \bibinfo {author} {\bibfnamefont {J.}~\bibnamefont
  {Stangl}}, \bibinfo {author} {\bibfnamefont {V.}~\bibnamefont {Zwiller}},
  \bibinfo {author} {\bibfnamefont {G.}~\bibnamefont {Bester}}, \bibinfo
  {author} {\bibfnamefont {A.}~\bibnamefont {Rastelli}},\ and\ \bibinfo
  {author} {\bibfnamefont {O.~G.}\ \bibnamefont {Schmidt}},\ }\bibfield
  {title} {\bibinfo {title} {A light-hole exciton in a quantum dot},\ }\href
  {https://doi.org/10.1038/nphys2799} {\bibfield  {journal} {\bibinfo
  {journal} {Nat. Phys.}\ }\textbf {\bibinfo {volume} {10}},\ \bibinfo {pages}
  {46} (\bibinfo {year} {2014})}\BibitemShut {NoStop}%
\bibitem [{\citenamefont {Csontosov\'a}\ and\ \citenamefont
  {Klenovsk\'y}(2020)}]{Csontosova2020}%
  \BibitemOpen
  \bibfield  {author} {\bibinfo {author} {\bibfnamefont {D.}~\bibnamefont
  {Csontosov\'a}}\ and\ \bibinfo {author} {\bibfnamefont {P.}~\bibnamefont
  {Klenovsk\'y}},\ }\bibfield  {title} {\bibinfo {title} {Theory of
  magneto-optical properties of neutral and charged excitons in {GaAs/AlGaAs}
  quantum dots},\ }\href {https://doi.org/10.1103/PhysRevB.102.125412}
  {\bibfield  {journal} {\bibinfo  {journal} {Phys. Rev. B}\ }\textbf {\bibinfo
  {volume} {102}},\ \bibinfo {pages} {125412} (\bibinfo {year}
  {2020})}\BibitemShut {NoStop}%
\bibitem [{\citenamefont {Millington-Hotze}\ \emph {et~al.}(2022)\citenamefont
  {Millington-Hotze}, \citenamefont {Manna}, \citenamefont {Covre~da Silva},
  \citenamefont {Rastelli},\ and\ \citenamefont
  {Chekhovich}}]{MillingtonHotze2022}%
  \BibitemOpen
  \bibfield  {author} {\bibinfo {author} {\bibfnamefont {P.}~\bibnamefont
  {Millington-Hotze}}, \bibinfo {author} {\bibfnamefont {S.}~\bibnamefont
  {Manna}}, \bibinfo {author} {\bibfnamefont {S.~F.}\ \bibnamefont {Covre~da
  Silva}}, \bibinfo {author} {\bibfnamefont {A.}~\bibnamefont {Rastelli}},\
  and\ \bibinfo {author} {\bibfnamefont {E.~A.}\ \bibnamefont {Chekhovich}},\
  }\bibfield  {title} {\bibinfo {title} {{Nuclear spin diffusion in the central
  spin system of a GaAs/AlGaAs quantum dot}},\ }\href
  {https://doi.org/10.48550/arXiv.2208.02037} {\bibfield  {journal} {\bibinfo
  {journal} {arXiv}\ ,\ \bibinfo {pages} {arXiv:2208.02037}} (\bibinfo {year}
  {2022})}\BibitemShut {NoStop}%
\bibitem [{\citenamefont {Xu}\ \emph {et~al.}(2009)\citenamefont {Xu},
  \citenamefont {Yao}, \citenamefont {Sun}, \citenamefont {Steel},
  \citenamefont {Bracker}, \citenamefont {Gammon},\ and\ \citenamefont
  {Sham}}]{Xu2009}%
  \BibitemOpen
  \bibfield  {author} {\bibinfo {author} {\bibfnamefont {X.}~\bibnamefont
  {Xu}}, \bibinfo {author} {\bibfnamefont {W.}~\bibnamefont {Yao}}, \bibinfo
  {author} {\bibfnamefont {B.}~\bibnamefont {Sun}}, \bibinfo {author}
  {\bibfnamefont {D.~G.}\ \bibnamefont {Steel}}, \bibinfo {author}
  {\bibfnamefont {A.~S.}\ \bibnamefont {Bracker}}, \bibinfo {author}
  {\bibfnamefont {D.}~\bibnamefont {Gammon}},\ and\ \bibinfo {author}
  {\bibfnamefont {L.~J.}\ \bibnamefont {Sham}},\ }\bibfield  {title} {\bibinfo
  {title} {Optically controlled locking of the nuclear field via coherent
  dark-state spectroscopy},\ }\href {https://doi.org/10.1038/nature08120}
  {\bibfield  {journal} {\bibinfo  {journal} {Nature}\ }\textbf {\bibinfo
  {volume} {459}},\ \bibinfo {pages} {1105} (\bibinfo {year}
  {2009})}\BibitemShut {NoStop}%
\bibitem [{\citenamefont {Jackson}\ \emph {et~al.}(2022)\citenamefont
  {Jackson}, \citenamefont {Haeusler}, \citenamefont {Zaporski}, \citenamefont
  {Bodey}, \citenamefont {Shofer}, \citenamefont {Clarke}, \citenamefont
  {Hugues}, \citenamefont {Atat\"ure}, \citenamefont {Le~Gall},\ and\
  \citenamefont {Gangloff}}]{Jackson2022}%
  \BibitemOpen
  \bibfield  {author} {\bibinfo {author} {\bibfnamefont {D.~M.}\ \bibnamefont
  {Jackson}}, \bibinfo {author} {\bibfnamefont {U.}~\bibnamefont {Haeusler}},
  \bibinfo {author} {\bibfnamefont {L.}~\bibnamefont {Zaporski}}, \bibinfo
  {author} {\bibfnamefont {J.~H.}\ \bibnamefont {Bodey}}, \bibinfo {author}
  {\bibfnamefont {N.}~\bibnamefont {Shofer}}, \bibinfo {author} {\bibfnamefont
  {E.}~\bibnamefont {Clarke}}, \bibinfo {author} {\bibfnamefont
  {M.}~\bibnamefont {Hugues}}, \bibinfo {author} {\bibfnamefont
  {M.}~\bibnamefont {Atat\"ure}}, \bibinfo {author} {\bibfnamefont
  {C.}~\bibnamefont {Le~Gall}},\ and\ \bibinfo {author} {\bibfnamefont {D.~A.}\
  \bibnamefont {Gangloff}},\ }\bibfield  {title} {\bibinfo {title} {Optimal
  purification of a spin ensemble by quantum-algorithmic feedback},\ }\href
  {https://doi.org/10.1103/PhysRevX.12.031014} {\bibfield  {journal} {\bibinfo
  {journal} {Phys. Rev. X}\ }\textbf {\bibinfo {volume} {12}},\ \bibinfo
  {pages} {031014} (\bibinfo {year} {2022})}\BibitemShut {NoStop}%
\bibitem [{\citenamefont {Zaporski}\ \emph {et~al.}(2023)\citenamefont
  {Zaporski}, \citenamefont {Shofer}, \citenamefont {Bodey}, \citenamefont
  {Manna}, \citenamefont {Gillard}, \citenamefont {Appel}, \citenamefont
  {Schimpf}, \citenamefont {Covre~da Silva}, \citenamefont {Jarman},
  \citenamefont {Delamare}, \citenamefont {Park}, \citenamefont {Haeusler},
  \citenamefont {Chekhovich}, \citenamefont {Rastelli}, \citenamefont
  {Gangloff}, \citenamefont {Atat{\"u}re},\ and\ \citenamefont
  {Le~Gall}}]{Zaporski2022}%
  \BibitemOpen
  \bibfield  {author} {\bibinfo {author} {\bibfnamefont {L.}~\bibnamefont
  {Zaporski}}, \bibinfo {author} {\bibfnamefont {N.}~\bibnamefont {Shofer}},
  \bibinfo {author} {\bibfnamefont {J.~H.}\ \bibnamefont {Bodey}}, \bibinfo
  {author} {\bibfnamefont {S.}~\bibnamefont {Manna}}, \bibinfo {author}
  {\bibfnamefont {G.}~\bibnamefont {Gillard}}, \bibinfo {author} {\bibfnamefont
  {M.~H.}\ \bibnamefont {Appel}}, \bibinfo {author} {\bibfnamefont
  {C.}~\bibnamefont {Schimpf}}, \bibinfo {author} {\bibfnamefont {S.~F.}\
  \bibnamefont {Covre~da Silva}}, \bibinfo {author} {\bibfnamefont
  {J.}~\bibnamefont {Jarman}}, \bibinfo {author} {\bibfnamefont
  {G.}~\bibnamefont {Delamare}}, \bibinfo {author} {\bibfnamefont
  {G.}~\bibnamefont {Park}}, \bibinfo {author} {\bibfnamefont {U.}~\bibnamefont
  {Haeusler}}, \bibinfo {author} {\bibfnamefont {E.~A.}\ \bibnamefont
  {Chekhovich}}, \bibinfo {author} {\bibfnamefont {A.}~\bibnamefont
  {Rastelli}}, \bibinfo {author} {\bibfnamefont {D.~A.}\ \bibnamefont
  {Gangloff}}, \bibinfo {author} {\bibfnamefont {M.}~\bibnamefont
  {Atat{\"u}re}},\ and\ \bibinfo {author} {\bibfnamefont {C.}~\bibnamefont
  {Le~Gall}},\ }\bibfield  {title} {\bibinfo {title} {Ideal refocusing of an
  optically active spin qubit under strong hyperfine interactions},\ }\bibfield
   {journal} {\bibinfo  {journal} {Nature Nanotechnology}\ }\href
  {https://doi.org/10.1038/s41565-022-01282-2} {10.1038/s41565-022-01282-2}
  (\bibinfo {year} {2023})\BibitemShut {NoStop}%
\bibitem [{\citenamefont {Latta}\ \emph {et~al.}(2009)\citenamefont {Latta},
  \citenamefont {H\"{o}gele}, \citenamefont {Zhao}, \citenamefont {Vamivakas},
  \citenamefont {Maletinsky}, \citenamefont {Kroner}, \citenamefont {Dreiser},
  \citenamefont {Carusotto}, \citenamefont {Badolato}, \citenamefont {Schuh},
  \citenamefont {Wegscheider}, \citenamefont {Atature},\ and\ \citenamefont
  {Imamo\ifmmode~\breve{g}\else \u{g}\fi{}lu}}]{Latta2009}%
  \BibitemOpen
  \bibfield  {author} {\bibinfo {author} {\bibfnamefont {C.}~\bibnamefont
  {Latta}}, \bibinfo {author} {\bibfnamefont {A.}~\bibnamefont {H\"{o}gele}},
  \bibinfo {author} {\bibfnamefont {Y.}~\bibnamefont {Zhao}}, \bibinfo {author}
  {\bibfnamefont {A.~N.}\ \bibnamefont {Vamivakas}}, \bibinfo {author}
  {\bibfnamefont {M.}~\bibnamefont {Maletinsky}}, \bibinfo {author}
  {\bibfnamefont {M.}~\bibnamefont {Kroner}}, \bibinfo {author} {\bibfnamefont
  {J.}~\bibnamefont {Dreiser}}, \bibinfo {author} {\bibfnamefont
  {I.}~\bibnamefont {Carusotto}}, \bibinfo {author} {\bibfnamefont
  {A.}~\bibnamefont {Badolato}}, \bibinfo {author} {\bibfnamefont
  {D.}~\bibnamefont {Schuh}}, \bibinfo {author} {\bibfnamefont
  {W.}~\bibnamefont {Wegscheider}}, \bibinfo {author} {\bibfnamefont
  {M.}~\bibnamefont {Atature}},\ and\ \bibinfo {author} {\bibfnamefont
  {A.}~\bibnamefont {Imamo\ifmmode~\breve{g}\else \u{g}\fi{}lu}},\ }\bibfield
  {title} {\bibinfo {title} {Confluence of resonant laser excitation and
  bidirectional quantum-dot nuclear-spin plarization},\ }\href
  {https://doi.org/10.1038/nphys1363} {\bibfield  {journal} {\bibinfo
  {journal} {Nat. Phys.}\ }\textbf {\bibinfo {volume} {5}},\ \bibinfo {pages}
  {758 } (\bibinfo {year} {2009})}\BibitemShut {NoStop}%
\bibitem [{\citenamefont {H\"{o}gele}\ \emph {et~al.}(2012)\citenamefont
  {H\"{o}gele}, \citenamefont {Kroner}, \citenamefont {Latta}, \citenamefont
  {Claassen}, \citenamefont {Carusotto}, \citenamefont {Bulutay},\ and\
  \citenamefont {Imamo\ifmmode~\breve{g}\else \u{g}\fi{}lu}}]{Hoegele2012}%
  \BibitemOpen
  \bibfield  {author} {\bibinfo {author} {\bibfnamefont {A.}~\bibnamefont
  {H\"{o}gele}}, \bibinfo {author} {\bibfnamefont {M.}~\bibnamefont {Kroner}},
  \bibinfo {author} {\bibfnamefont {C.}~\bibnamefont {Latta}}, \bibinfo
  {author} {\bibfnamefont {M.}~\bibnamefont {Claassen}}, \bibinfo {author}
  {\bibfnamefont {I.}~\bibnamefont {Carusotto}}, \bibinfo {author}
  {\bibfnamefont {C.}~\bibnamefont {Bulutay}},\ and\ \bibinfo {author}
  {\bibfnamefont {A.}~\bibnamefont {Imamo\ifmmode~\breve{g}\else
  \u{g}\fi{}lu}},\ }\bibfield  {title} {\bibinfo {title} {Dynamic nuclear spin
  polarization in the resonant laser excitation of an {InGaAs} quantum dot},\
  }\href {https://doi.org/10.1103/PhysRevLett.108.197403} {\bibfield  {journal}
  {\bibinfo  {journal} {Phys. Rev. Lett.}\ }\textbf {\bibinfo {volume} {108}},\
  \bibinfo {pages} {197403} (\bibinfo {year} {2012})}\BibitemShut {NoStop}%
\bibitem [{\citenamefont {Gangloff}\ \emph {et~al.}(2021)\citenamefont
  {Gangloff}, \citenamefont {Zaporski}, \citenamefont {Bodey}, \citenamefont
  {Bachorz}, \citenamefont {Jackson}, \citenamefont {{\'E}thier-Majcher},
  \citenamefont {Lang}, \citenamefont {Clarke}, \citenamefont {Hugues},
  \citenamefont {Le~Gall},\ and\ \citenamefont {Atat{\"u}re}}]{Gangloff2021}%
  \BibitemOpen
  \bibfield  {author} {\bibinfo {author} {\bibfnamefont {D.~A.}\ \bibnamefont
  {Gangloff}}, \bibinfo {author} {\bibfnamefont {L.}~\bibnamefont {Zaporski}},
  \bibinfo {author} {\bibfnamefont {J.~H.}\ \bibnamefont {Bodey}}, \bibinfo
  {author} {\bibfnamefont {C.}~\bibnamefont {Bachorz}}, \bibinfo {author}
  {\bibfnamefont {D.~M.}\ \bibnamefont {Jackson}}, \bibinfo {author}
  {\bibfnamefont {G.}~\bibnamefont {{\'E}thier-Majcher}}, \bibinfo {author}
  {\bibfnamefont {C.}~\bibnamefont {Lang}}, \bibinfo {author} {\bibfnamefont
  {E.}~\bibnamefont {Clarke}}, \bibinfo {author} {\bibfnamefont
  {M.}~\bibnamefont {Hugues}}, \bibinfo {author} {\bibfnamefont
  {C.}~\bibnamefont {Le~Gall}},\ and\ \bibinfo {author} {\bibfnamefont
  {M.}~\bibnamefont {Atat{\"u}re}},\ }\bibfield  {title} {\bibinfo {title}
  {Witnessing quantum correlations in a nuclear ensemble via an electron spin
  qubit},\ }\href {https://doi.org/10.1038/s41567-021-01344-7} {\bibfield
  {journal} {\bibinfo  {journal} {Nat. Phys.}\ }\textbf {\bibinfo {volume}
  {17}},\ \bibinfo {pages} {1247} (\bibinfo {year} {2021})}\BibitemShut
  {NoStop}%


\setcounter{firstbib}{\value{NAT@ctr}}

\end{thebibliography}

\begin{thebibliography}{34}%

\setcounter{NAT@ctr}{\value{firstbib}}


\makeatletter
\providecommand \@ifxundefined [1]{%
 \@ifx{#1\undefined}
}%
\providecommand \@ifnum [1]{%
 \ifnum #1\expandafter \@firstoftwo
 \else \expandafter \@secondoftwo
 \fi
}%
\providecommand \@ifx [1]{%
 \ifx #1\expandafter \@firstoftwo
 \else \expandafter \@secondoftwo
 \fi
}%
\providecommand \natexlab [1]{#1}%
\providecommand \enquote  [1]{``#1''}%
\providecommand \bibnamefont  [1]{#1}%
\providecommand \bibfnamefont [1]{#1}%
\providecommand \citenamefont [1]{#1}%
\providecommand \href@noop [0]{\@secondoftwo}%
\providecommand \href [0]{\begingroup \@sanitize@url \@href}%
\providecommand \@href[1]{\@@startlink{#1}\@@href}%
\providecommand \@@href[1]{\endgroup#1\@@endlink}%
\providecommand \@sanitize@url [0]{\catcode `\\12\catcode `\$12\catcode
  `\&12\catcode `\#12\catcode `\^12\catcode `\_12\catcode `\%12\relax}%
\providecommand \@@startlink[1]{}%
\providecommand \@@endlink[0]{}%
\providecommand \url  [0]{\begingroup\@sanitize@url \@url }%
\providecommand \@url [1]{\endgroup\@href {#1}{\urlprefix }}%
\providecommand \urlprefix  [0]{URL }%
\providecommand \Eprint [0]{\href }%
\providecommand \doibase [0]{https://doi.org/}%
\providecommand \selectlanguage [0]{\@gobble}%
\providecommand \bibinfo  [0]{\@secondoftwo}%
\providecommand \bibfield  [0]{\@secondoftwo}%
\providecommand \translation [1]{[#1]}%
\providecommand \BibitemOpen [0]{}%
\providecommand \bibitemStop [0]{}%
\providecommand \bibitemNoStop [0]{.\EOS\space}%
\providecommand \EOS [0]{\spacefactor3000\relax}%
\providecommand \BibitemShut  [1]{\csname bibitem#1\endcsname}%
\let\auto@bib@innerbib\@empty
\bibitem [{\citenamefont {Oshiyama}\ and\ \citenamefont
  {Ohnishi}(1986)}]{Oshiyama1986}%
  \BibitemOpen
  \bibfield  {author} {\bibinfo {author} {\bibfnamefont {A.}~\bibnamefont
  {Oshiyama}}\ and\ \bibinfo {author} {\bibfnamefont {S.}~\bibnamefont
  {Ohnishi}},\ }\bibfield  {title} {\bibinfo {title} {{DX center: Crossover of
  deep and shallow states in Si-doped ${\mathrm{Al}}_{\mathrm{x}}$${\mathrm{Ga}}_{1\mathrm{\ensuremath{-}}\mathrm{x}}$As}},\ }\href {https://doi.org/10.1103/PhysRevB.33.4320} {\bibfield  {journal}
  {\bibinfo  {journal} {Phys. Rev. B}\ }\textbf {\bibinfo {volume} {33}},\
  \bibinfo {pages} {4320} (\bibinfo {year} {1986})}\BibitemShut {NoStop}%
\bibitem [{\citenamefont {Mooney}(1990)}]{Mooney1990}%
  \BibitemOpen
  \bibfield  {author} {\bibinfo {author} {\bibfnamefont {P.~M.}\ \bibnamefont
  {Mooney}},\ }\bibfield  {title} {\bibinfo {title} {{Deep donor levels (DX
  centers) in III-V semiconductors}},\ }\href
  {https://doi.org/10.1063/1.345628} {\bibfield  {journal} {\bibinfo  {journal}
  {Journal of Applied Physics}\ }\textbf {\bibinfo {volume} {67}},\ \bibinfo
  {pages} {R1} (\bibinfo {year} {1990})}\BibitemShut {NoStop}%
\bibitem [{\citenamefont {Zhai}\ \emph {et~al.}(2020)\citenamefont {Zhai},
  \citenamefont {L{\"o}bl}, \citenamefont {Nguyen}, \citenamefont {Ritzmann},
  \citenamefont {Javadi}, \citenamefont {Spinnler}, \citenamefont {Wieck},
  \citenamefont {Ludwig},\ and\ \citenamefont {Warburton}}]{Zhai2020}%
  \BibitemOpen
  \bibfield  {author} {\bibinfo {author} {\bibfnamefont {L.}~\bibnamefont
  {Zhai}}, \bibinfo {author} {\bibfnamefont {M.~C.}\ \bibnamefont {L{\"o}bl}},
  \bibinfo {author} {\bibfnamefont {G.~N.}\ \bibnamefont {Nguyen}}, \bibinfo
  {author} {\bibfnamefont {J.}~\bibnamefont {Ritzmann}}, \bibinfo {author}
  {\bibfnamefont {A.}~\bibnamefont {Javadi}}, \bibinfo {author} {\bibfnamefont
  {C.}~\bibnamefont {Spinnler}}, \bibinfo {author} {\bibfnamefont {A.~D.}\
  \bibnamefont {Wieck}}, \bibinfo {author} {\bibfnamefont {A.}~\bibnamefont
  {Ludwig}},\ and\ \bibinfo {author} {\bibfnamefont {R.~J.}\ \bibnamefont
  {Warburton}},\ }\bibfield  {title} {\bibinfo {title} {{Low-noise GaAs quantum
  dots for quantum photonics}},\ }\href
  {https://doi.org/10.1038/s41467-020-18625-z} {\bibfield  {journal} {\bibinfo
  {journal} {Nat. Commun.}\ }\textbf {\bibinfo {volume} {11}},\ \bibinfo
  {pages} {4745} (\bibinfo {year} {2020})}\BibitemShut {NoStop}%
\bibitem [{\citenamefont {Heyn}\ \emph {et~al.}(2009)\citenamefont {Heyn},
  \citenamefont {Stemmann}, \citenamefont {Koppen}, \citenamefont {Strelow},
  \citenamefont {Kipp}, \citenamefont {Grave}, \citenamefont {Mendach},\ and\
  \citenamefont {Hansen}}]{Heyn2009}%
  \BibitemOpen
  \bibfield  {author} {\bibinfo {author} {\bibfnamefont {C.}~\bibnamefont
  {Heyn}}, \bibinfo {author} {\bibfnamefont {A.}~\bibnamefont {Stemmann}},
  \bibinfo {author} {\bibfnamefont {T.}~\bibnamefont {Koppen}}, \bibinfo
  {author} {\bibfnamefont {C.}~\bibnamefont {Strelow}}, \bibinfo {author}
  {\bibfnamefont {T.}~\bibnamefont {Kipp}}, \bibinfo {author} {\bibfnamefont
  {M.}~\bibnamefont {Grave}}, \bibinfo {author} {\bibfnamefont
  {S.}~\bibnamefont {Mendach}},\ and\ \bibinfo {author} {\bibfnamefont
  {W.}~\bibnamefont {Hansen}},\ }\bibfield  {title} {\bibinfo {title} {Highly
  uniform and strain-free {{GaAs}} quantum dots fabricated by filling of
  self-assembled nanoholes},\ }\href {https://doi.org/10.1063/1.3133338}
  {\bibfield  {journal} {\bibinfo  {journal} {Appl. Phys. Lett.}\ }\textbf
  {\bibinfo {volume} {94}},\ \bibinfo {pages} {183113} (\bibinfo {year}
  {2009})}\BibitemShut {NoStop}%
\bibitem [{\citenamefont {Atkinson}\ \emph {et~al.}(2012)\citenamefont
  {Atkinson}, \citenamefont {Zallo},\ and\ \citenamefont
  {Schmidt}}]{Atkinson2012}%
  \BibitemOpen
  \bibfield  {author} {\bibinfo {author} {\bibfnamefont {P.}~\bibnamefont
  {Atkinson}}, \bibinfo {author} {\bibfnamefont {E.}~\bibnamefont {Zallo}},\
  and\ \bibinfo {author} {\bibfnamefont {O.~G.}\ \bibnamefont {Schmidt}},\
  }\bibfield  {title} {\bibinfo {title} {Independent wavelength and density
  control of uniform {GaAs/AlGaAs} quantum dots grown by infilling
  self-assembled nanoholes},\ }\href
  {https://doi.org/ttp://dx.doi.org/10.1063/1.4748183} {\bibfield  {journal}
  {\bibinfo  {journal} {J. Appl. Phys.}\ }\textbf {\bibinfo {volume} {112}},\
  \bibinfo {pages} {054303} (\bibinfo {year} {2012})}\BibitemShut {NoStop}%
\bibitem [{\citenamefont {El~Khalifi}\ \emph {et~al.}(1989)\citenamefont
  {El~Khalifi}, \citenamefont {Gil}, \citenamefont {Mathieu}, \citenamefont
  {Fukunaga},\ and\ \citenamefont {Nakashima}}]{ElKhalifi1989}%
  \BibitemOpen
  \bibfield  {author} {\bibinfo {author} {\bibfnamefont {Y.}~\bibnamefont
  {El~Khalifi}}, \bibinfo {author} {\bibfnamefont {B.}~\bibnamefont {Gil}},
  \bibinfo {author} {\bibfnamefont {H.}~\bibnamefont {Mathieu}}, \bibinfo
  {author} {\bibfnamefont {T.}~\bibnamefont {Fukunaga}},\ and\ \bibinfo
  {author} {\bibfnamefont {H.}~\bibnamefont {Nakashima}},\ }\bibfield  {title}
  {\bibinfo {title} {Dependence of the light-hole---heavy-hole splitting on
  layer thickness and substrate orientation in {{GaAs-(GaAl) As}} single
  quantum wells},\ }\href {https://doi.org/10.1103/PhysRevB.39.13533}
  {\bibfield  {journal} {\bibinfo  {journal} {Phys. Rev. B}\ }\textbf {\bibinfo
  {volume} {39}},\ \bibinfo {pages} {13533} (\bibinfo {year}
  {1989})}\BibitemShut {NoStop}%
\bibitem [{\citenamefont {Timofeev}\ \emph {et~al.}(1996)\citenamefont
  {Timofeev}, \citenamefont {Bayer}, \citenamefont {Forchel},\ and\
  \citenamefont {Potemski}}]{Timofeev1996}%
  \BibitemOpen
  \bibfield  {author} {\bibinfo {author} {\bibfnamefont {V.~B.}\ \bibnamefont
  {Timofeev}}, \bibinfo {author} {\bibfnamefont {M.}~\bibnamefont {Bayer}},
  \bibinfo {author} {\bibfnamefont {A.}~\bibnamefont {Forchel}},\ and\ \bibinfo
  {author} {\bibfnamefont {M.}~\bibnamefont {Potemski}},\ }\bibfield  {title}
  {\bibinfo {title} {Mixing of excitonic states containing light and heavy
  holes in an isolated {{GaAs/AlGaAs}} quantum well in a magnetic field},\
  }\href {https://doi.org/10.1134/1.567159} {\bibfield  {journal} {\bibinfo
  {journal} {Journal of Experimental and Theoretical Physics Letters}\ }\textbf
  {\bibinfo {volume} {64}},\ \bibinfo {pages} {57} (\bibinfo {year}
  {1996})}\BibitemShut {NoStop}%
\bibitem [{\citenamefont {Checkhovich}\ \emph {et~al.}(2013)\citenamefont
  {Checkhovich}, \citenamefont {Glazov}, \citenamefont {Krysa}, \citenamefont
  {Hopkinson}, \citenamefont {Senellart}, \citenamefont {Lema\^{i}tre},
  \citenamefont {Skolnick},\ and\ \citenamefont
  {Tartakovskii}}]{Chekhovich2013NPhys}%
  \BibitemOpen
  \bibfield  {author} {\bibinfo {author} {\bibfnamefont {E.~A.}\ \bibnamefont
  {Checkhovich}}, \bibinfo {author} {\bibfnamefont {M.~M.}\ \bibnamefont
  {Glazov}}, \bibinfo {author} {\bibfnamefont {A.~B.}\ \bibnamefont {Krysa}},
  \bibinfo {author} {\bibfnamefont {M.}~\bibnamefont {Hopkinson}}, \bibinfo
  {author} {\bibfnamefont {P.}~\bibnamefont {Senellart}}, \bibinfo {author}
  {\bibfnamefont {A.}~\bibnamefont {Lema\^{i}tre}}, \bibinfo {author}
  {\bibfnamefont {M.~S.}\ \bibnamefont {Skolnick}},\ and\ \bibinfo {author}
  {\bibfnamefont {A.~I.}\ \bibnamefont {Tartakovskii}},\ }\bibfield  {title}
  {\bibinfo {title} {Element-sensitive measurement of the hole-nuclear spin
  interaction in quantum dots},\ }\href {https://doi.org/10.1038/nphys2514}
  {\bibfield  {journal} {\bibinfo  {journal} {Nat. Phys.}\ }\textbf {\bibinfo
  {volume} {9}},\ \bibinfo {pages} {74} (\bibinfo {year} {2013})}\BibitemShut
  {NoStop}%
\bibitem [{\citenamefont {Slichter}(1990)}]{SlichterBook}%
  \BibitemOpen
  \bibfield  {author} {\bibinfo {author} {\bibfnamefont {C.~P.}\ \bibnamefont
  {Slichter}},\ }\href@noop {} {\emph {\bibinfo {title} {{Principles of
  Magnetic Resonance}}}}\ (\bibinfo  {publisher} {Springer},\ \bibinfo {year}
  {1990})\BibitemShut {NoStop}%
\bibitem [{\citenamefont {Huber}\ \emph {et~al.}(2019)\citenamefont {Huber},
  \citenamefont {Lehner}, \citenamefont {Csontosov\'a}, \citenamefont {Reindl},
  \citenamefont {Schuler}, \citenamefont {Covre~da Silva}, \citenamefont
  {Klenovsk\'y},\ and\ \citenamefont {Rastelli}}]{Huber2019}%
  \BibitemOpen
  \bibfield  {author} {\bibinfo {author} {\bibfnamefont {D.}~\bibnamefont
  {Huber}}, \bibinfo {author} {\bibfnamefont {B.~U.}\ \bibnamefont {Lehner}},
  \bibinfo {author} {\bibfnamefont {D.}~\bibnamefont {Csontosov\'a}}, \bibinfo
  {author} {\bibfnamefont {M.}~\bibnamefont {Reindl}}, \bibinfo {author}
  {\bibfnamefont {S.}~\bibnamefont {Schuler}}, \bibinfo {author} {\bibfnamefont
  {S.~F.}\ \bibnamefont {Covre~da Silva}}, \bibinfo {author} {\bibfnamefont
  {P.}~\bibnamefont {Klenovsk\'y}},\ and\ \bibinfo {author} {\bibfnamefont
  {A.}~\bibnamefont {Rastelli}},\ }\bibfield  {title} {\bibinfo {title}
  {{{Single-particle-picture breakdown in laterally weakly confining GaAs
  quantum dots}}},\ }\href {https://doi.org/10.1103/PhysRevB.100.235425}
  {\bibfield  {journal} {\bibinfo  {journal} {Phys. Rev. B}\ }\textbf {\bibinfo
  {volume} {100}},\ \bibinfo {pages} {235425} (\bibinfo {year}
  {2019})}\BibitemShut {NoStop}%
\bibitem [{\citenamefont {Yuan}\ \emph {et~al.}(2018)\citenamefont {Yuan},
  \citenamefont {Weyhausen-Brinkmann}, \citenamefont {Mart{\'i}n-S{\'a}nchez},
  \citenamefont {Piredda}, \citenamefont {K{\v{r}}{\'a}pek}, \citenamefont
  {Huo}, \citenamefont {Huang}, \citenamefont {Schimpf}, \citenamefont
  {Schmidt}, \citenamefont {Edlinger}, \citenamefont {Bester}, \citenamefont
  {Trotta},\ and\ \citenamefont {Rastelli}}]{Yuan2018}%
  \BibitemOpen
  \bibfield  {author} {\bibinfo {author} {\bibfnamefont {X.}~\bibnamefont
  {Yuan}}, \bibinfo {author} {\bibfnamefont {F.}~\bibnamefont
  {Weyhausen-Brinkmann}}, \bibinfo {author} {\bibfnamefont {J.}~\bibnamefont
  {Mart{\'i}n-S{\'a}nchez}}, \bibinfo {author} {\bibfnamefont {G.}~\bibnamefont
  {Piredda}}, \bibinfo {author} {\bibfnamefont {V.}~\bibnamefont
  {K{\v{r}}{\'a}pek}}, \bibinfo {author} {\bibfnamefont {Y.}~\bibnamefont
  {Huo}}, \bibinfo {author} {\bibfnamefont {H.}~\bibnamefont {Huang}}, \bibinfo
  {author} {\bibfnamefont {C.}~\bibnamefont {Schimpf}}, \bibinfo {author}
  {\bibfnamefont {O.~G.}\ \bibnamefont {Schmidt}}, \bibinfo {author}
  {\bibfnamefont {J.}~\bibnamefont {Edlinger}}, \bibinfo {author}
  {\bibfnamefont {G.}~\bibnamefont {Bester}}, \bibinfo {author} {\bibfnamefont
  {R.}~\bibnamefont {Trotta}},\ and\ \bibinfo {author} {\bibfnamefont
  {A.}~\bibnamefont {Rastelli}},\ }\bibfield  {title} {\bibinfo {title}
  {Uniaxial stress flips the natural quantization axis of a quantum dot for
  integrated quantum photonics},\ }\href
  {https://doi.org/10.1038/s41467-018-05499-5} {\bibfield  {journal} {\bibinfo
  {journal} {Nature Commun.}\ }\textbf {\bibinfo {volume} {9}},\ \bibinfo
  {pages} {3058} (\bibinfo {year} {2018})}\BibitemShut {NoStop}%
\bibitem [{\citenamefont {Huo}\ \emph {et~al.}(2013)\citenamefont {Huo},
  \citenamefont {Rastelli},\ and\ \citenamefont {Schmidt}}]{HuoAPL2013}%
  \BibitemOpen
  \bibfield  {author} {\bibinfo {author} {\bibfnamefont {Y.~H.}\ \bibnamefont
  {Huo}}, \bibinfo {author} {\bibfnamefont {A.}~\bibnamefont {Rastelli}},\ and\
  \bibinfo {author} {\bibfnamefont {O.~G.}\ \bibnamefont {Schmidt}},\
  }\bibfield  {title} {\bibinfo {title} {Ultra-small excitonic fine structure
  splitting in highly symmetric quantum dots on {GaAs} {(001)} substrate},\
  }\href {https://doi.org/http://dx.doi.org/10.1063/1.4802088} {\bibfield
  {journal} {\bibinfo  {journal} {Appl. Phys. Lett.}\ }\textbf {\bibinfo
  {volume} {102}},\ \bibinfo {eid} {152105} (\bibinfo {year}
  {2013})}\BibitemShut {NoStop}%
\bibitem [{\citenamefont {Waeber}\ \emph {et~al.}(2016)\citenamefont {Waeber},
  \citenamefont {Hopkinson}, \citenamefont {Farrer}, \citenamefont {Ritchie},
  \citenamefont {Nilsson}, \citenamefont {Stevenson}, \citenamefont {Bennett},
  \citenamefont {Shields}, \citenamefont {Burkard}, \citenamefont
  {Tartakovskii}, \citenamefont {Skolnick},\ and\ \citenamefont
  {Chekhovich}}]{Waeber2016}%
  \BibitemOpen
  \bibfield  {author} {\bibinfo {author} {\bibfnamefont {A.~M.}\ \bibnamefont
  {Waeber}}, \bibinfo {author} {\bibfnamefont {M.}~\bibnamefont {Hopkinson}},
  \bibinfo {author} {\bibfnamefont {I.}~\bibnamefont {Farrer}}, \bibinfo
  {author} {\bibfnamefont {D.~A.}\ \bibnamefont {Ritchie}}, \bibinfo {author}
  {\bibfnamefont {J.}~\bibnamefont {Nilsson}}, \bibinfo {author} {\bibfnamefont
  {R.~M.}\ \bibnamefont {Stevenson}}, \bibinfo {author} {\bibfnamefont {A.~J.}\
  \bibnamefont {Bennett}}, \bibinfo {author} {\bibfnamefont {A.~J.}\
  \bibnamefont {Shields}}, \bibinfo {author} {\bibfnamefont {G.}~\bibnamefont
  {Burkard}}, \bibinfo {author} {\bibfnamefont {A.~I.}\ \bibnamefont
  {Tartakovskii}}, \bibinfo {author} {\bibfnamefont {M.~S.}\ \bibnamefont
  {Skolnick}},\ and\ \bibinfo {author} {\bibfnamefont {E.~A.}\ \bibnamefont
  {Chekhovich}},\ }\bibfield  {title} {\bibinfo {title} {{Few-second-long
  correlation times in a quantum dot nuclear spin bath probed by frequency-comb
  nuclear magnetic resonance spectroscopy}},\ }\href
  {https://doi.org/10.1038/nphys3686} {\bibfield  {journal} {\bibinfo
  {journal} {Nat. Phys.}\ }\textbf {\bibinfo {volume} {12}},\ \bibinfo {pages}
  {688} (\bibinfo {year} {2016})}\BibitemShut {NoStop}%
\bibitem [{\citenamefont {Urbaszek}\ \emph {et~al.}(2007)\citenamefont
  {Urbaszek}, \citenamefont {Braun}, \citenamefont {Amand}, \citenamefont
  {Krebs}, \citenamefont {Belhadj}, \citenamefont {Lema\'{\i}tre},
  \citenamefont {Voisin},\ and\ \citenamefont {Marie}}]{Urbaszek2007}%
  \BibitemOpen
  \bibfield  {author} {\bibinfo {author} {\bibfnamefont {B.}~\bibnamefont
  {Urbaszek}}, \bibinfo {author} {\bibfnamefont {P.-F.}\ \bibnamefont {Braun}},
  \bibinfo {author} {\bibfnamefont {T.}~\bibnamefont {Amand}}, \bibinfo
  {author} {\bibfnamefont {O.}~\bibnamefont {Krebs}}, \bibinfo {author}
  {\bibfnamefont {T.}~\bibnamefont {Belhadj}}, \bibinfo {author} {\bibfnamefont
  {A.}~\bibnamefont {Lema\'{\i}tre}}, \bibinfo {author} {\bibfnamefont
  {P.}~\bibnamefont {Voisin}},\ and\ \bibinfo {author} {\bibfnamefont
  {X.}~\bibnamefont {Marie}},\ }\bibfield  {title} {\bibinfo {title} {Efficient
  dynamical nuclear polarization in quantum dots: Temperature dependence},\
  }\href {https://doi.org/10.1103/PhysRevB.76.201301} {\bibfield  {journal}
  {\bibinfo  {journal} {Phys. Rev. B}\ }\textbf {\bibinfo {volume} {76}},\
  \bibinfo {pages} {201301} (\bibinfo {year} {2007})}\BibitemShut {NoStop}%
\bibitem [{\citenamefont {Villas-B\^oas}\ \emph {et~al.}(2005)\citenamefont
  {Villas-B\^oas}, \citenamefont {Ulloa},\ and\ \citenamefont
  {Govorov}}]{VillasBoas2005}%
  \BibitemOpen
  \bibfield  {author} {\bibinfo {author} {\bibfnamefont {J.~M.}\ \bibnamefont
  {Villas-B\^oas}}, \bibinfo {author} {\bibfnamefont {S.~E.}\ \bibnamefont
  {Ulloa}},\ and\ \bibinfo {author} {\bibfnamefont {A.~O.}\ \bibnamefont
  {Govorov}},\ }\bibfield  {title} {\bibinfo {title} {Decoherence of rabi
  oscillations in a single quantum dot},\ }\href
  {https://doi.org/10.1103/PhysRevLett.94.057404} {\bibfield  {journal}
  {\bibinfo  {journal} {Phys. Rev. Lett.}\ }\textbf {\bibinfo {volume} {94}},\
  \bibinfo {pages} {057404} (\bibinfo {year} {2005})}\BibitemShut {NoStop}%
\bibitem [{\citenamefont {Adachi}(2009)}]{Adachi2009}%
  \BibitemOpen
  \bibfield  {author} {\bibinfo {author} {\bibfnamefont {S.}~\bibnamefont
  {Adachi}},\ }\href@noop {} {\emph {\bibinfo {title} {{Properties of
  Semiconductor Alloys: Group-$IV$, $III$-$V$ and $II$-$VI$ Semiconductors}}}}\
  (\bibinfo  {publisher} {Wiley},\ \bibinfo {year} {2009})\BibitemShut
  {NoStop}%
\bibitem [{\citenamefont {Wenckebach}(2008)}]{Wenckebach2008}%
  \BibitemOpen
  \bibfield  {author} {\bibinfo {author} {\bibfnamefont {W.~T.}\ \bibnamefont
  {Wenckebach}},\ }\bibfield  {title} {\bibinfo {title} {The solid effect},\
  }\href {https://doi.org/10.1007/s00723-008-0121-9} {\bibfield  {journal}
  {\bibinfo  {journal} {Applied Magnetic Resonance}\ }\textbf {\bibinfo
  {volume} {34}},\ \bibinfo {pages} {227} (\bibinfo {year} {2008})}\BibitemShut
  {NoStop}%
\bibitem [{\citenamefont {Ragunathan}(2019)}]{Ragunathan2019Thesis}%
  \BibitemOpen
  \bibfield  {author} {\bibinfo {author} {\bibfnamefont {G.}~\bibnamefont
  {Ragunathan}},\ }\emph {\bibinfo {title} {{Nuclear Spin Phenomena in III-V
  and II-VI Semiconductor Quantum Dots}}},\ \href
  {https://etheses.whiterose.ac.uk/25466/1/191028GauThesisFinalCorrections.pdf}
  {Ph.D. thesis},\ \bibinfo  {school} {University of Sheffield} (\bibinfo
  {year} {2019})\BibitemShut {NoStop}%
\bibitem [{\citenamefont {Janzen}(1973)}]{Janzen1973}%
  \BibitemOpen
  \bibfield  {author} {\bibinfo {author} {\bibfnamefont {W.}~\bibnamefont
  {Janzen}},\ }\bibfield  {title} {\bibinfo {title} {Adiabatic rapid passage
  {{NMR}} signal shape and passage conditions in solids},\ }\href
  {https://doi.org/https://doi.org/10.1016/0022-2364(73)90104-2} {\bibfield
  {journal} {\bibinfo  {journal} {Journal of Magnetic Resonance (1969)}\
  }\textbf {\bibinfo {volume} {12}},\ \bibinfo {pages} {71} (\bibinfo {year}
  {1973})}\BibitemShut {NoStop}%
\bibitem [{\citenamefont {Goldman}\ \emph {et~al.}(1975)\citenamefont
  {Goldman}, \citenamefont {Jacquinot}, \citenamefont {Chapellier},\ and\
  \citenamefont {Chau}}]{Goldman1975}%
  \BibitemOpen
  \bibfield  {author} {\bibinfo {author} {\bibfnamefont {M.}~\bibnamefont
  {Goldman}}, \bibinfo {author} {\bibfnamefont {J.}~\bibnamefont {Jacquinot}},
  \bibinfo {author} {\bibfnamefont {M.}~\bibnamefont {Chapellier}},\ and\
  \bibinfo {author} {\bibfnamefont {V.~H.}\ \bibnamefont {Chau}},\ }\bibfield
  {title} {\bibinfo {title} {Nonlinear effects in spin temperature},\ }\href
  {https://doi.org/https://doi.org/10.1016/0022-2364(75)90220-6} {\bibfield
  {journal} {\bibinfo  {journal} {Journal of Magnetic Resonance (1969)}\
  }\textbf {\bibinfo {volume} {18}},\ \bibinfo {pages} {22} (\bibinfo {year}
  {1975})}\BibitemShut {NoStop}%
\bibitem [{\citenamefont {Cowan}(1998)}]{CowanBook}%
  \BibitemOpen
  \bibfield  {author} {\bibinfo {author} {\bibfnamefont {G.}~\bibnamefont
  {Cowan}},\ }\href@noop {} {\emph {\bibinfo {title} {{Statistical Data
  Analysis}}}}\ (\bibinfo  {publisher} {Clarendon Press},\ \bibinfo {year}
  {1998})\BibitemShut {NoStop}%
\end{thebibliography}
\end{document}